\documentclass{aims}
\usepackage{amsmath}
  \usepackage{paralist}
  \usepackage{graphics} 
  \usepackage{epsfig} 
\usepackage{graphicx}  \usepackage{epstopdf}
 \usepackage[colorlinks=true]{hyperref}
\hypersetup{urlcolor=blue, citecolor=red}
\usepackage{cite}

\def\currentvolume{X}
 \def\currentissue{X}
  \def\currentyear{200X}
   \def\currentmonth{XX}
    \def\ppages{X--XX}

\theoremstyle{definition}

\newtheorem{remark}{Remark}

\newcommand{\be}{\begin{equation}}
\newcommand{\ee}{\end{equation}}

\newcommand{\LL}{\Lambda}

\renewcommand{\aa}{{a}}
\newcommand{\as}{{a}}
\newcommand{\z}{\mathbf{z}}
\newcommand{\fI}{i}
\newcommand{\fR}{r}
\newcommand{\fS}{s}
\newcommand{\fE}{e}
\newcommand{\fD}{d}
\newcommand{\h}{h}

\title[Modelling lockdown in epidemic outbreaks with uncertainty] 
      {Modelling lockdown measures in epidemic outbreaks using selective socio-economic containment with uncertainty}

\author[Giacomo Albi and Lorenzo Pareschi and Mattia Zanella]{}


 \email{giacomo.albi@univr.it}
 \email{lorenzo.pareschi@unife.it}
 \email{mattia.zanella@unipv.it}

\thanks{$^*$ Corresponding author: Giacomo Albi}

\begin{document}
\maketitle

\centerline{\scshape Giacomo Albi $^*$}
\medskip
{\footnotesize
 \centerline{Department of Computer Science}
   \centerline{University of Verona, Italy}
} 

\medskip

\centerline{\scshape Lorenzo Pareschi}
\medskip
{\footnotesize
 \centerline{Department of Mathematics and Computer Science}
   \centerline{University of Ferrara, Italy}
}

\medskip

\centerline{\scshape Mattia Zanella}
\medskip
{\footnotesize
 \centerline{Department of Mathematics "F. Casorati"}
   \centerline{University of Pavia, Italy}
}

\bigskip


\begin{abstract}
After the introduction of drastic containment measures aimed at stopping the epidemic contagion from SARS-CoV2, many governments have adopted a strategy based on a periodic relaxation of such measures in the face of a severe economic crisis caused by  lockdowns. Assessing the impact of such openings in relation to the risk of a resumption of the spread of the disease is an extremely difficult problem due to the many unknowns concerning the actual number of people infected, the actual reproduction number and infection fatality rate of the disease. In this work, starting from a compartmental model with a social structure and stochastic inputs, we derive models with multiple feedback controls depending on the social activities that allow to assess the impact of a selective relaxation of the containment measures in the presence of uncertain data. Specific contact patterns in the home, work, school and other locations have been considered. Results from different scenarios concerning the first wave of the epidemic in some major countries, including Germany, France, Italy, Spain, the United Kingdom and the United States, are presented and discussed. 
\end{abstract}

\section{Introduction}
It is now clear that the end of the pandemic will not immediately correspond to the disappearance of SARS-CoV2 and that until a vaccination campaign is completed on a global scale we will have to deal with several measures of social distancing and containment. This is why various intermediate phases have been carefully considered, with some activities that can be resumed, regulating the reintegration of workers, for example through indicators measuring the impact of work activities on potential infections, increasing prevention measures, or through so-called immunity passports. It is essential to build scenarios that will help us understand how the situation might evolve in the future.

Among the many controversial aspects are, for example, the reopening of schools, sport activities and other social activities at different levels, which, while having less economic impact, have a very high social cost.  Indeed, it is clear that it is difficult for the population to sustain an excessively long period of lockdown. It is therefore of primary importance to analyze the impact of relaxing the control measures put in place by many countries in order to make them more sustainable on the socio-economic front, keeping the reproductive rate of the epidemic under control and without incurring health risks \cite{Gatto, DPTZ, FIR, Tang}.

The problem is clearly very challenging, traditional epidemiological models based on the assumption of homogeneous population mixing are inadequate, since the whole social and economic structure of the country is involved \cite{KMK, H00,CHALL89,H96,FP,IMP}. On the other hand, interventions involving the whole population allow to use mathematical descriptions in analogy with classical statistical physics drawing on the statistical characteristics of a very large system of interacting individuals \cite{Albi1, Albi2, Albi3, BFK, DPT,Wang}. 

A further problem that cannot be ignored is the uncertainty present in the official data provided by the different countries in relation to the number of infected people. The heterogeneity of the procedures used to carry out the disease positivity tests, the delays in recording and reporting the results, and the large percentage of asymptomatic patients (in varying percentages depending on the studies and the countries but estimated by WHO at an average of around $80\%$ of cases) make the construction of predictive scenarios affected by high uncertainty \cite{JRGL,MKZC,Zhang_etal}. As a consequence, the actual number of infected and  recovered people is typically underestimated, causing fatal delays in the implementation of public health policies facing the propagation of epidemic fronts. 

In this research, we try to make a contribution to these problems starting from a description of the spread of the epidemic based on a compartmental model with social structure in the presence of uncertain data.  The presence of a social characteristic such as the age of individuals is, in fact, essential in the case of the COVID-19 outbreak to characterize the heterogeneity of the impact of infection in relation to age. In addition, the model allows to take into account the specific nature of the different activities involved through appropriate interaction functions derived from  experimental interaction matrices \cite{Betal, Metal,PCJ,GFMDC12} and to systematically include the uncertainty present in the data \cite{Capaldi_etal, CCVH, Chowell, JRGL, MKZC, Roberts}.

The latter property is achieved by increasing the dimensionality of the problem adding the possible sources of uncertainty from the very beginning of the modelling.
Hence, we extrapolate statistics by looking at the so-called quantities of interest, i.e. statistical quantities that can be obtained from the solution and that give some global information with respect to the input parameters. Several techniques can be adopted for the approximation of the quantities of interest. Here, following \cite{APZ1} we adopt stochastic Galerkin methods that allow to reduce the problem to a set of deterministic equations for the numerical evaluation of the solution in presence of uncertainties \cite{X, PareschiUQ, DPZ}.

The main assumption made in this study is that the control measures adopted by the different countries cannot be described by the standard compartmental model but must necessarily be seen as external actions carried out by policy makers in order to reduce the epidemic peak. Most current research in this direction has focused on control procedures aimed at optimizing the use of vaccinations and medical treatments \cite{BGO, BBSG, SCCC10, CG,Donofrio, LGC12} and only recently the problem has been tackled from the perspective of non-pharmaceutical interventions \cite{APZ1, LML10, MRPL, Franco, IC, DGKP}. For this purpose we derive new models based on multiple feedback controls that act selectively on each specific contact function and therefore social activity. Based on the data in \cite{PCJ} this allows to analyze the impact of containment measures in a differentiated way on family, work, school, and other activities. 

In our line of approach, the classical epidemiological parameters that define the rate of reproduction of the infectious disease are therefore estimated only in the regime prior to the first lockdown and define an estimate of the reproductive rate in the absence of control. At this stage the estimation mainly serves to calibrate the model parameters and its variability will then be considered in the intrinsic uncertainty of these values. In particular, this makes it possible to introduce the role of the asymptomatic population without adding additional compartments but directly via the stochastic component in the number of infected persons. The control action is then estimated in the first lockdown phase using the data available. On the modelling front, we next focus our interest on the phase following the first lockdown period, in which social characteristics become essential to quantify the impact of possible government decisions. 

This makes it possible to carry out a systematic analysis for different countries and to observe the different behaviour of the control action in line with the dynamics observed and the measures taken by different governments. Of course, a realistic comparison between countries is an extremely difficult problem that would require a complex phase of renormalization of the data according to the different recording and acquisition methods used.  In an attempt to provide comparative results altered as little as possible by assumptions that cannot be justified, we decided to adopt the same criteria for each country and therefore the scenarios presented, although based on realistic values, maintain a primarily qualitative rather than quantitative nature.

We present different simulation scenarios for various countries where the first wave of the epidemic showed some similarities, including Germany, France, Italy, Spain, the United Kingdom and the United States analyzing the effect of relaxing the lockdown measures in a selective way on the various social activities. 
Although the choice of which specific activities to reopen remains mainly a political decision, numerical simulations show that a progressive loosening strategy in subsequent phases, as adopted by some governments, may be capable to keep the epidemic under control by restarting various productive activities. 

The rest of the manuscript is organized as follows. In Section \ref{sect:model} we present compartmental models with social structure and with uncertainties where interaction matrices depend on various social activities. Next, a selective control mimicking  containment measures in relation to a specific social activity and in presence of model uncertainty is derived in Section \ref{sect:control}. Finally, in Section \ref{sect:numerics} we propose various numerical experiments based on several countries highlighting the importance of the social structure to evaluate possible relaxations in relation to specific social activities. 

\section{The epidemiological model}\label{sect:model}
The starting model in our discussion is a SEIRD-type compartmental model with a social structure and uncertain inputs. The presence of a social structure is in fact essential in deriving appropriate sustainable control techniques from the population for a protracted period, as in the case of the recent COVID-19 epidemic. In addition we include the effects on the dynamics of uncertain data, such as the initial conditions on the number of infected people or the interaction and recovery rates. This permits to include the role of the asymptomatic population directly in the uncertainty.

\subsection{A socially structured compartmental model with uncertainty}
The heterogeneity of the social structure, which impacts the diffusion of the infective disease, is characterized by $\aa\in \LL \subset \mathbb{R_+}$ representing the age of the individual \cite{H96,H00}. We assume that the rapid spread of the disease and the low mortality rate allows to ignore changes in the social structure, such as the aging process, births and deaths. Furthermore, we introduce the random vector $\z = (z_1,\dots,z_{d_z})\in\mathbb{R}^{d_z}$ whose components are assumed to be independent real valued random variables taking into account various possible sources of uncertainty in the model. We assume to know the probability density $p(\z): \mathbb R^{d_z} \rightarrow \mathbb R^{d_z}_+$ characterizing the distribution of $\z$. 

We denote by $s(\z,\as,t)$, $e(\z,\as,t)$, $i(\z,\as,t)$, $r(\z,\as,t)$ and $d(\z,\as,t)$ the densities at time $t\ge 0$ of susceptible, exposed, infectious, recovered and dead individuals, respectively in relation to their age $a$ and the source of uncertainty $\z$. The density of individuals of a given age $a$ and the total population number $N$ are deterministic conserved quantities in time, i.e.
\[
s(\z,\as,t)  + e(\z,\as,t) + i(\z,\as,t) + r(\z,\as,t)+ d(\z,\as,t) = f(\as), \qquad \int_{\LL} f(\as)d\as = N. 
\]
Hence, the quantities 
\begin{equation}
\begin{split}
S(\z,t)&=\int_{\LL}\fS (\z,\as,t)\,d\as,\quad E(\z,t)=\int_{\LL}\fE (\z,\as,t)\,d\as,\quad I(\z,t)=\int_{\LL}\fI (\z,\as,t)\,d\as,\\
R(\z,t)&=\int_{\LL}\fR (\z,\as,t)\,d\as,\quad D(\z,t)=\int_{\LL}\fD (\z,\as,t)\,d\as,
\end{split}
\end{equation}
denote the uncertain fractions of the population that are susceptible, exposed, infectious, recovered and dead respectively. 

In a situation where changes in the social features act on a slower scale with respect to the spread of the disease, the socially structured compartmental model with uncertainties follows the dynamics
\begin{equation}\label{eq:SIR_u}
\begin{split}
\frac{d}{dt} \fS(\z,\as,t)&= - \fS(\z,\as,t)\sum_{j\in \mathcal A} \int_{\LL} \beta_j(\z,\as,\as_*)\dfrac{\fI(\z,\as_*,t)}{N}\ d\as_* \\
\frac{d}{dt} \fE(\z,\as,t)&=  \fS(\z,\as,t)\sum_{j\in \mathcal A} \int_{\LL} \beta_j(\z,\as,\as_*)\dfrac{\fI(\z,\as_*,t)}{N}\ d\as_* - \sigma (\z,\as) \fE(\z,\as,t) \\
\frac{d}{dt} \fI(\z,\as,t) &= \sigma (\z,\as) \fE(\z,\as,t)  - (\gamma(\z,\as)+\alpha(\z,\as)) \fI(\z,\as,t) \\
\frac{d}{dt} \fR(\z,\as,t) &= \gamma (\z,\as) \fI(\z,\as,t)\\
\frac{d}{dt} \fD(\z,\as,t) &= \alpha (\z,\as) \fI(\z,\as,t)
\end{split}
\end{equation}
with initial condition $\fS(\z,\as,0) = \fS_0(\z,\as)$, $\fE(\z,\as,0) = \fE_0(\z,\as)$, $\fI(\z,\as,0) = \fI_0(\z,\as)$,  $\fR(\z,\as,0) = \fR_0(\z,\as)$ and $\fD(\z,\as,0) = \fD_0(\z,\as)$. In \eqref{eq:SIR_u} we assume age-dependent contact rates $\beta_j(\z,\aa,\aa_*) \geq 0$, $j\in\mathcal A$, representing transmission rates among individuals related to a specific activity characterized by the set $\mathcal A$, such as home, work, school, etc., $\gamma(\z,\aa) \geq 0$ is the recovery rate which may be age dependent, $\sigma (\z,\as) \geq 0$ is the transition rate of exposed individuals to the infected class, and $\alpha (\z,\as) \geq 0$ is the disease-induced death rate of infectious individuals.

In the following, we introduce the usual normalization scaling 
\[
\begin{split}
\frac{\fS(\z,\aa,t)}{N}&\to \fS(\z,\aa,t),\quad\frac{\fE(\z,\aa,t)}{N}\to \fE(\z,\aa,t),\quad\frac{\fI(\z,\aa,t)}{N}\to \fI(\z,\aa,t),\\ 
\frac{\fR(\z,\aa,t)}{N}&\to \fR(\z,\aa,t),\quad\frac{\fD(\z,\aa,t)}{N}\to \fD(\z,\aa,t),\quad\int_\Lambda f(\z,\aa) d\aa= 1, 
\end{split}
\]
and observe that the quantities $S(t)$, $E(t)$, $I(t)$, $R(t)$ and $D(t)$ satisfy the uncertain SEIRD dynamics
\begin{equation}\label{eq:SIRm}
\begin{split}
\frac{d}{dt} S(\z,t)&= - \sum_{j\in \mathcal A} \int_{\Lambda\times\Lambda}\!\!\!\! \beta_j(\z,\aa,\aa_*)\fS(\z,\aa,t)\fI(\z,\aa_*,t)\,d\aa_*d\aa \\
\frac{d}{dt} E(\z,t) &=  \sum_{j\in \mathcal A} \int_{\Lambda\times\Lambda}\!\!\!\! \beta_j(\z,\aa,\aa_*)\fS(\z,\aa,t)\fI(\z,\aa_*,t)\,d\aa_*d\aa - \int_\Lambda\sigma (\z,\aa) \fE(\z,\aa,t)d\aa\\
\frac{d}{dt} I(\z,t) &= \int_\Lambda\sigma (\z,\aa) \fE(\z,\aa,t)\,d\aa  - \int_\Lambda (\gamma(\z,\as)+\alpha(\z,\as)) \fI(\z,\as,t)\,d\aa \\
\frac{d}{dt} R(\z,t) &= \int_\Lambda\gamma(\z,\as) \fI(\z,\as,t)\,d\aa\\
\frac{d}{dt} D(\z,t) &= \int_\Lambda\alpha(\z,\as) \fI(\z,\as,t)\,d\aa.
\end{split}
\end{equation}
We refer to \cite{H00, H96, FP, IMP} for analytical results concerning model \eqref{eq:SIR_u} and \eqref{eq:SIRm} in a deterministic setting. 

Before entering the discussion of the control problem that formalizes the action of a policy maker aimed at reducing the epidemic impact, we discuss the role of the uncertainty in the model and how it relates to other compartmental models including the asymptomatic population.

\subsection{Relationship to compartmental models including undetected infectious}
One of the main difficulties in mathematical modelling of the COVID-19 epidemic is due to the presence of a large number of undetected (asymptomatic) infected individuals. This has motivated the construction of various models in which the  infected population is subdivided into further compartments with different roles in the spread of the disease \cite{Giordano, Gatto, Tang, IC}. To clarify the relationships to such models, let us consider model \eqref{eq:SIRm} in absence of a social structure and social activities 
\begin{equation}
\begin{split}
\frac{d}{dt} S(z,t)&= - \beta(z) S(z,t)I(z,t) \\
\frac{d}{dt} E(z,t) &= \beta(z)S(z,t)I(z,t)-\sigma(z) E(z,t),\\
\frac{d}{dt} I(z,t) &= \sigma(z) E(z,t)-(\gamma(z)+\alpha(z)) I(z,t),\\
\frac{d}{dt} R(z,t) &= \gamma(z)I(z,t),\\
\frac{d}{dt} D(z,t) &= \alpha(z) I(z,t),
\end{split}
\label{eq:homo3}
\end{equation}
and with a one-dimensional random input $z \in \mathbb{R}$ distributed as $p(z)$. 
 Furthermore, for a function $F(z,t)$ we will denote its expected value as $\bar F(t) = \mathbb{E}[F(\cdot,t)]$. Now, starting from a discrete probability density function
\[
p_k=P\left\{Z=z_k\right\},\qquad \sum_{k=1}^n p_k = 1,
\]
we have $\bar F(t) = \sum_{k=1}^n p_k F_k$, with $F_k=F(z_k)$. 
Taking the expectation in \eqref{eq:homo3}, we can write
\begin{equation}
\begin{split}
\frac{d}{dt} \bar S(t)&= - \bar S(t) \sum_{k=1}^n \tilde \beta_k  p_k I_k(t) \\
\frac{d}{dt} \bar E(t) &= \bar S(t) \sum_{k=1}^n \tilde \beta_k  p_k I_k(t)-\bar E(t)\sum_{k=1}^n \tilde\sigma_k p_k,\\
\frac{d}{dt} \bar I(t) &= \bar E(t)\sum_{k=1}^n \tilde\sigma_k p_k-\sum_{k=1}^n (\gamma_k+\alpha_k) p_k I_k(t),\\
\frac{d}{dt} \bar R(t) &= \sum_{k=1}^n \gamma_k p_k I_k(t),\\
\frac{d}{dt} \bar D(t) &= \sum_{k=1}^n \alpha_k p_k I_k(t),
\end{split}
\label{eq:homopar}
\end{equation}
with $\tilde \beta_k = S_k \beta_k/\bar S $, $\tilde \sigma_k = E_k \sigma_k/\bar E $, $k=1,\ldots,n$. For example, in the case $n=2$, by identifying $I_d=p_1 I_1$ and $I_u=p_2 I_2$ with the compartments of detected and undetected infectious individuals, assuming $\tilde \sigma_k=\tilde \sigma$, and denoting $p_1=\rho$ we can write
\begin{equation}
\begin{split}
\frac{d}{dt} \bar S(t)&= - \bar S(t) \left(\tilde \beta_1  I_d(t)+\tilde \beta_2  I_u(t)\right) \\
\frac{d}{dt} \bar E(t)&=  \bar S(t) \left(\tilde \beta_1  I_d(t)+\tilde \beta_2  I_u(t)\right) - \tilde \sigma \bar E(t) \\
\frac{d}{dt}  I_d(t) &= \tilde \sigma\rho \bar E(t) -(\gamma_1 +\alpha_1) I_d(t),\\
\frac{d}{dt}  I_u(t) &= \tilde \sigma(1-\rho) \bar E(t) -(\gamma_2 +\alpha_2) I_u(t),\\
\frac{d}{dt} \bar R(t) &= \gamma_1 I_d(t)+\gamma_2 I_u(t),\\
\frac{d}{dt} \bar D(t) &= \alpha_1 I_d(t)+\alpha_2 I_u(t),
\end{split}
\label{eq:SPIAR}
\end{equation}
which has the same structure of a SEIARD-type compartmental model including the undetected (or the asymptomatic) class \cite{Gatto, Tang}.

\section{Multiple control of structured compartmental model}\label{sect:control}

In order to characterize the action of a policy maker introducing a control over the system based on selective containment measures in relation to a specific social activity  we consider the following optimal control setting 
\begin{equation}\label{eq:func_z}
\min_{{\mathbf u}\in \mathcal U}J({\mathbf u}):= \int_0^T\!\!\mathcal R [\psi(S,I)(\cdot,t)] dt+\dfrac{1}{2}\sum_{j\in \mathcal A}\int_0^T\!\!\!\int_{\LL\times\LL}\!\!\!\! \nu_j(a,t)|u_j(\as,\as_*,t)|^2 d\as\, d\as_*  dt, 
\end{equation}
where ${\mathbf u}=(u_1,\ldots,u_L)$ is a vector of controls acting locally on the interaction between individuals of ages $a$ and $a_*$, the function $\nu_j(a,t) > 0$ is a selective penalization term and $\mathcal R[\cdot]$ is a suitable statistical operator taking into account the presence of the uncertainties, and $\psi(S,I)$, s.t. $\partial_I \psi(S,I)\geq 0$, is a function characterizing the policy maker's perception of the impact of the epidemic. 

Examples of such operator that are of interest in epidemic modelling are the expectation with respect to uncertainties
\begin{equation}
\label{eq:R1}
\mathcal R [\psi(S,I)(\cdot,t)]=\mathbb E[\psi(S,I)(\cdot,t)]= \int_{\mathbb R^{d_z}} \psi(S,I)(\z,t) \; p(\z)d\z
\end{equation} 
or relying on deterministic data which underestimate the number of infected
\begin{equation} 
\label{eq:R2}
\mathcal R [\psi(S,I)(\cdot,t)]=\psi(S,I)(\z_0,t),
\end{equation}
where $\z_0$ is a given value such that $I(\z_0,t) \leq I(\z,t)$, $\forall\, \z\in \mathbb{R}^{d_z}$ and $t>0$.
Concerning the perception function, in the sequel we will consider two relevant examples given by a convex function underestimating the number of infected 
\begin{equation} 
\label{eq:psi1}
\psi(S,I)(\z,t)=C\frac{I^q(\z,t)}{q},\qquad q \geq 1,
\end{equation}
and a concave function overestimating such number
\begin{equation} 
\label{eq:psi2}
\psi(S,I)(\z,t)=C\frac{\ln(1+\tau I(\z,t))}{\tau S(\z,t)},\quad \tau > 0,
\end{equation}
with $C>0$ a suitable renormalization constant. The function in \eqref{eq:psi1} has been introduced in \cite{APZ1} and we will rely on the same arguments in deriving the corresponding feedback controlled model, whereas the function in \eqref{eq:psi2} permits to recover as feedback controlled models well-known epidemic models with nonlinear transmission rates \cite{Franco, Maini}.

In \eqref{eq:func_z} the set $\mathcal U \subseteq \mathbb{R}^L$ is the space of admissible controls $u_j$, $j\in \mathcal A$ defined as 
\[
\mathcal U=\left\{{\mathbf u}\in\mathbb{R}^L\,|\, 0 \leq \mathcal{I}(u_j)(\as,t) \leq\min\{M,\min_{\z}\mathcal{I}(\beta_j)(\z,\as,t)\},\,\, \forall\, (\as,t),\, M>0\right\},
\] 
where
\[
\begin{split}
\mathcal{I}(u_j)(\as,t)=\frac1{I(\z,t)}{\int_{\LL} u_j(\as,\as_*,t)\fI(\z,\as_*,t)\,da_*},\\ \mathcal{I}(\beta_j)(\z,\as,t) = \frac1{I(\z,t)}{\int_{\LL} \beta_j(\z,\as,\as_*)\fI(\z,\as_*,t)\,da_*}.
\end{split}
\]
Note that,  here we are considering less restrictive conditions on the space of admissible controls than in \cite{APZ1}. 
The above minimization is subject to the following dynamics
\begin{equation}\label{eq:SIR_z}
\begin{split}
\frac{d}{dt} \fS(\z,\as,t)&= - \fS(\z,\as,t)\sum_{j\in \mathcal A} \int_{\LL} (\beta_j(\z,\as,\as_*) -u_j(\as,\as_*,t))\dfrac{\fI(\z,\as_*,t)}{N}\ d\as_* \\
\frac{d}{dt} \fE(\z,\as,t)&=  \fS(\z,\as,t)\sum_{j\in \mathcal A} \int_{\LL} (\beta_j(\z,\as,\as_*) -u_j(\as,\as_*,t))\dfrac{\fI(\z,\as_*,t)}{N}\ d\as_*\\
&\quad - \sigma (\z,\as) \fE(\z,\as,t) \\
\frac{d}{dt} \fI(\z,\as,t) &= \sigma (\z,\as) \fE(\z,\as,t)  - (\gamma(\z,\as)+\alpha(\z,\as)) \fI(\z,\as,t),
\end{split}
\end{equation}
where for simplicity we omitted the equations for $\fR(\z,\as,t)$ and $\fD(\z,\as,t)$ since they do not affect directly the above system.  

Solving the above optimization problem, however, is generally quite complicated and computationally demanding when there are uncertainties as it involves solving simultaneously the forward problem \eqref{eq:func_z}- \eqref{eq:SIR_z} and the backward problem derived from the optimality conditions \cite{APZ1}. Furthermore, the assumption that the policy maker follows an optimal strategy over a long time horizon seems rather unrealistic in the case of a rapidly spreading disease such as the COVID-19 epidemic.  

\subsection{Feedback controlled compartmental models with uncertainty}\label{sect:ins}
In this section we consider short time horizon strategies which permits to derive suitable feedback controlled models. These strategies are suboptimal with respect the original problem \eqref{eq:func_z}-\eqref{eq:SIR_z} but they have proved to be very successful in several social modeling problems \cite{Albi1, Albi2, Albi3, DPT}. To this aim, we consider a short time horizon of length $h>0$ and formulate a time discretize optimal control problem through the functional $J_h(u)$ restricted to the interval $[t,t+\h]$, as follows
\begin{equation}\label{eq:func_I}
\min_{{\mathbf u}\in\mathcal U} J_h({\mathbf u}) = {\mathcal R}[\psi(S(\cdot,t),I(\cdot,t+\h))]+\frac{1}{2}\sum_{j\in \mathcal A}\int_{\Lambda\times\Lambda}\nu_j(a,t)|u_j(\as,\as_*,t)|^2d\as d\as_*,
\end{equation}
subject to
\begin{equation}
\label{eq:SIR_I}
\begin{split}
\fS(\z,\as,t+\h) &=  \fS(\z,\as,t) - \h{\fS(\z,\as,t)}\sum_{j\in \mathcal A} \!\int_{\LL}\!\! \left(\beta_j(\z,\as,\as_*) - u_j(\as,\as_*,t)\right)\fI(\z,\as_*,t) d\as_* \\
\fE(\z,\as,t+\h) &= \fE(\z,\as,t)  + \h{\fS(\z,\as,t)} \sum_{j\in \mathcal A}\!\int_{\LL}\!\! \left(\beta_j(\z,\as,\as_*) - u_j(\as,\as_*,t)\right)\fI(\z,\as_*,t) d\as_*\\
& \quad - \h\sigma(\z,\as) \fE(\z,\as,t),\\
\fI(\z,\as,t+\h) &= \fI(\z,\as,t)  + \h\sigma(\z,\as) \fE(\z,\as,t+h)-h(\gamma(\z,\as)+\alpha(\z,\as)) \fI(\z,\as,t).
\end{split}
\end{equation}
Recalling that the macroscopic information on the infected is
\[
\begin{split}
I(\z,t+\h)&= I(\z,t)+ \h \int_{\LL} \sigma(\z,\as) \fE(\z,\as,t+h)\,d\as - h \int_{\LL} (\gamma(\z,\as)+\alpha(\z,\as)) \fI(\z,\as,t)\,d\as
\end{split}
\] 
we can derive the minimizer of $J_h$ computing $\nabla_{{\mathbf u}} J_h({\mathbf u})\equiv 0$ or equivalently 
\[
\frac{\partial J_h({\mathbf u})}{\partial u_j} =0,\quad j\in \mathcal A.
\]
Using \eqref{eq:func_I} we can compute 
\[
{\mathcal R}\left[\frac{\partial \psi(S(\cdot,t),I(\cdot,t+\h))}{\partial u_j}\right] = {\nu_j(a,t)}u_j (\as,\as_*,t),
\]
where we assumed ${\partial {\mathcal R}\left[\psi(\cdot,\cdot)\right]}/{\partial u_j}={\mathcal R}\left[{\partial \psi(\cdot,\cdot)}/{\partial u_j}\right]$, to obtain the following nonlinear identities
\begin{align}\label{eq:ic}
{\nu_j(a,t)}u_j (\as,\as_*,t)={\h^2}{\mathcal R}[\sigma(\cdot,\as)\fS(\cdot,\as,t) i(\cdot,\as_*,t)\partial_I\psi(S(\cdot,t),I(\cdot,t+\h))].
\end{align}
The above assumption on $\mathcal R[\cdot]$ is clearly satisfied by \eqref{eq:R1} and \eqref{eq:R2}. Introducing the scaling $\nu_j(a,t) = \h^2 \kappa_j(\as,\as_*,t)$ we obtain the instantaneous control
\begin{equation}\label{eq:ic0}
u_j (\as,\as_*,t)=\frac{ 1}{\kappa_j(\as,\as_*)}{\mathcal R}[\sigma(\cdot,\as)\fS(\cdot,\as,t) i(\cdot,\as_*,t)\partial_I\psi(S(\cdot,t),I(\cdot,t+\h))].
\end{equation}
Now, passing to the limit for $h\to 0$ into the discrete system \eqref{eq:SIR_I}  we obtain the feedback controlled system \eqref{eq:SIR_z} with the instantaneous control term \eqref{eq:ic0}.

Let us now, report explicit expressions of the control term for the perception function \eqref{eq:psi1} and \eqref{eq:psi2}. In the convex case we have
\begin{equation}\label{eq:ic1}
u_j (\as,\as_*,t)=\frac{C}{\kappa_j(\as,\as_*)}{\mathcal R}[\sigma(\cdot,\as)\fS(\cdot,\as,t) i(\cdot,\as_*,t)I^{q-1}(\cdot,t)],
\end{equation}
whereas in the logarithmic case 
\begin{equation}\label{eq:ic2}
u_j (\as,\as_*,t)=\frac{C}{\kappa_j(\as,\as_*)}{\mathcal R}\left[\sigma(\cdot,\as)\fS(\cdot,\as,t) i(\cdot,\as_*,t)
\frac{1}{S(\cdot,t)(1+\tau I(\cdot,t))}\right].
\end{equation}
In the sequel we will restrict our attention to feedback controlled models of the form \eqref{eq:ic1} for $q=1$, namely there is no bias in the perception of the infectious disease from the policy maker, and where ${\mathcal R}[\cdot]$ is given by \eqref{eq:R2} corresponding to the number of reported cases.
 
\begin{remark}
To understand the action of the feedback controls \eqref{eq:ic1}-\eqref{eq:ic2}, let us consider the simplest case of a standard SEIRD model without age dependence, specific social interactions and uncertainty. In this simplified setting, it is easy to verify that the corresponding feedback controlled model has the same SEIRD structure with the modified transmission rate
\be
\tilde\beta=\beta-\frac{\sigma S(t)I(t)\partial_I \psi(S(t),I(t))}{\kappa}, 
\ee
that takes the specific form
\[
\tilde\beta=\beta-\frac{C \sigma S(t) I(t)^q}{\kappa} = \beta\left(1-\frac{S(t) I(t)^q}{\kappa}\right),
\]
in the case \eqref{eq:psi1} assuming $C=\beta/\sigma$, and
\[
\tilde\beta = \beta-\frac{C\sigma I(t)}{\kappa(1+\tau I(t))} = \frac{\beta}{1+\tau I(t)},
\]
in the case \eqref{eq:psi2} taking $\tau=1/\kappa$, $C=\beta/\sigma$. These correspond to the nonlinear incidence transmissions considered in \cite{APZ1, XLiu} and \cite{Franco, Maini, CS78}, respectively. Other nonlinear incidence  rates may be obtained similarly by considering different perception functions (see \cite{XLiu} and the references therein).
\end{remark}

\subsection{Application to the COVID-19 outbreak}
\label{sect:numerics}

In this section, we first present a comparison between different control strategies considering both a SEIR and SIR compartmentalization. Subsequently, we focus on the application of the feedback controlled models with uncertain data, that takes into account the presence of unreported symptomatic and asymptomatic cases, to the first wave of the COVID-19 pandemic in different countries. 
Details of the stochastic Galerkin method used to deal efficiently with uncertain data may be found in \cite{APZ1, PareschiUQ}. The data concerning the actual number of infected, recovered and deaths in the various country have been taken from the Johns Hopkins University Github repository \cite{Dong} ad for the specific case of Italy from the Github repository of the Italian Civil Protection Department \footnote{Presidenza del Consiglio dei Ministri, Dipartimento della Protezione Civile. GitHub: COVID-19 Italia - Monitoraggio situazione, \texttt{https://github.com/pcm-dpc/COVID-19}, 2020}. The social interaction functions $\beta_j$ have been reconstructed from the dataset of age and location specific contact matrices related to home, work, school and other activities in \cite{PCJ}. 
Finally, the demographic characteristics of the population for the various country have been taken from the United Nations World Populations Prospects \footnote{\tt https://population.un.org/wpp/}. Other sources of data which have been used include the Coronavirus disease (COVID-2019) situation reports of the WHO \footnote{{\tt https://www.who.int/emergencies/diseases/novel-coronavirus-2019/}} and the Statistic and Research Coronavirus Pandemic (COVID-19) from OWD \footnote{{\tt https://ourworldindata.org/coronavirus}}.

\subsection{Containment in homogeneous social mixing}

In the first test case, we will not attempt to analyze the data in a quantitative setting, but will compare the behaviour of the feedback controlled models with different controls of the form defined in \eqref{eq:ic1}-\eqref{eq:ic2}. Furthermore, to simplify the modeling we neglect any dependence on uncertainties and we consider the case of homogeneous social mixing. 

\begin{figure}
\includegraphics[scale = 0.25]{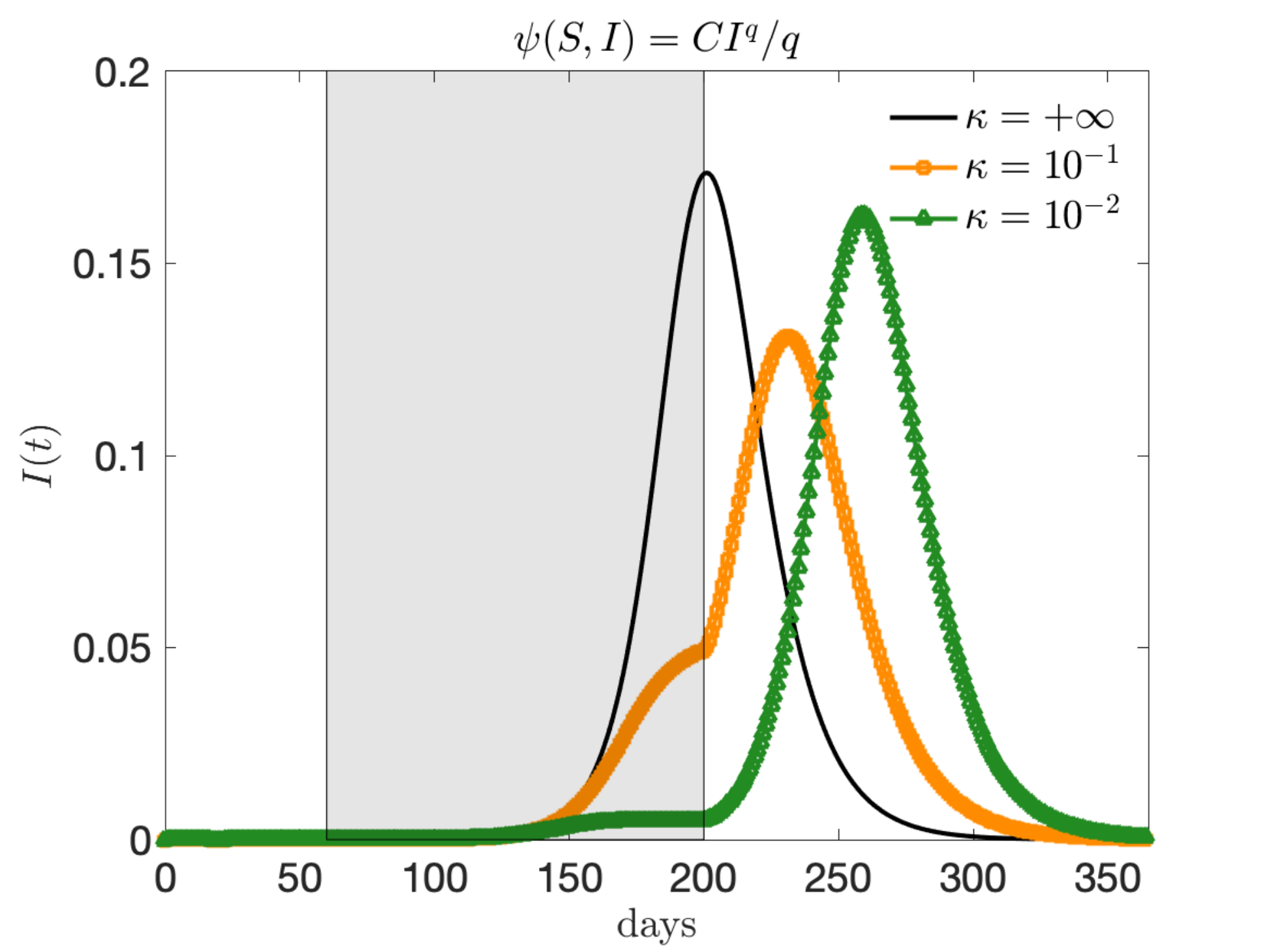}
\includegraphics[scale = 0.25]{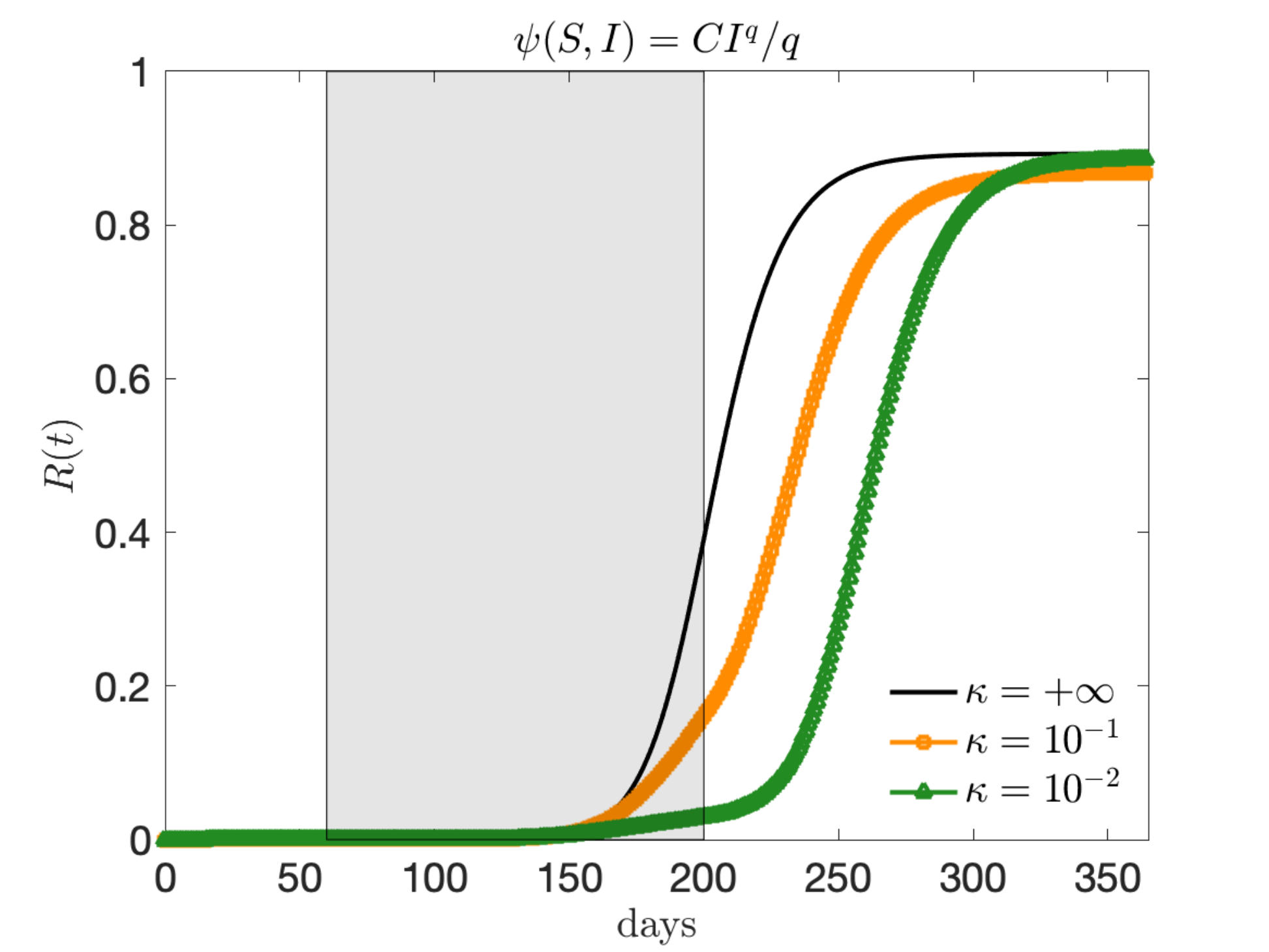}\\
\includegraphics[scale = 0.25]{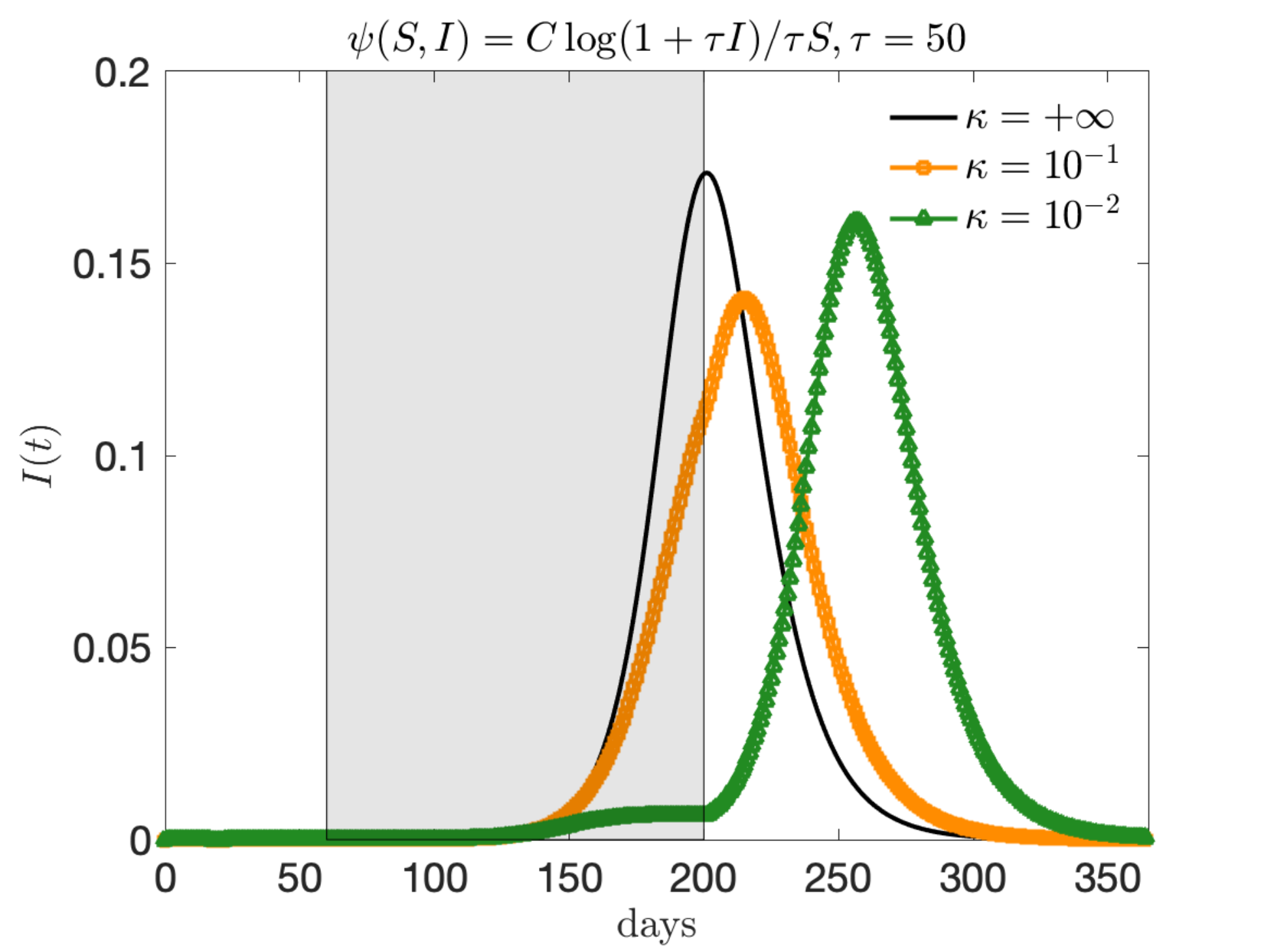}
\includegraphics[scale = 0.25]{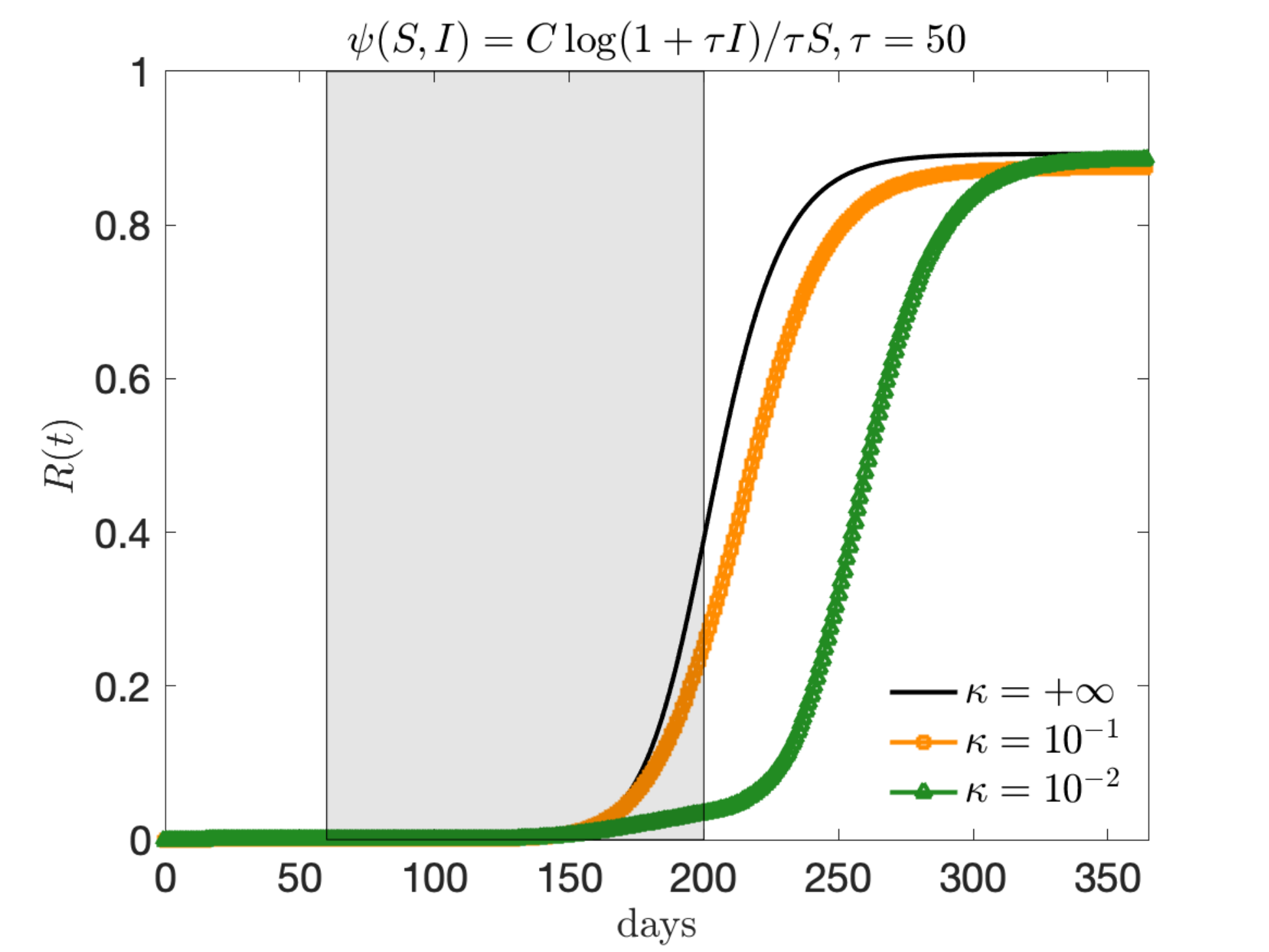} \\
 \includegraphics[scale = 0.25]{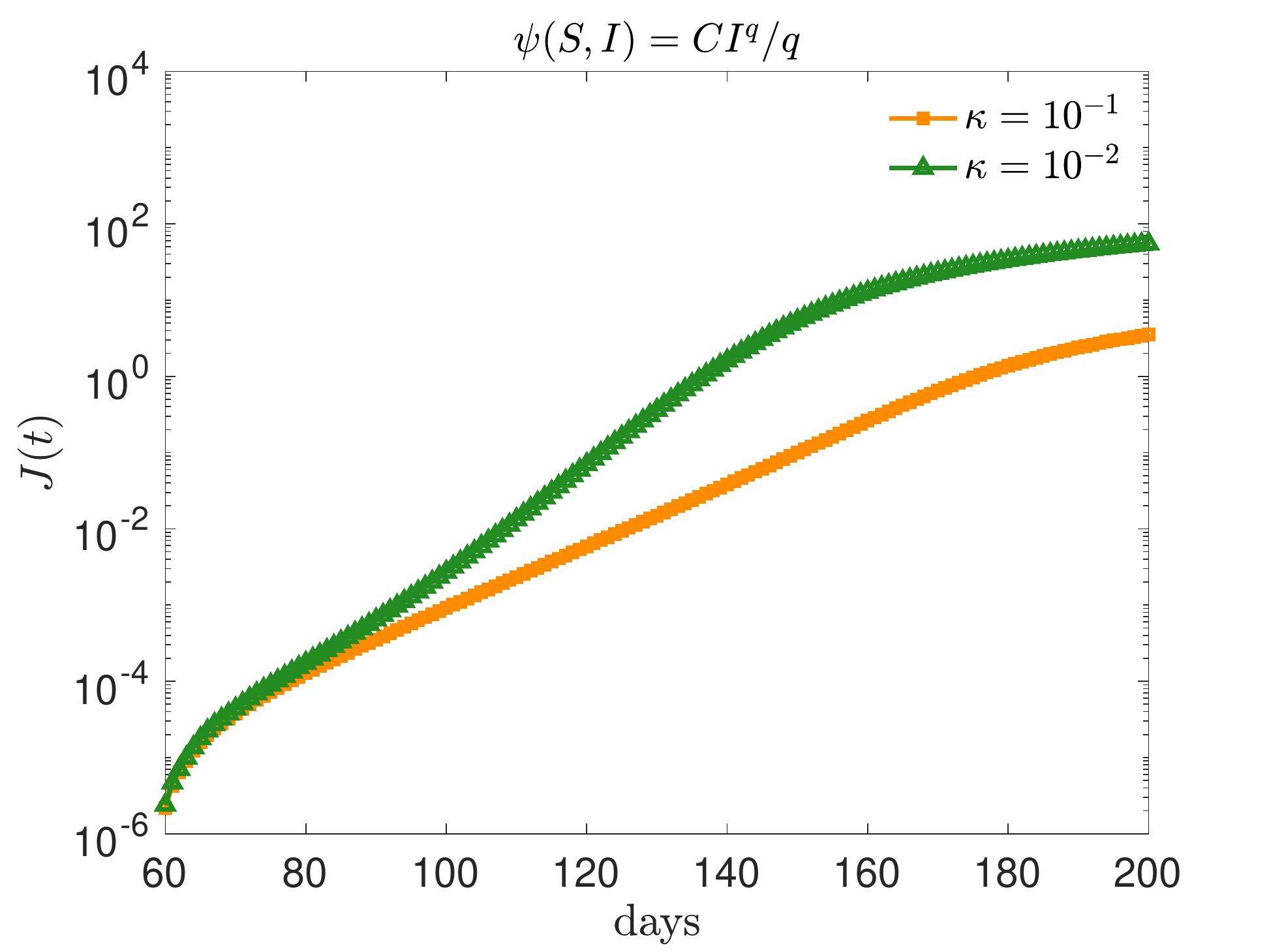}
  \includegraphics[scale = 0.25]{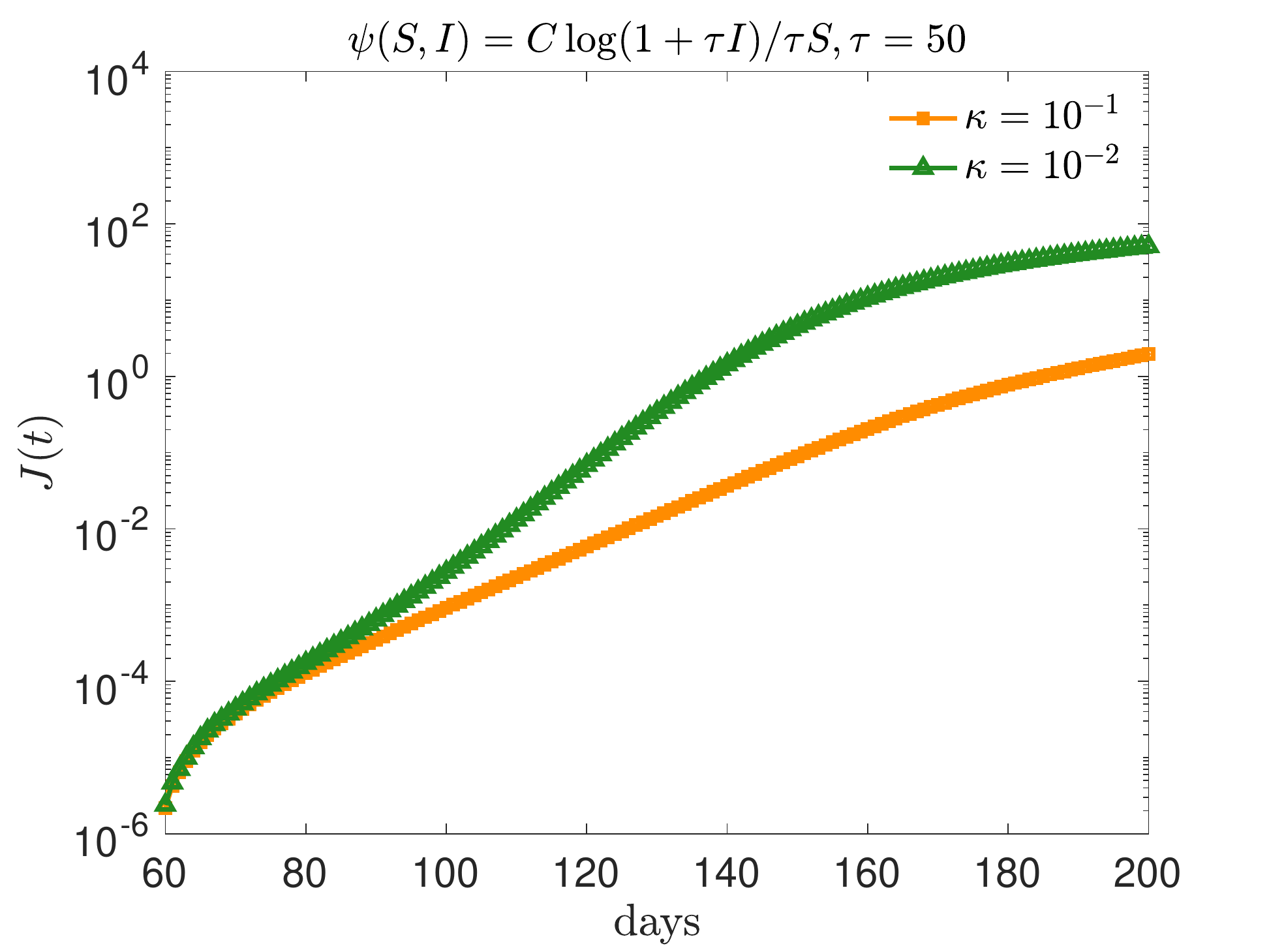}
\caption{Evolution fo the fraction of infected (left) and recovered (right) based on the two different feedback controls defined in \eqref{eq:ic1} with $q=1$ (first row) and \eqref{eq:ic2} (second row) for the SEIR model with homogeneous mixing. We considered different penalizations $\kappa = 10^{-2},10^{-1}$. The choice $\kappa = +\infty$ corresponds to the unconstrained case. Bottom figure: evaluation of the cost functional $J$ for the introduced controls.  }
\label{fig:SEIR_1}
\end{figure}

In the SEIR case we consider a population of size $N = 60 \cdot 10^6$ where at time $t = 0$ the initial number of exposed is given by $E(0) = \frac{1}{N}$ and the number of susceptibles is $S(0) =\frac{N-1}{N}$, whereas $I(0) = R(0) = 0$. To exemplify the possible evolution of the pandemic we consider $\beta = 0.25$, $\gamma = 0.1$, corresponding to a recovery rate of 10 days, so that $R_0 = 2.5$. Furthermore, we assume a latency period of $3.32$ days, leading to $\sigma \approx 0.3012$, see \cite{Gatto}. 

In Figure \ref{fig:SEIR_1} we report the dynamics of infected and recovered based on the activation of the control in the time frame $t \in [60,200]$, meaning that the control is activated after $60$ days the first exposed and after 200 days we suppose that all the restrictions are cancelled. We may easily observe how the delation of social restrictions leads for both controls to a restart of the epidemic and therefore to a second wave of infection. Both the controls have comparable costs but the perception function $\psi(S,I) = CI^q/q$ is more capable to flatten the curve of infection. After the deactivation of the control we may observe how the number of recovered for large times does not significantly change with respect to the unconstrained dynamics.

\begin{figure}
\includegraphics[scale = 0.25]{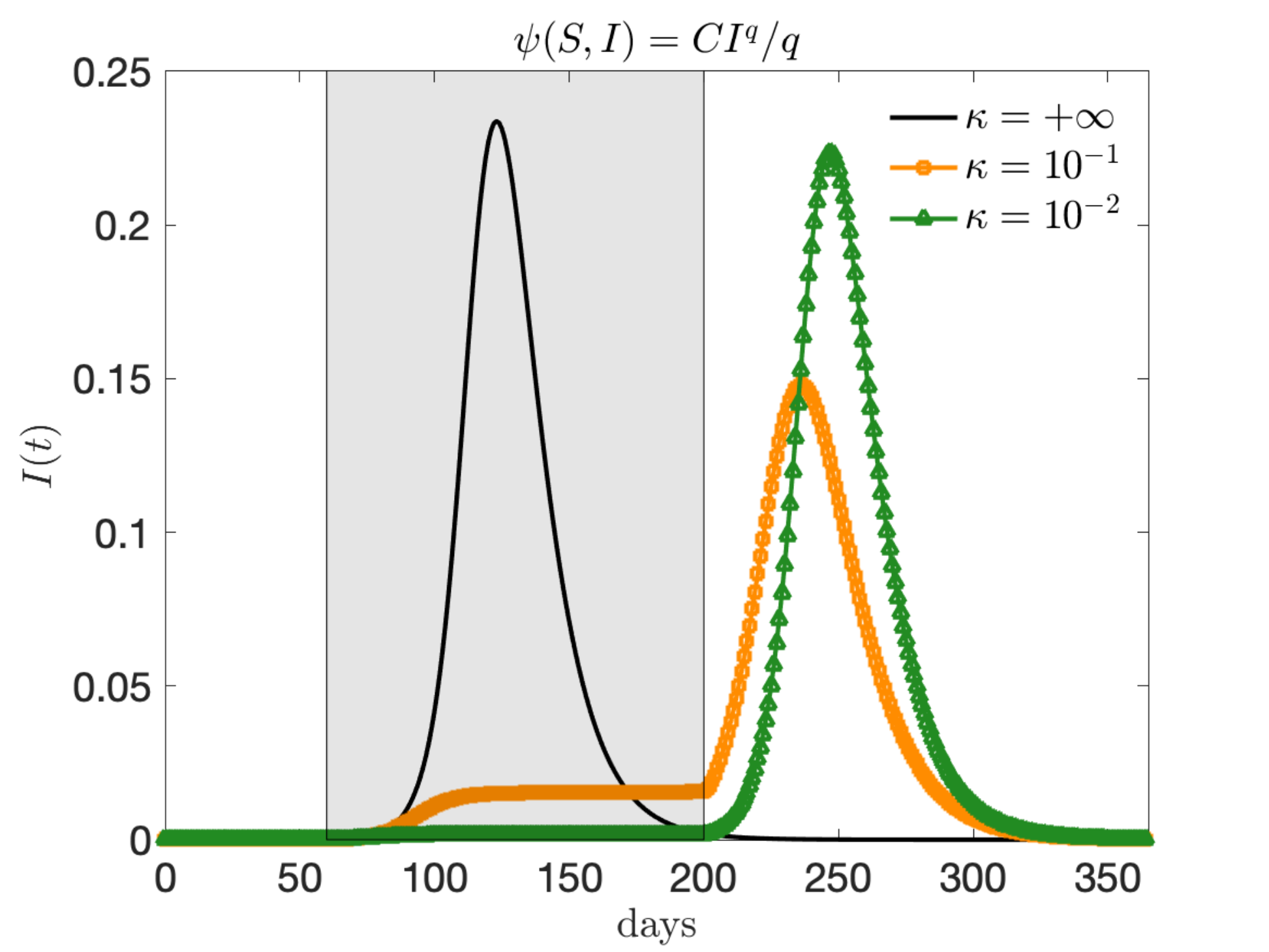}
\includegraphics[scale = 0.25]{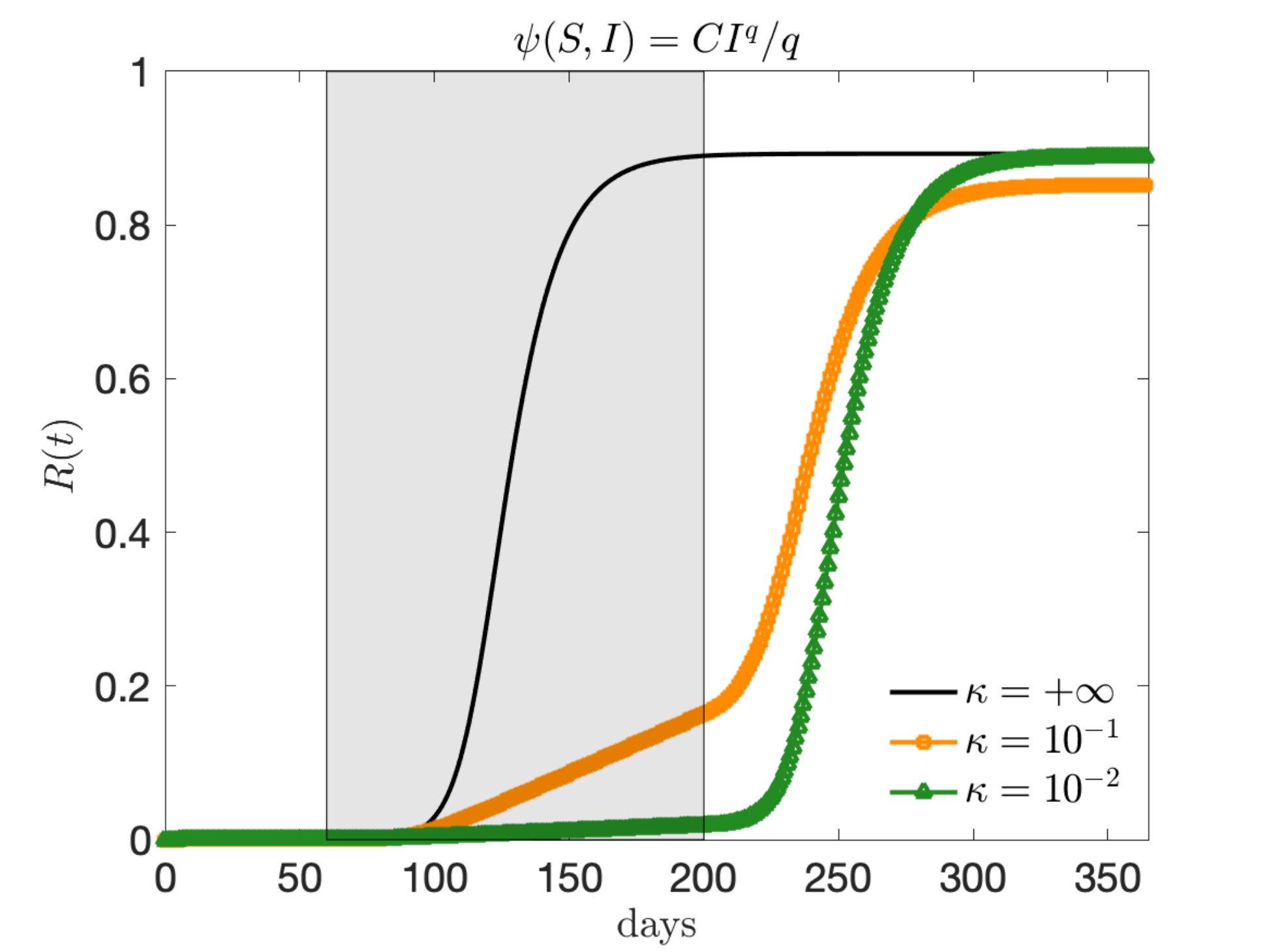}\\
\includegraphics[scale = 0.25]{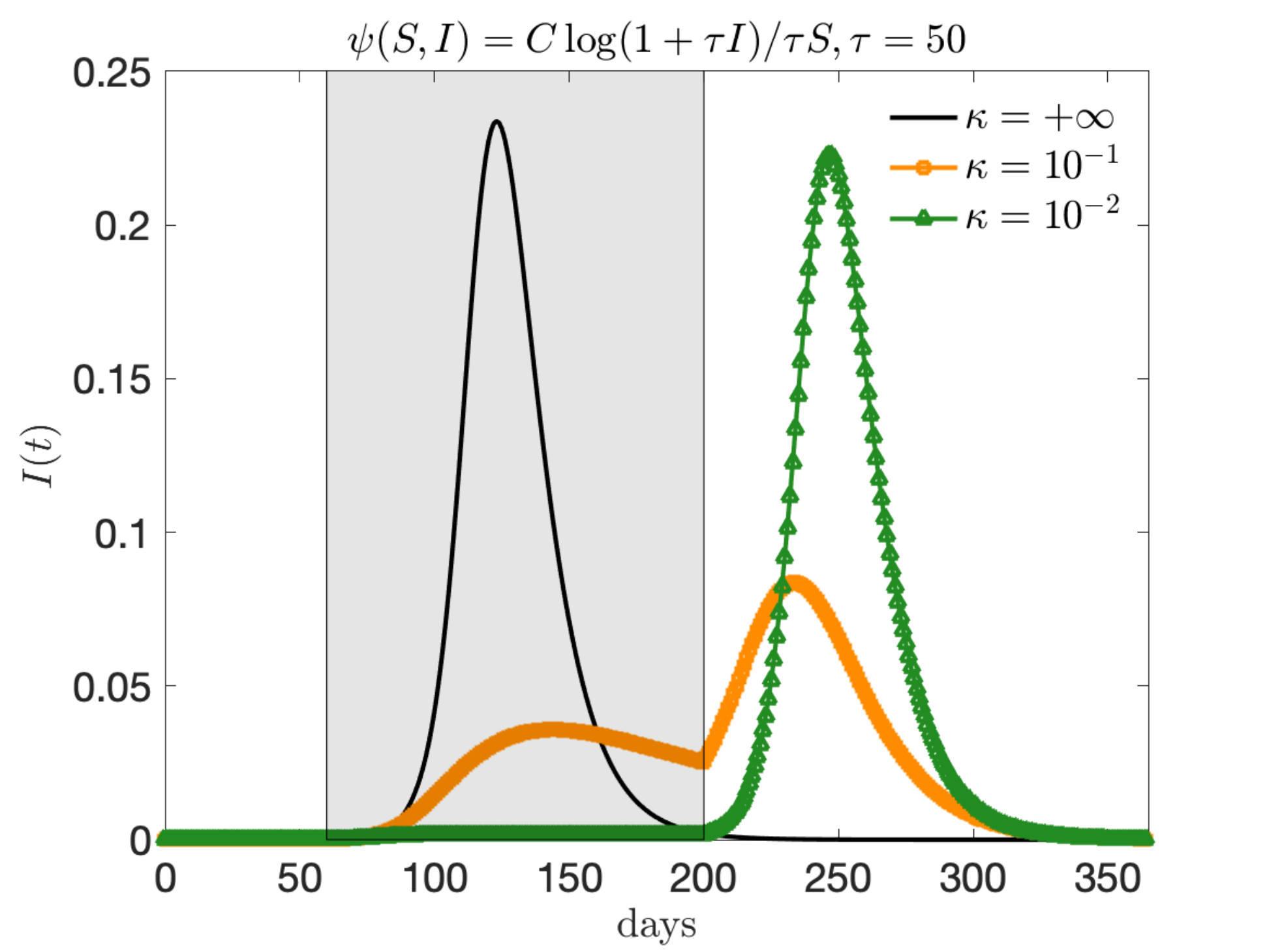}
\includegraphics[scale = 0.25]{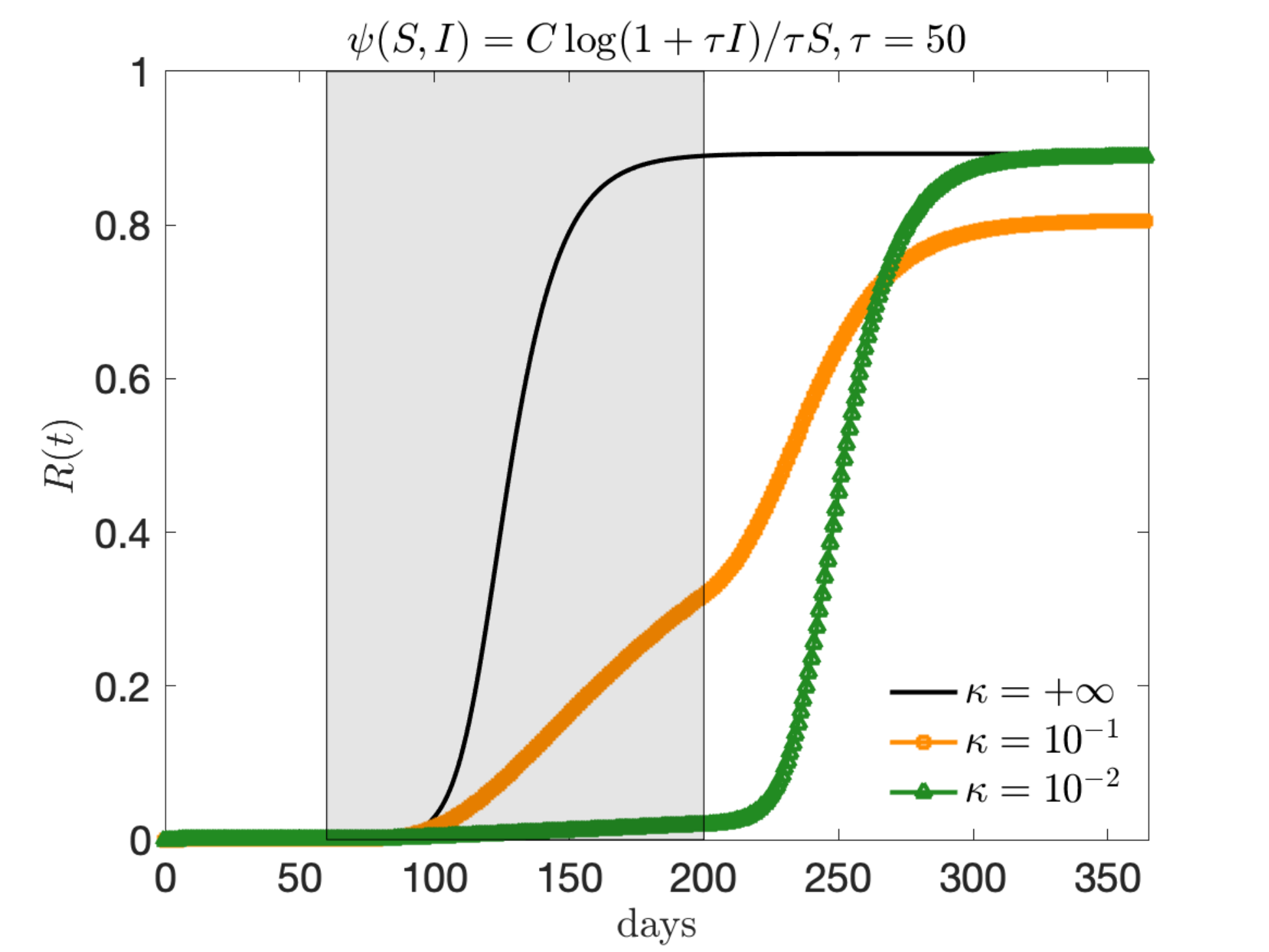}\\
 \includegraphics[scale = 0.25]{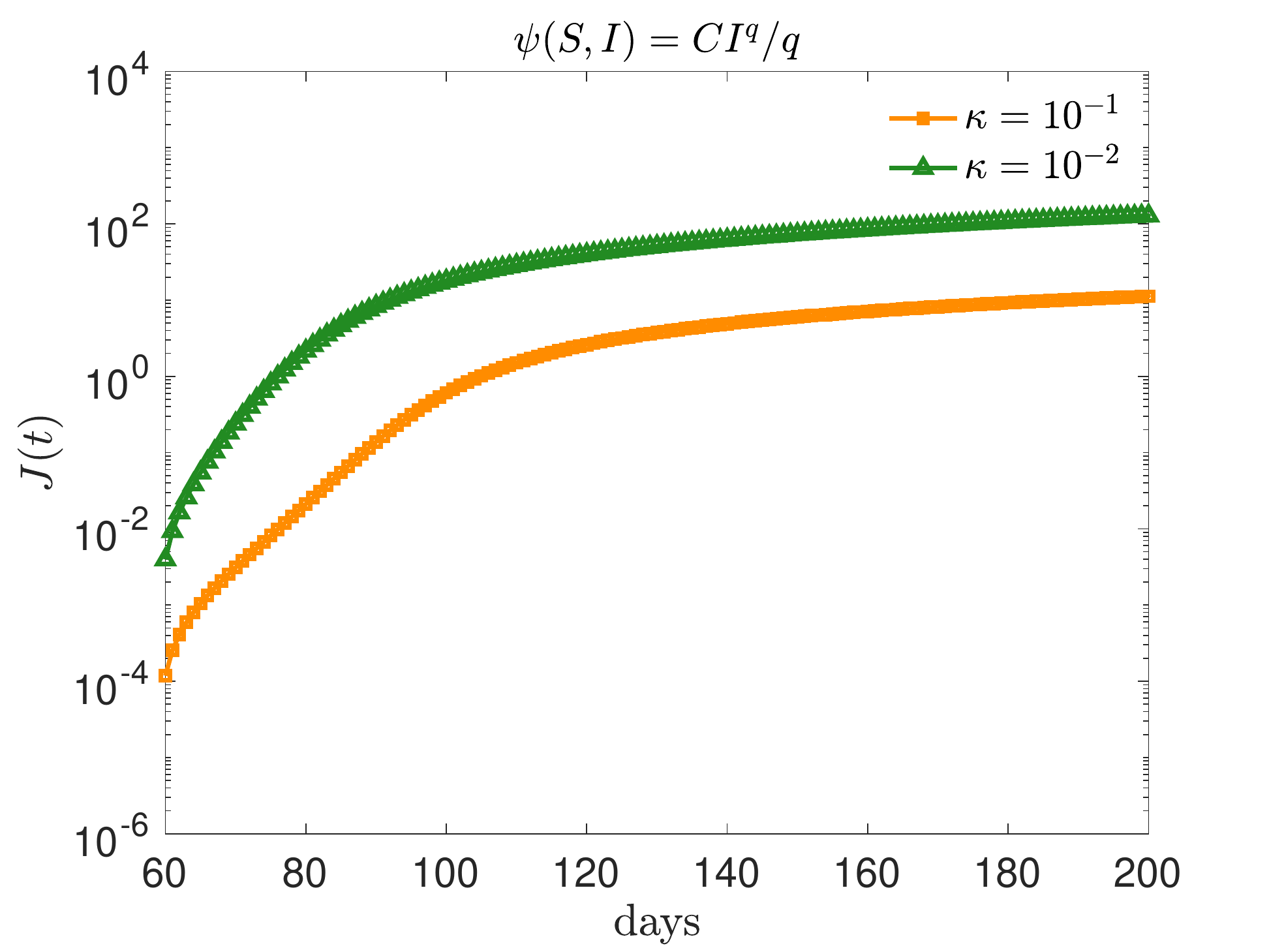}
  \includegraphics[scale = 0.25]{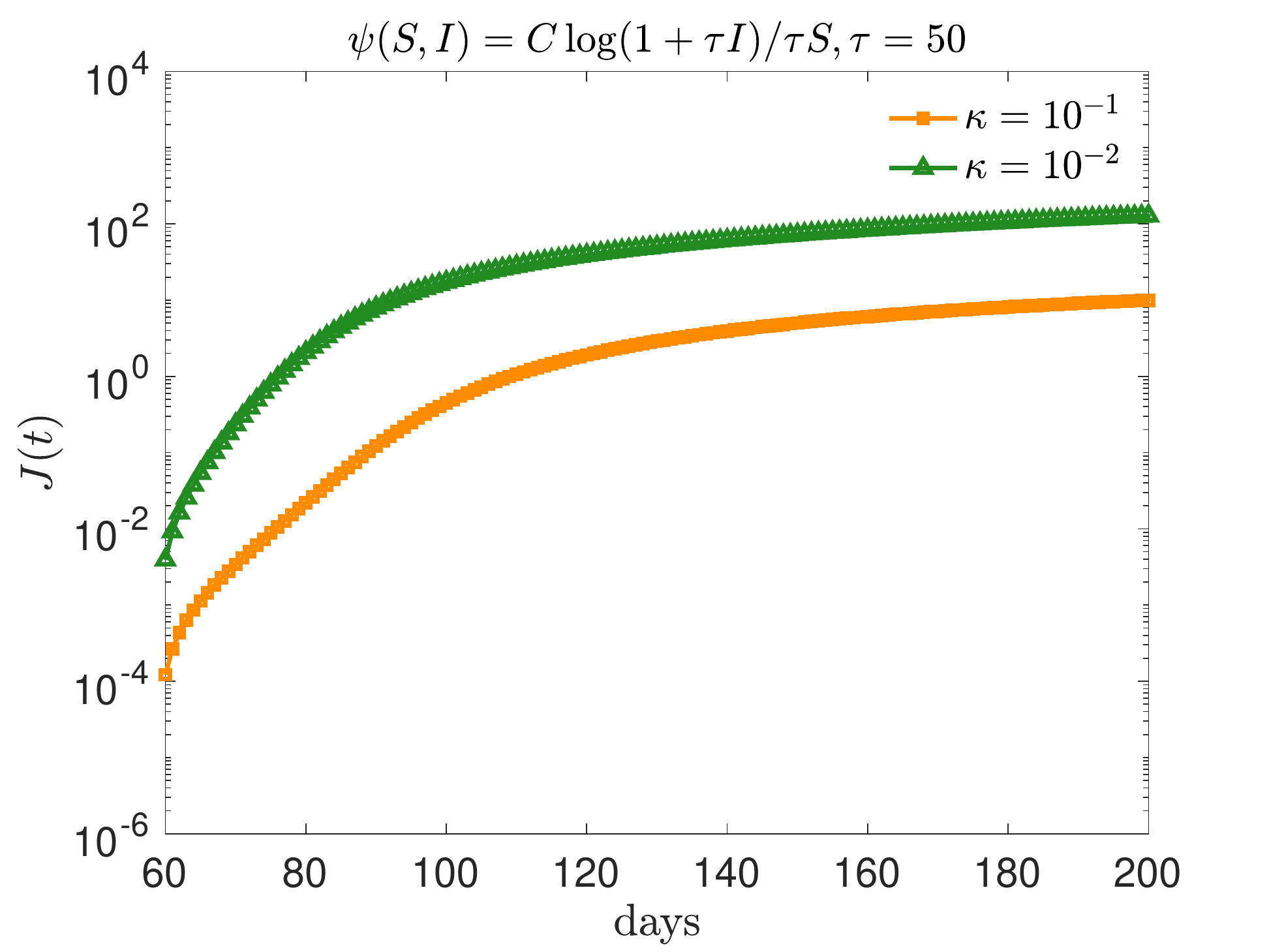}  
\caption{Evolution fo the fraction of infected (left) and recovered (right) based on the two different feedback controls defined in \eqref{eq:ic1} with $q=1$ (first row) and \eqref{eq:ic2} (second row) for the SIR model with homogeneous mixing. We considered different penalizations $\kappa = 10^{-2},10^{-1}$. The choice $\kappa = +\infty$ corresponds to the unconstrained case. Bottom figure: evaluation of the cost functional $J$ for the introduced controls. }
\label{fig:SIR_2}
\end{figure}

We perform a similar test in the case of SIR compartimentalization. Therefore we assume a population of the same size $N = 60 \cdot 10^6$ of the previous test where at time $t = 0$ the initial number of infected is $I(0) = \frac{1}{N}$ and $S(0) = \frac{N-1}{N}$. We considered epidemiological parameters that are compatible with the ones considered above and leading to $R_0 = 2.5$, i.e. $\beta = 0.25$ and $\gamma = 0.1$. 
We remark that in presence of a SIRD-type compartmentalization we obtain a feedback control compatible with \eqref{eq:ic1} with $\sigma = 1$ for a perception function $\psi(S,I ) = CI^q/q$, whereas in the logarithmic case we can derive a feedback control compartible with \eqref{eq:ic2} with $\sigma = 1$, we point the interested reader to \cite{APZ1} for more details. In Figure \ref{fig:SIR_2} we report the dynamics of infected and recovered with an activation of the control in the time interval $[60,200]$. Interestingly enough, the logarithmic perception function is in this case more effective in the reduction of the number of recovered, that is the total number of infected of the population.

\subsection{Model calibration}
Estimating epidemiological parameters is a very difficult problem that can be addressed with different approaches \cite{Capaldi_etal, Chowell, Roberts}. In the case of COVID-19 due to the limited number of data and their great heterogeneity is an even bigger problem that can easily lead to wrong results. Here, we restrict ourselves to identifying the deterministic parameters of the model through a suitable fitting procedure, considering the possible uncertainties due to such estimation as part of the subsequent uncertainty quantification process. For this reason in the sequel we will neglect the presence of the exposed population and thus consider the feedback controlled SIR model. 

More precisely, we have adopted the following two-level approach in estimating the parameters. In the phase preceding the lockdown we estimated the epidemic parameters, and hence the model reproduction number $R_0$, in an uncontrolled regime. This estimate was then kept in the subsequent lockdown phase where we estimated as a function of time the value of the control penalty parameter. Both these two calibration steps were analyzed under the assumption of homogeneous mixing. 

Thua, we solved two separate constrained optimization problems. First we estimated $\beta_e>0$ and $\gamma_e>0$ in each country by solving in the uncontrolled time interval $t\in [t_0,t_u]$ a least square problem based on minimizing the relative $L^2$ norm of the difference between the reported number of infected $\hat I(t)$ and recovered $\hat R(t)$, and the theoretical evolution of the unconstrained model $I(t)$ and $R(t)$. In details, we considered the following minimization problem 
\[
\min_{\beta,\gamma \in \mathbb R_+} \left[ (1-\theta)\| I(t)  - \hat I(t)\|_{L^2([t_0,t_u])} + \theta \| R(t) - \hat R(t) \|_{L^2([t_0,t_u])} \right], 
\]
where $\theta \in [0,1]$ is a penalization parameter and $\|  \cdot \|_{L^2([t,s])}$ denotes the relative $L^2$ norm over the time horizon $[t,s]$.
It is worth to remark that the lack of reliable informations concerning the recovered in early stages of the disease suggests to adapt the model mainly to the curve of infectious and to introduce the uncertainty in the reproductive number using this estimated value as an upper bound of the reproduction number. 

Due to the heterogeneity of the data between the different countries, 
 we constrained the value of $\beta\in [0,1]$ and the value of $\gamma \in \left[ \dfrac{1}{24},\dfrac{1}{10}\right]$. Indeed, according to clinical studies, time to viral clearance during the early phases of the epidemic, i.e. the time from the first positive test to the first negative test, can approximately span from $10$ to $24$ days, see \cite{Chen2020,Gatto,Lavezzo2020}.
 At the end of this optimization procedure, we obtain the values $\beta_e$, $\gamma_e$ for each country reported in Table \eqref{tab:kcont}. The results have been obtained by averaging the optimization outputs with penalization factors $\theta = 10^{-2}$ and $\theta = 10^{-6}$, respectively. The choice of small values for $\theta$ is due to the increased heterogeneity in data for recovered in this early stages of the epidemic. 

Next, we estimate the penalization $\kappa=\kappa(t)>0$ in time by solving in the controlled time interval $t\in (t_u,t_c]$ for a sequence of unitary time steps $t_i$  the corresponding least square problems in $[t_i-k_l,t_i+k_r]$, $k_l, k_r\geq 1$ integers, and where for the evolution we consider the values $\beta_e$ and $\gamma_e$ estimated in the first optimization step using the curve of infectious. In details we solve the following minimization problem 
\[
\min_{\kappa(t_i) \in \mathbb R_+} \left[(1-\theta) \| I(t) - \hat I(t) \|_{L^2([t_i-k_l,t_i+k_r])} + \theta \|R(t) - \hat R(t) \|_{L^2([t_i-k_l,t_i+k_r])} \right],
\]
over a window of seven days ($k_l=3$, $k_r=4$) for regularization along one week of available data. For consistency we performed the same optimization process used to estimate $\beta$ and $\gamma$, namely using two different penalization factors and then averaging the results. These optimization problems have been solved testing different optimization methods in combination with adaptive solvers for the system of ODEs. The results reported have been obtained using the Matlab functions {\tt fmincon} in combination with {\tt ode45}.
 
\begin{figure}
\centering
\includegraphics[scale =.3]{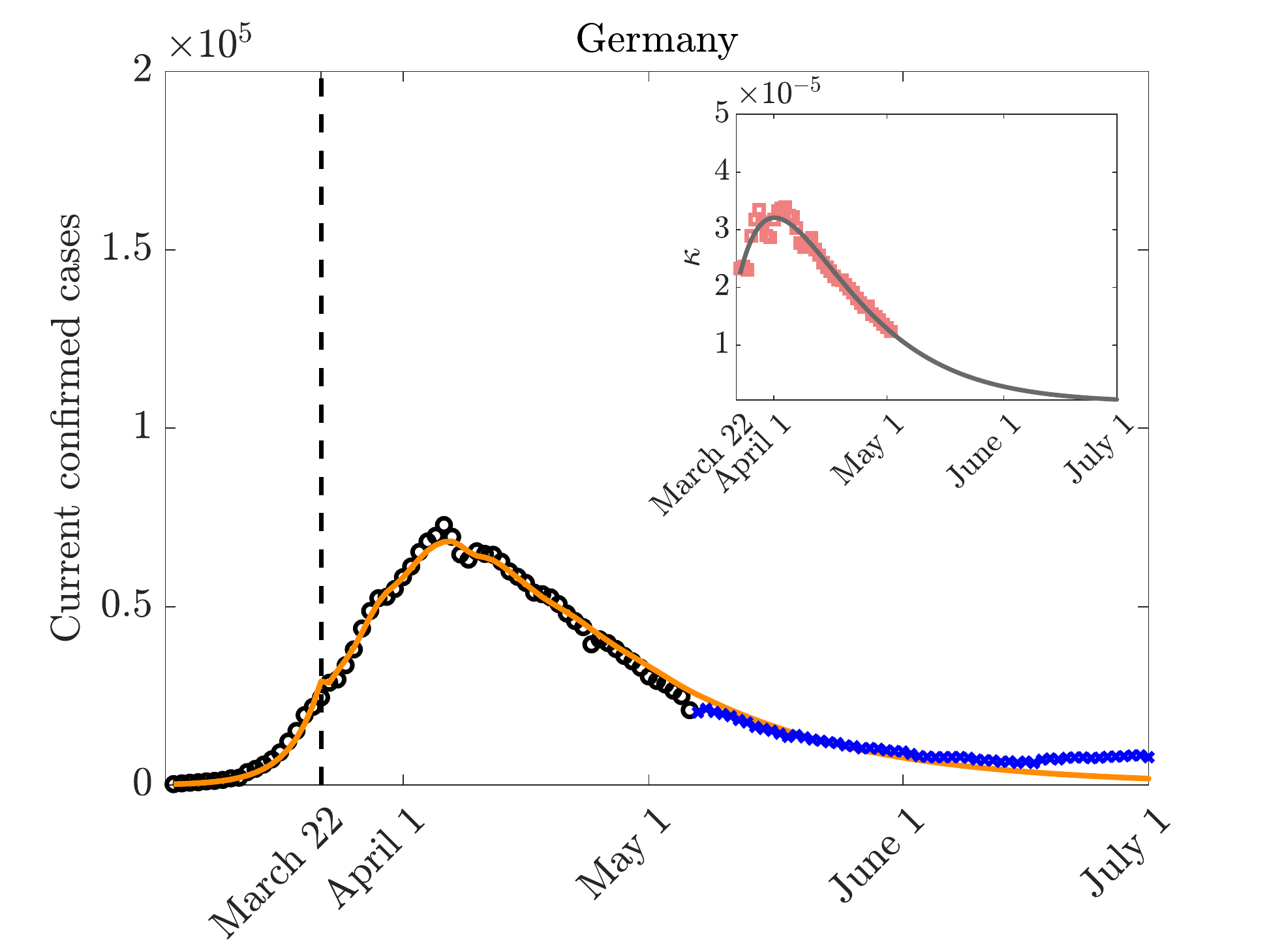}
\includegraphics[scale =.3]{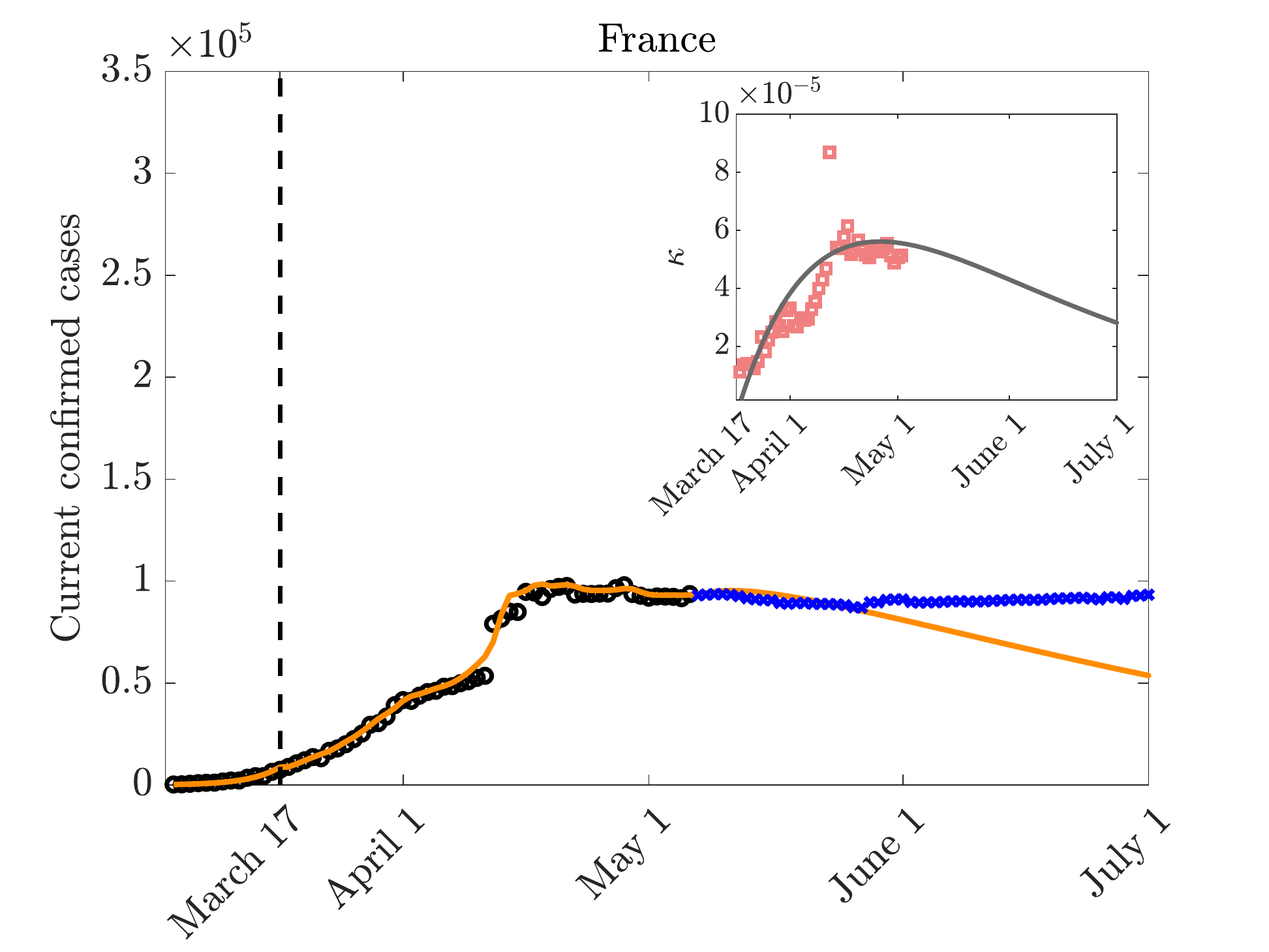}\\
\includegraphics[scale =.3]{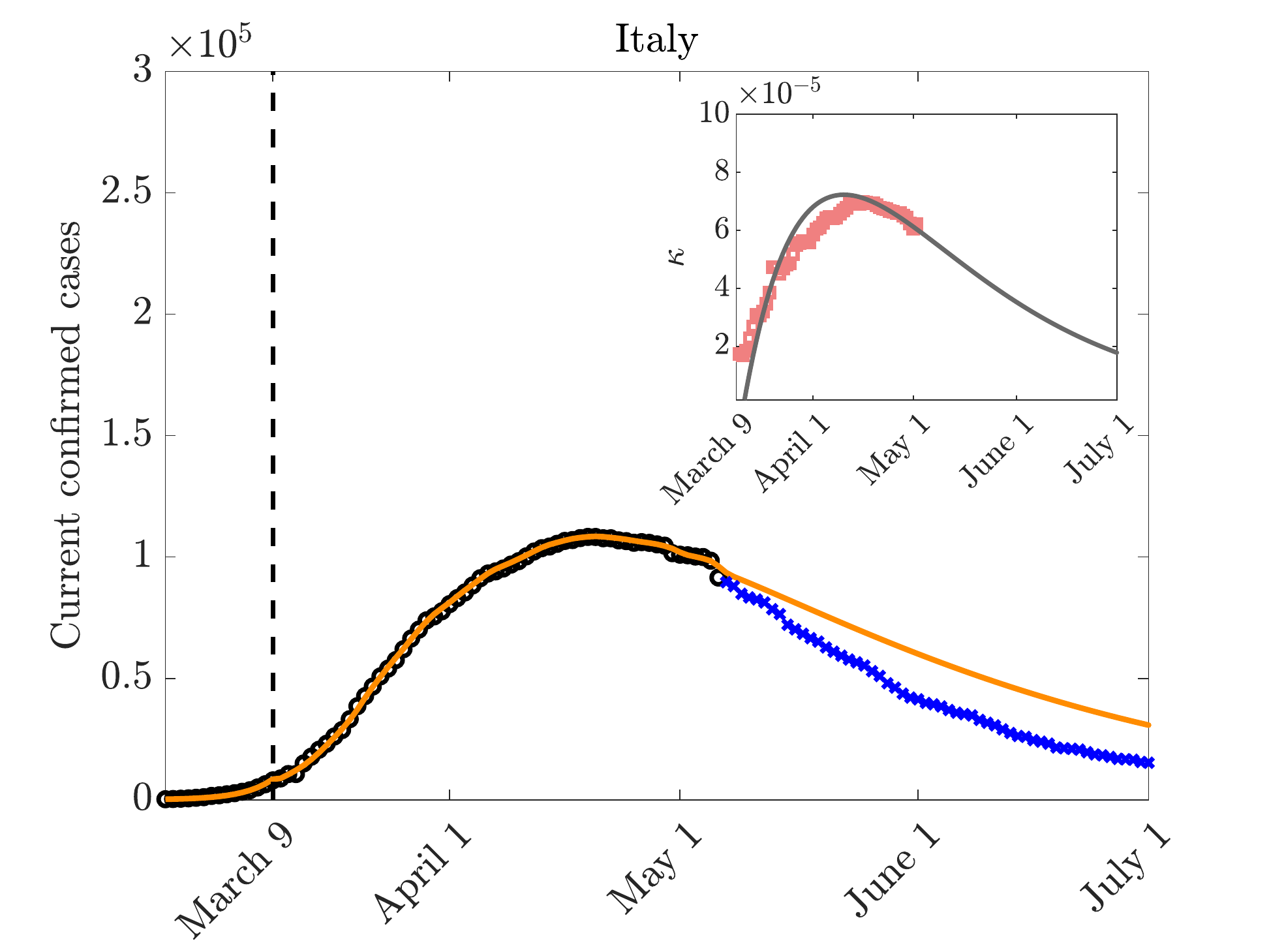} 
\includegraphics[scale =.3]{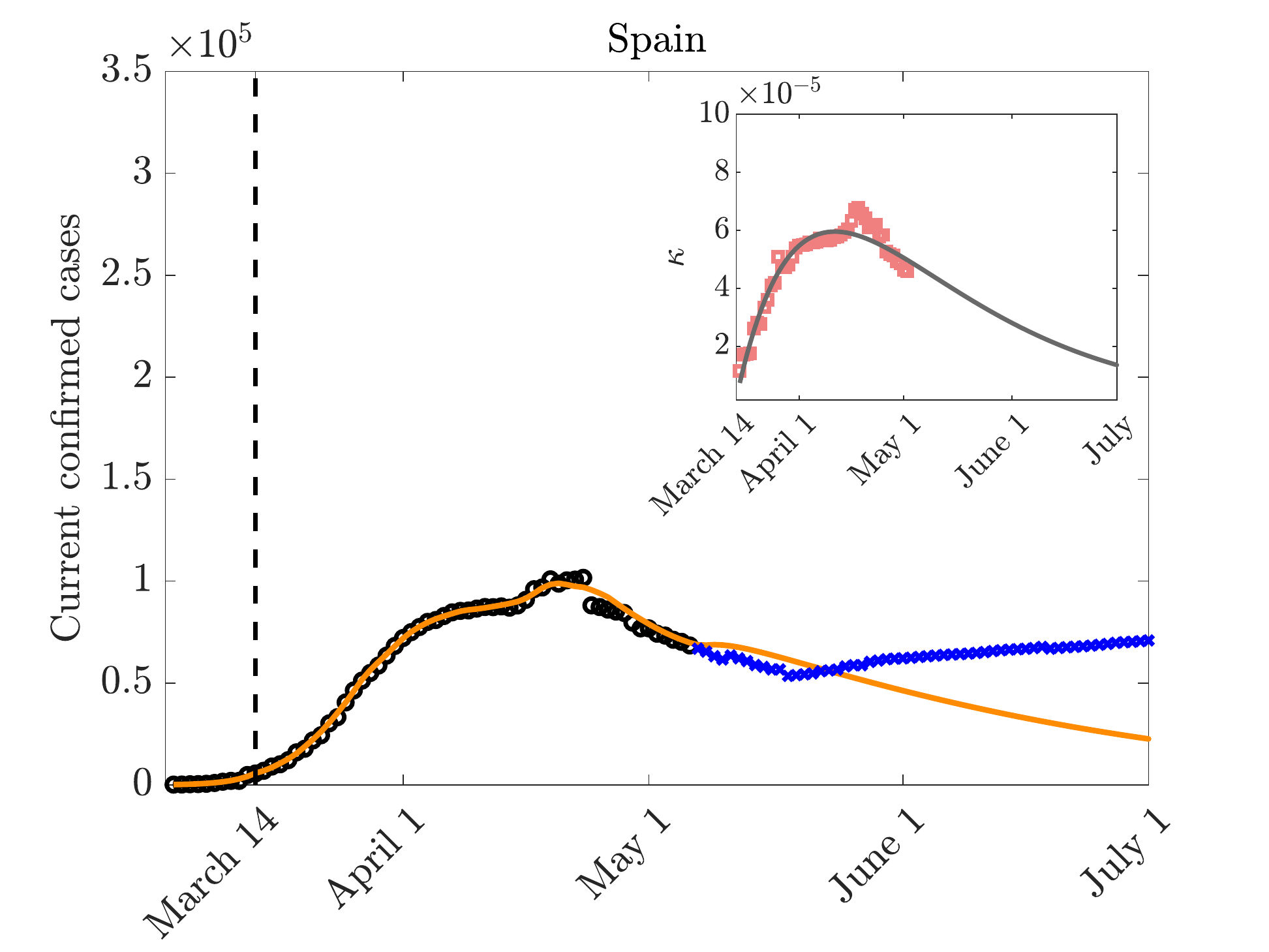}\\
\includegraphics[scale =.3]{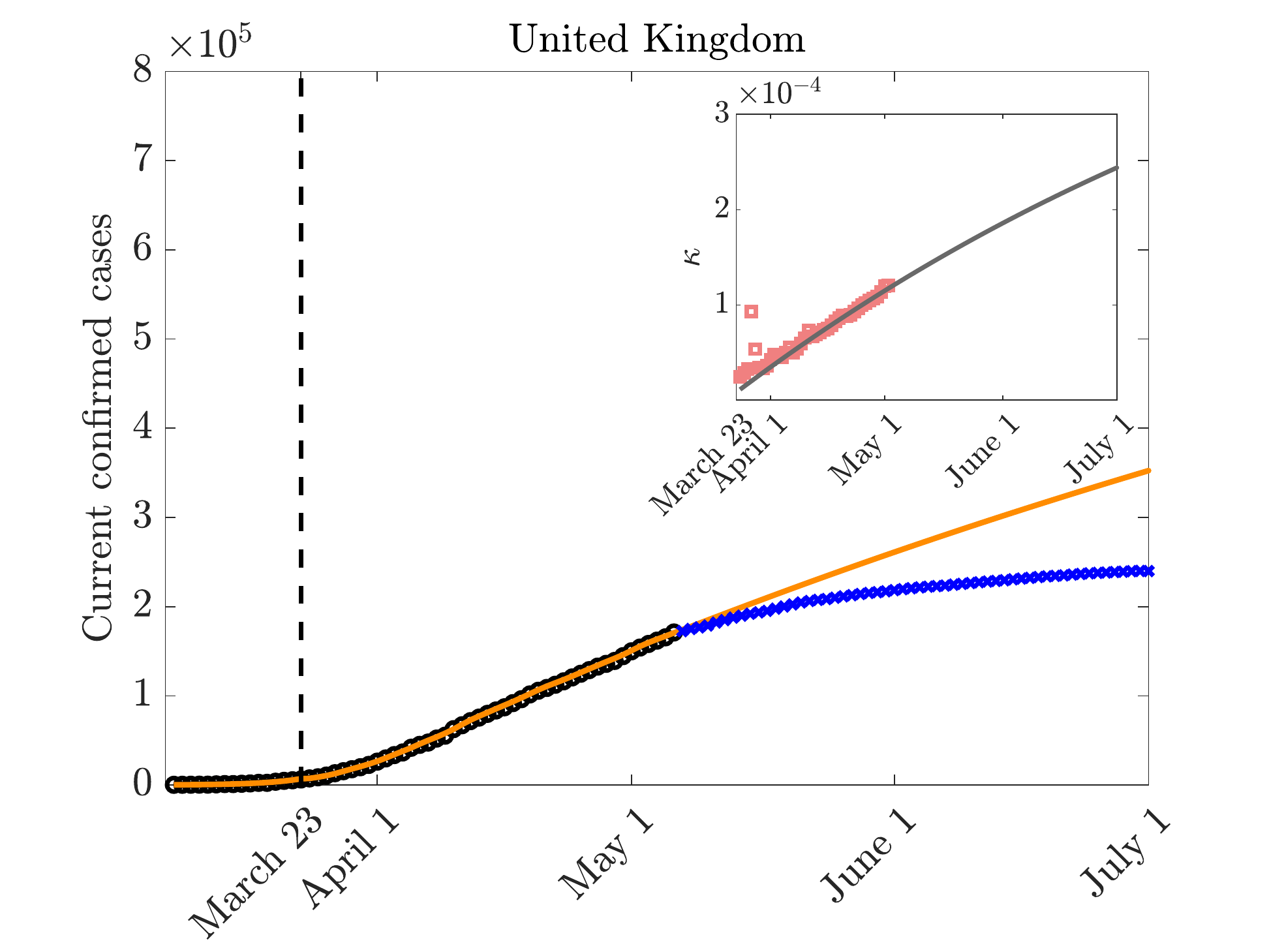}
\includegraphics[scale =.3]{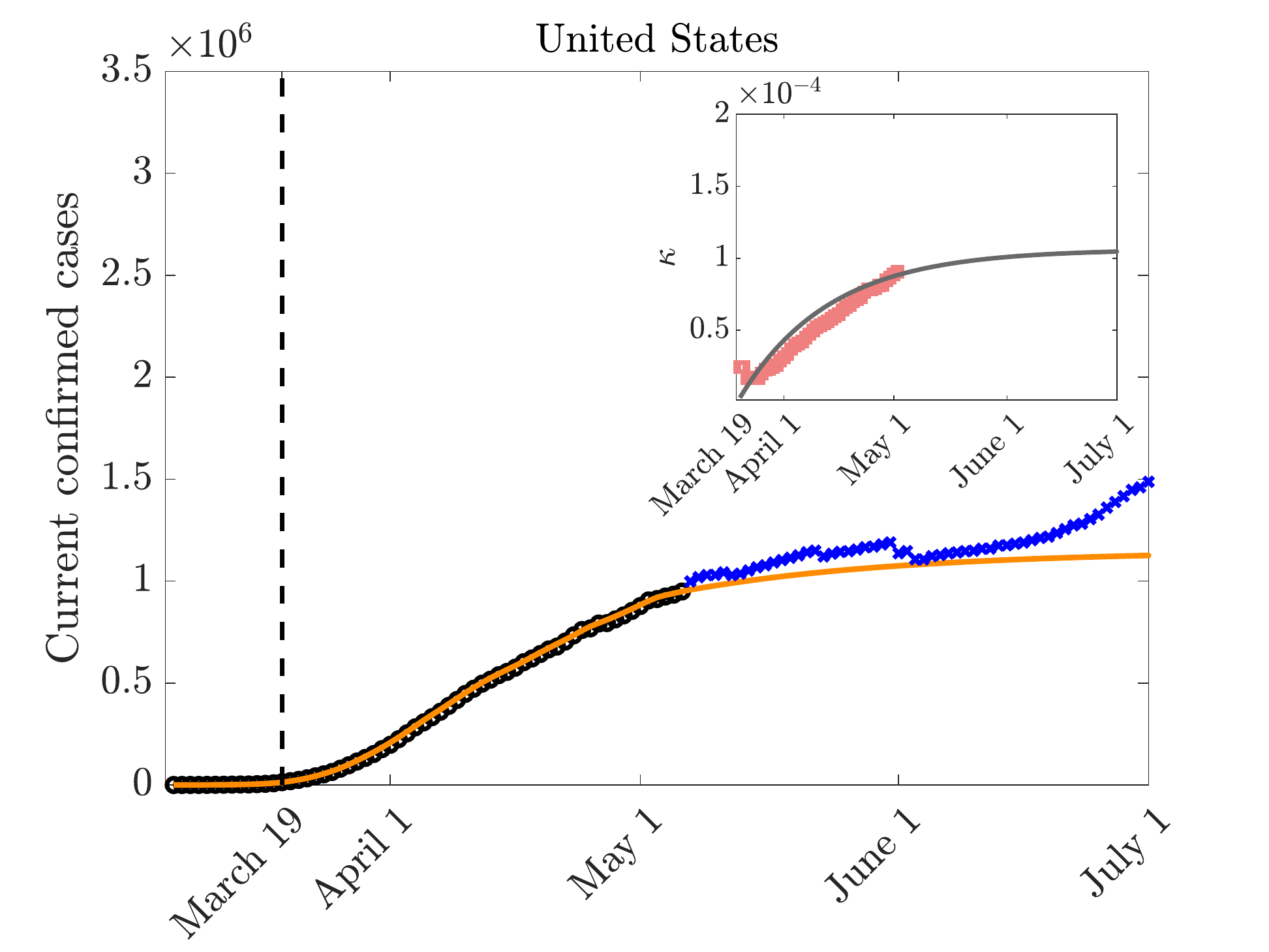}
\caption{Model behavior with fitting parameters and actual trends in the number of reported infectious using the estimated control penalization terms after lockdown over time in the various countries.}
\label{fig:datak}
\end{figure}

The corresponding time dependent values for the controls as well as results of the model fitting with the actual trends of infectious are reported in Figure \ref{fig:datak}. The trends have been computed using a weighted least square fitting with the model function $k(t)=ae^{bt}(1-e^{ct})$. 

For some countries, like France, Spain and Italy after an initial adjustment phase the penalty term converged towards a peak and has just started to decrease. This is consistent with a situation in which data concerning the number of reported infectious needs a certain period of time before being affected by the lockdown policy and can also be considered as an indicator of an unstable situation where reducing control could lead to a potential restart of the infectious curve. The penalty terms for the US and the UK clearly indicates that the pandemic was still in its growing phase. In the case of Germany the dynamics highlight a significative decrease in the penalization term, this fact is coherent with the timely implementation of social distancing measures. Note that, see figure \ref{fig:datak}, the behavior of the model is able to fairly realistically describe the observed data for a time window of about one month after calibration. On the other hand, a larger time window, up to the end of June, clearly presents significant deviations from the expected behavior due to the restart of the pandemic wave as in France, Spain and the US or a drop down in the number of cases as in Italy and the UK.

\begin{table}[t]
\begin{center}
\begin{tabular}{c | c | c |  c | c | c| c}
& $\underset{\textrm{Mar 5-Mar 22}}{\textrm{Germany}}$ & $\underset{\textrm{Mar 5-Mar 17}}{\textrm{France}}$& $\underset{\textrm{Feb 24-Mar 9}}{\textrm{Italy}}$ & $\underset{\textrm{Mar 5-Mar 14}}{\textrm{Spain}}$ &  $\underset{\textrm{Mar 8-Mar 23}}{\textrm{UK}}$  & $\underset{\textrm{Mar 7-Mar 19}}{\textrm{US}}$ \\ 
\hline\hline
& & & & & &\\[-.3cm]
$\beta_e$ & $0.3134$ & $0.3164$ & $0.3101$& $0.3686$ & $0.2698$ & $0.3716$ \\
\hline
$\gamma_e$ & $0.0483$ & $0.0483$ & $0.0494$ & $0.0400$ & $0.0482$ & $0.0481$\\
\hline
$R_0^e$ & $6.487$ & $6.5525$ & $6.2710$ & $9.2150$ & $5.6010$ & $7.7155$
\end{tabular}
\end{center}
\caption{Model fitting parameters in estimating attack values for the COVID-19 outbreak before lockdown in various countries.}
\label{tab:kcont}
\end{table}

\subsection{Estimating actual infection trends with uncertain data}
\begin{figure}
\begin{center}
\includegraphics[scale=0.45]{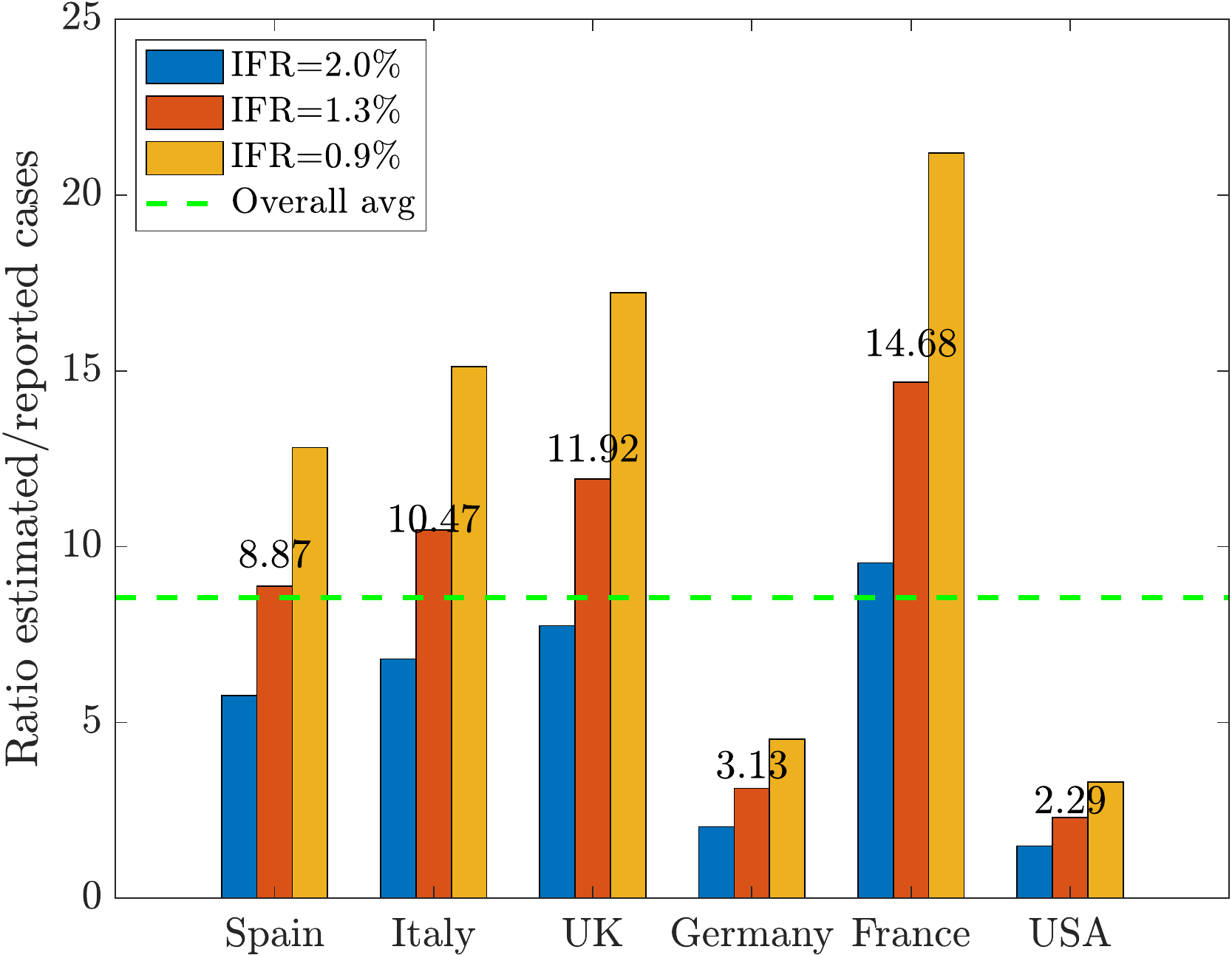}
\end{center}
\caption{Estimated disagreement in the total number of cases based on an IFR of $1.3\%$ in the range $0.9\%$-$2.0\%$. The uncertainty is measured as the estimated values divided by the reported cases. The country specific values are given on the top of each red bar, the average value of $c=8.56$ is reported as a dashed green line.}
\label{fig:inc}
\end{figure}

Next we focus on the influence of uncertain quantities on the controlled system with homogeneous mixing. According to recent results on the diffusion of COVID-19 in many countries the number of infected, and therefore recovered, is largely underestimated on the official reports, see e.g. \cite{JRGL,MKZC}. One possible way to understand this is based on a renormalization process of the reported data based on the estimated infection fatality rate (IFR) of Covid-19. Although estimating the true IFR is generally hazardous while an epidemic is underway, some studies have estimated an overall IFR around $1.3\%$ with an age dependent credible interval \cite{RHJZ,RPc}. In the sequel we consider a range spanning between $0.9\%-2.0\%$. On the contrary the current fatality rate (CFR) may vary strongly from country to country accordingly to the differences in the number of people tested, demographics, health care system. One way to have in insight in the uncertainty of data is to use the estimated IFR ranges as normalization factors for the current data reported of total cases $I_{\rm tot}$. This is done computing an estimated number of total confirmed cases as $\hat I_{\rm tot} = 100 \times D_r/{\rm IFR}$, where $D_r$ is the total number of confirmed deaths. The results of the variations $\hat I_{\rm tot}/I_{\rm tot}$ for the various countries are summarized in Figure \ref{fig:inc} and are directly proportional to the CFR of the country. We are aware that the estimate obtained is certainly coarse, nevertheless it allows to get an idea of the disagreement between the data observed and expected in the various countries and therefore to be able to define a common scenario between the various countries.    

In order to have an insight on global impact of uncertain parameters we consider a two-dimensional uncertainty $\z = (z_1,z_2)$ with independent components such that 
\be\label{eq:ir_z1}
I(\z,0) = I_0(1+\mu z_1),\qquad R(\z,0) = R_0(1+\mu z_1), \qquad \mu>0
\ee
and
\be\label{eq:beta_z2}
\beta(\z) = \beta_e-\alpha_\beta z_2,\qquad \gamma(\z)=\gamma_e + \alpha_\gamma z_2,\qquad \alpha_\beta,\alpha_\gamma>0 
\ee
where $z_1$, $z_2$ are chosen distributed as symmetric Beta functions in $[0,1]$, $i_0$ and $r_0$ are the initial number of reported cases and recovered taken from \cite{Chen2020} and $\beta_e$, $\gamma_e$ are the fitted values given in Table \ref{tab:kcont}. In the following we will consider $\mu=2(c-1)$ common for all countries such that $\mathbb E[I(\z,0)]=c I(0)$, $\mathbb E[R(\z,0)]=c R(0)$ where $c=8.56$, the average value from Figure \ref{fig:inc}. 

From a computational viewpoint we adopted the method developed in \cite{APZ1} based on  a stochastic Galerkin approach. The feedback controlled model has been computed using an estimation of the total number of susceptible and infected reported, namely we have the control term 
\begin{equation}
u(t)=-\frac1{k(t)} S_r(t)I_r(t),
\label{eq:r0e}
\end{equation}
where $S_r(t)$ and $I_r(t)$ are the model solution obtained from the registered data, and thus $I_r(t)$ represents a lower bound for the uncertain solution $I(\z,t)$. 

\begin{figure}
\centering
\hspace{-0.5cm}
\includegraphics[scale =.3]{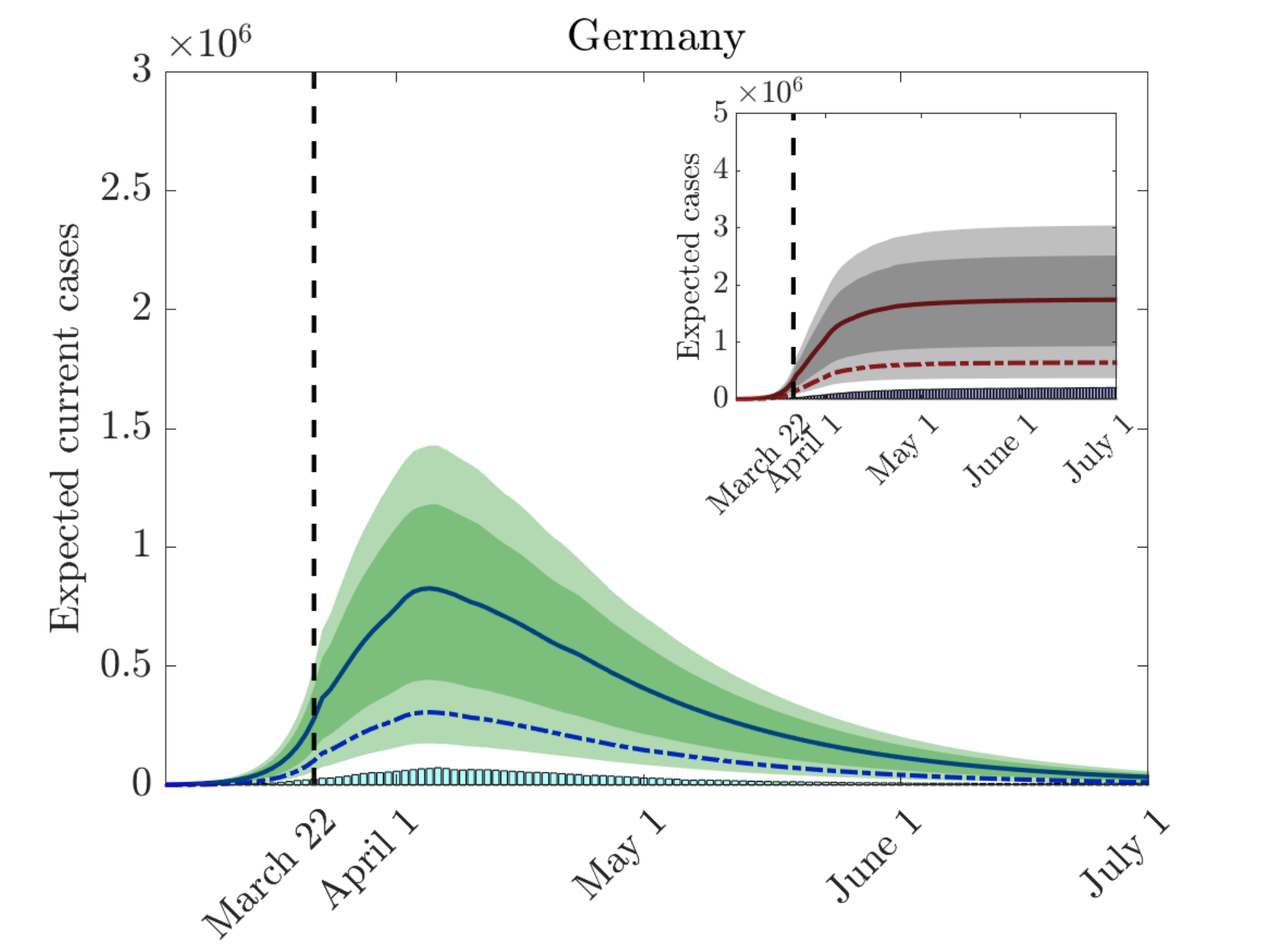} 
\includegraphics[scale =.3]{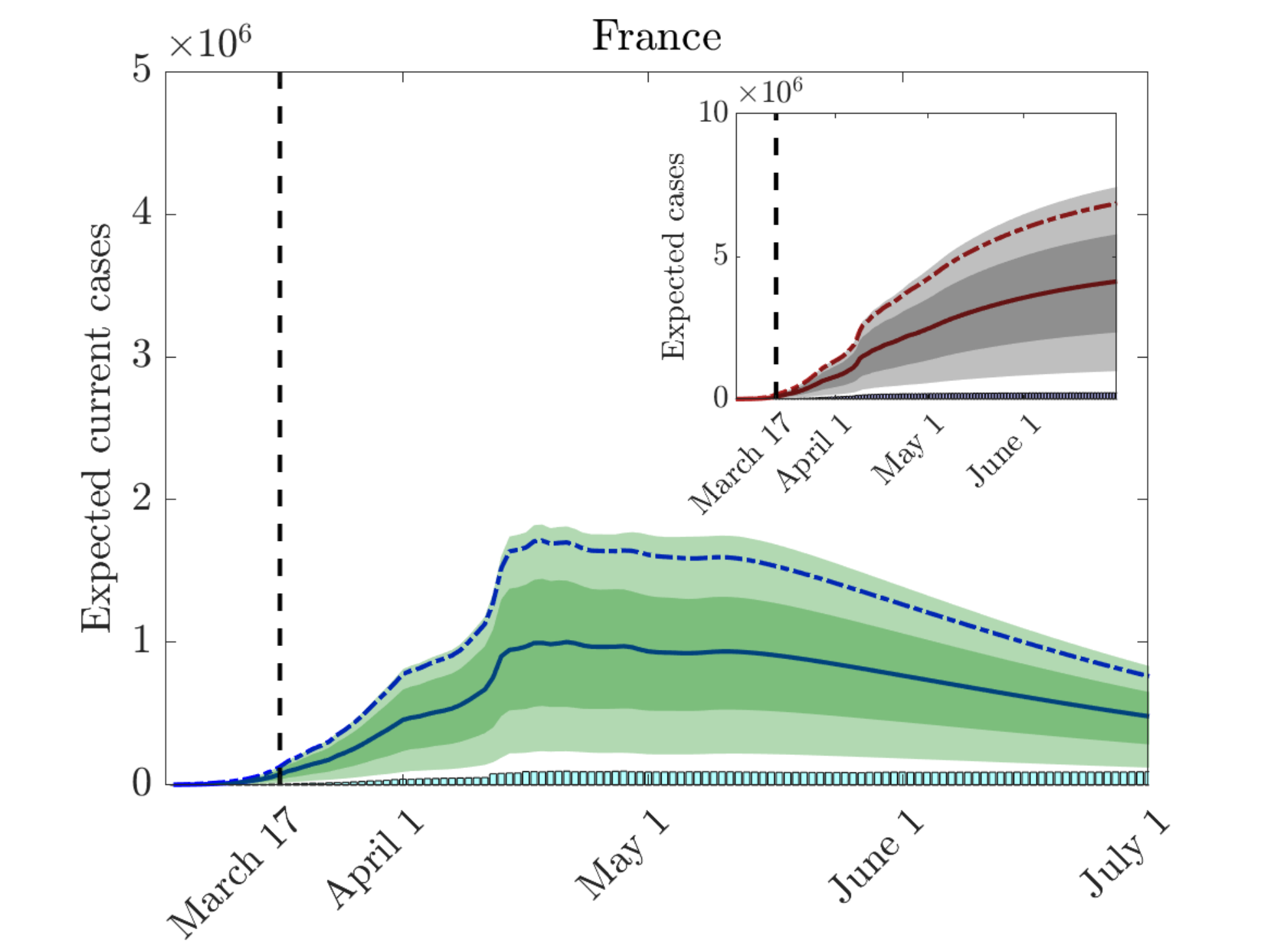}\\
\hspace{-0.5cm}
\includegraphics[scale =.3]{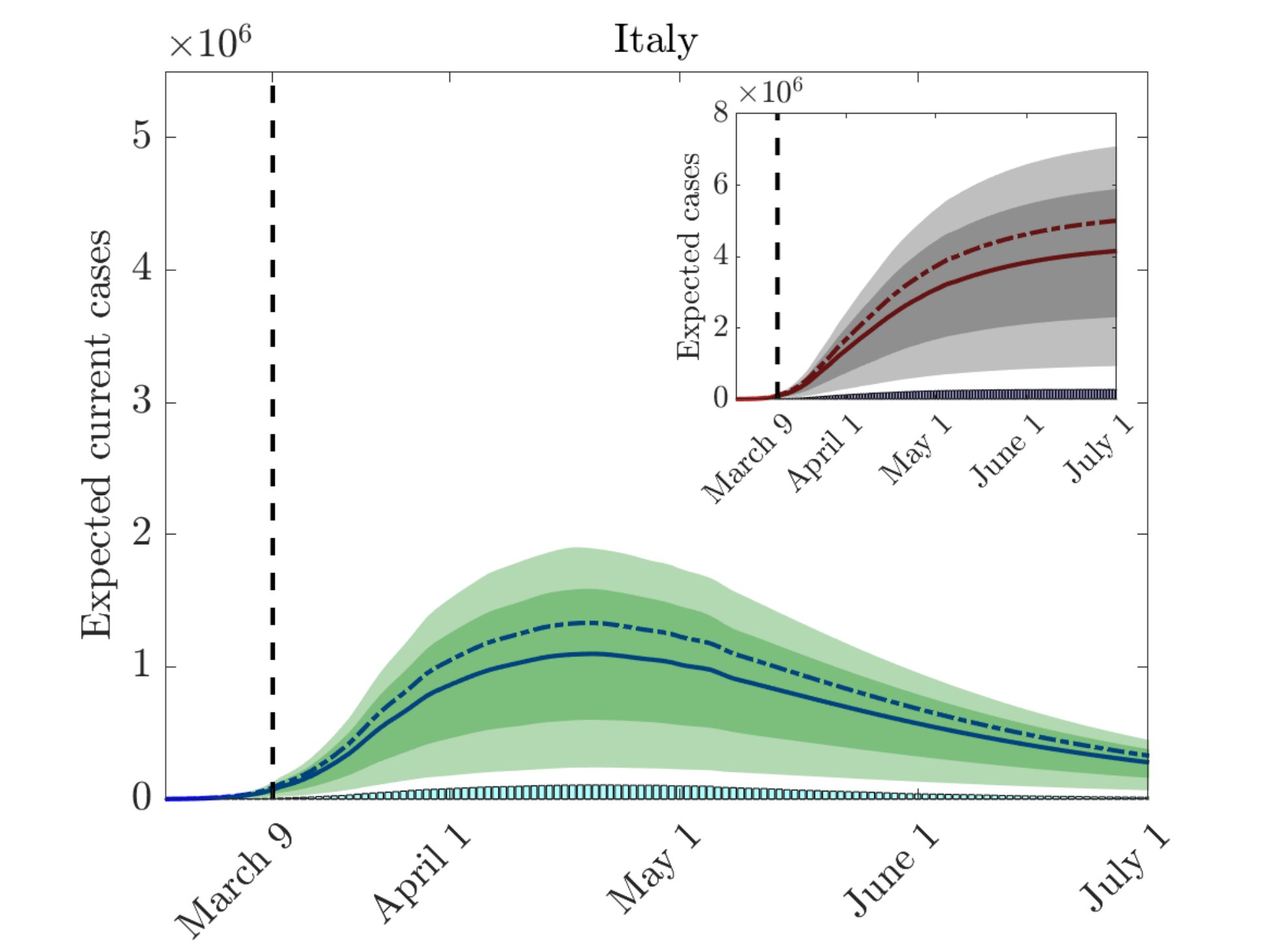} 
\includegraphics[scale =.3]{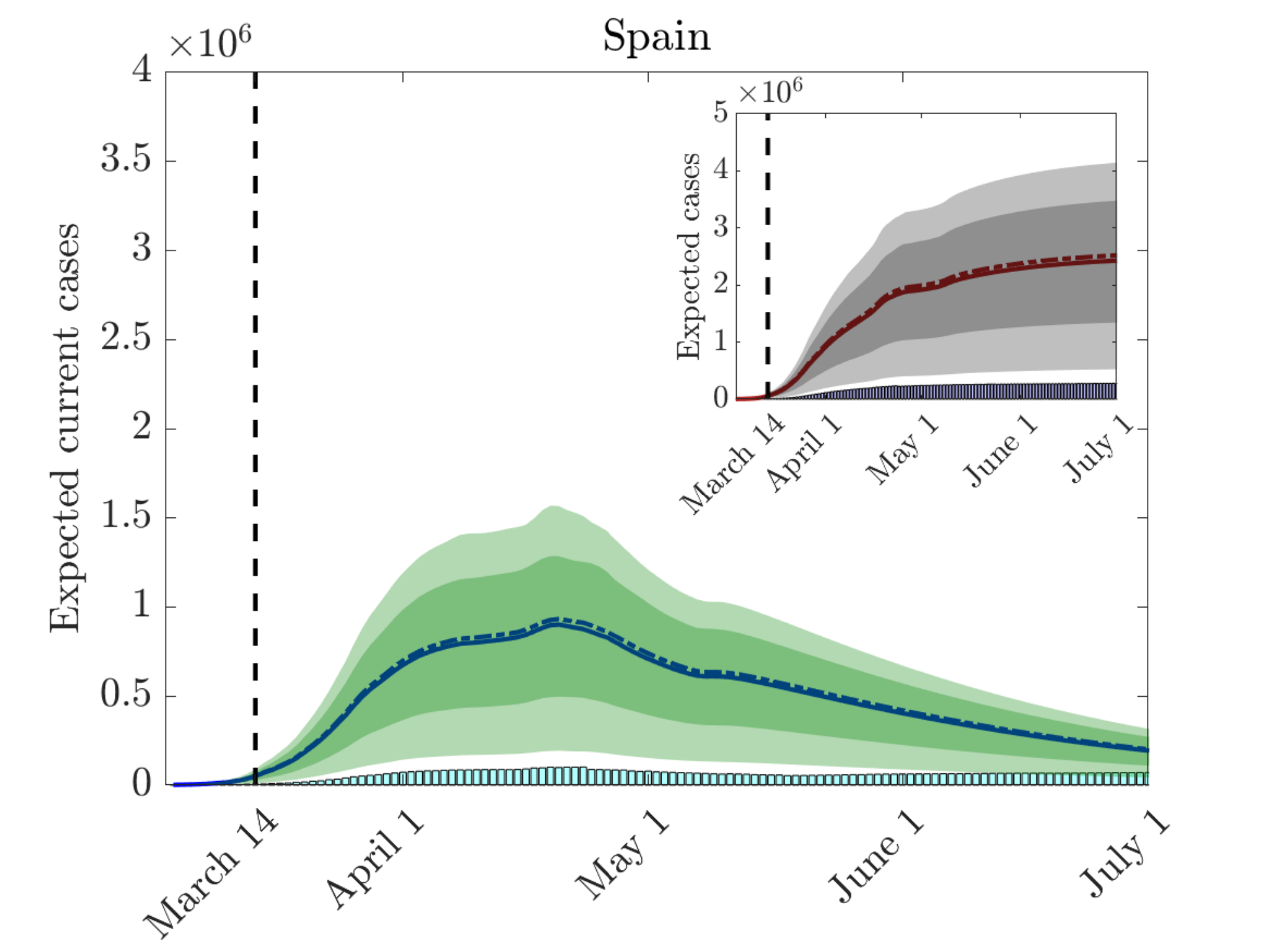} \\
\hspace{-0.5cm}
\includegraphics[scale =.3]{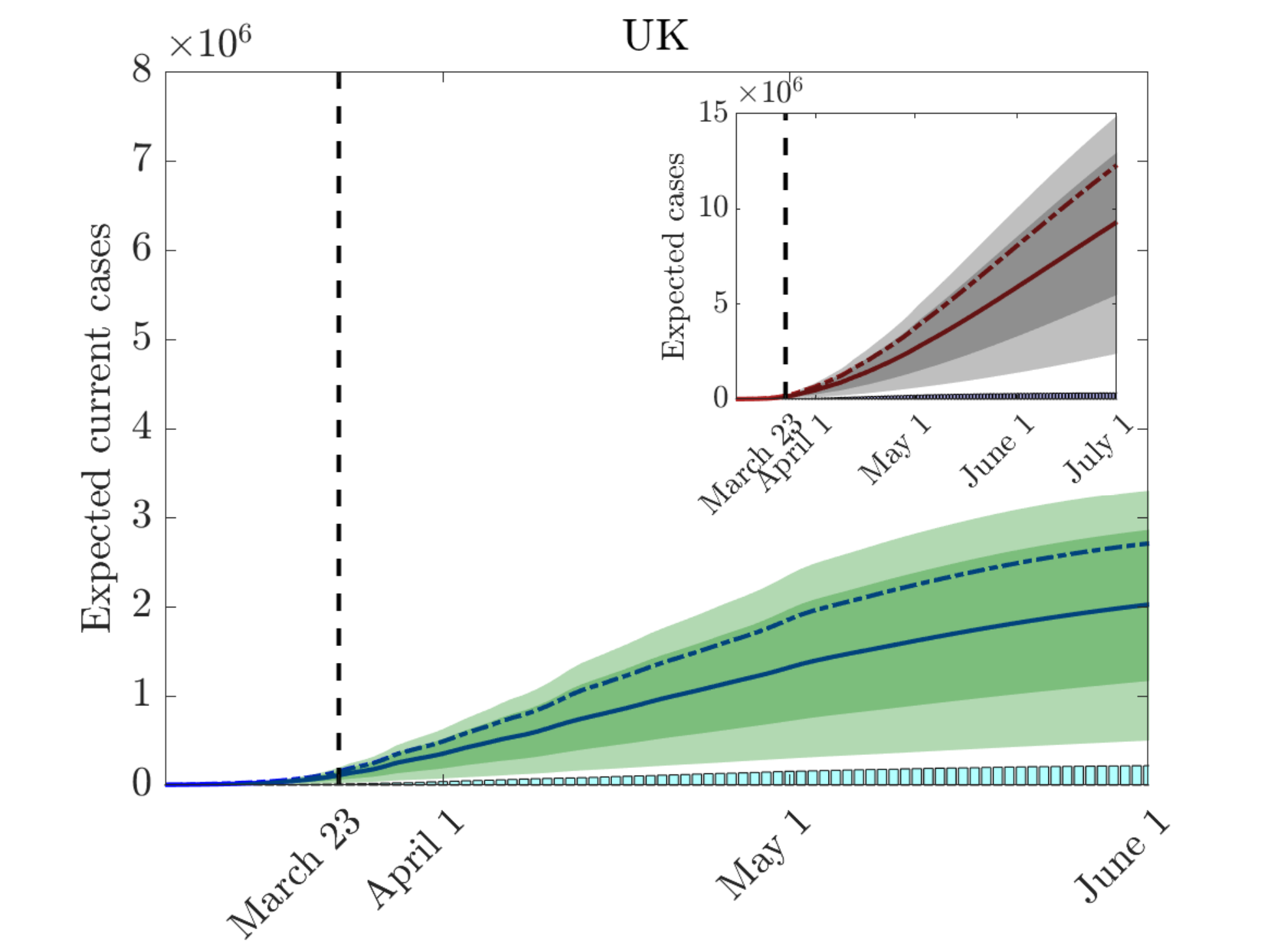} 
\includegraphics[scale =.3]{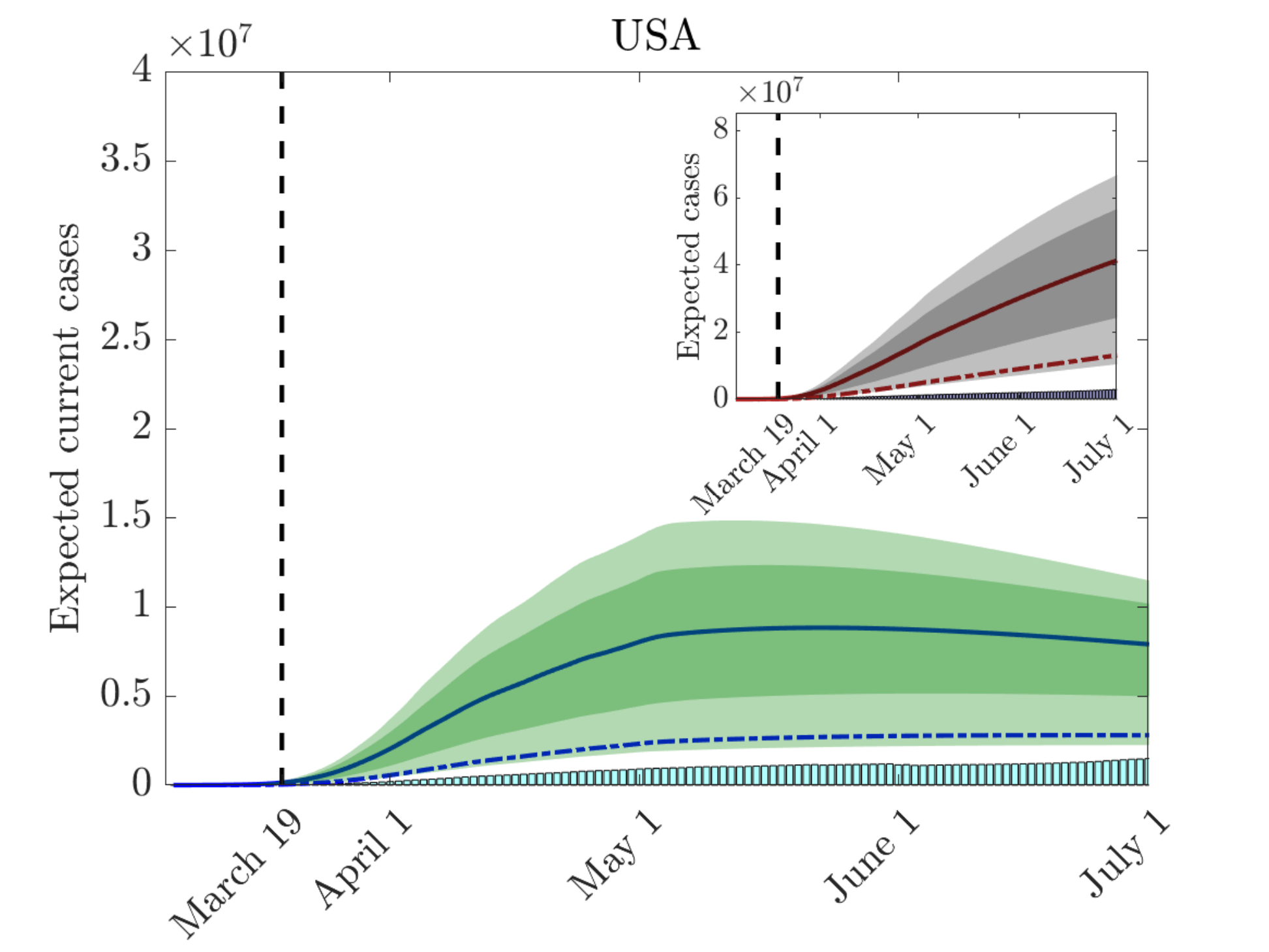}
\caption{Evolution of current and total cases for each country with uncertain initial data as in \eqref{eq:ir_z1} based on the average uncertainty between countries. The $95\%$ and $50\%$ confidence levels  are represented as shaded and darker shaded areas respectively. The dash-dotted lines denote the expected trends with a country dependent uncertainty from Figure \ref{fig:inc}. }
\label{fig:infected_UQ}
\end{figure}

In Figure \ref{fig:infected_UQ} we report the results concerning the evolution of estimated current infectious cases from the beginning of the pandemic in the reference countries using $z_1\sim B(10,10)$ and $\alpha_{\beta}=\alpha_{\gamma}=0$. In the inset figures the evolution of total cases is reported. The expected number of infectious is  plotted with blue continuous line. Furthermore, to highlight the country-dependent underestimation of cases we report with dash-dotted lines both the expected evolutions, where the uncertain parameter $c>0$ varies from country to country accordingly to the numbers on the top of the red bars in Figure \ref{fig:inc}. 

\begin{figure}
\centering
\hspace{-0.5cm}
\includegraphics[scale =.22]{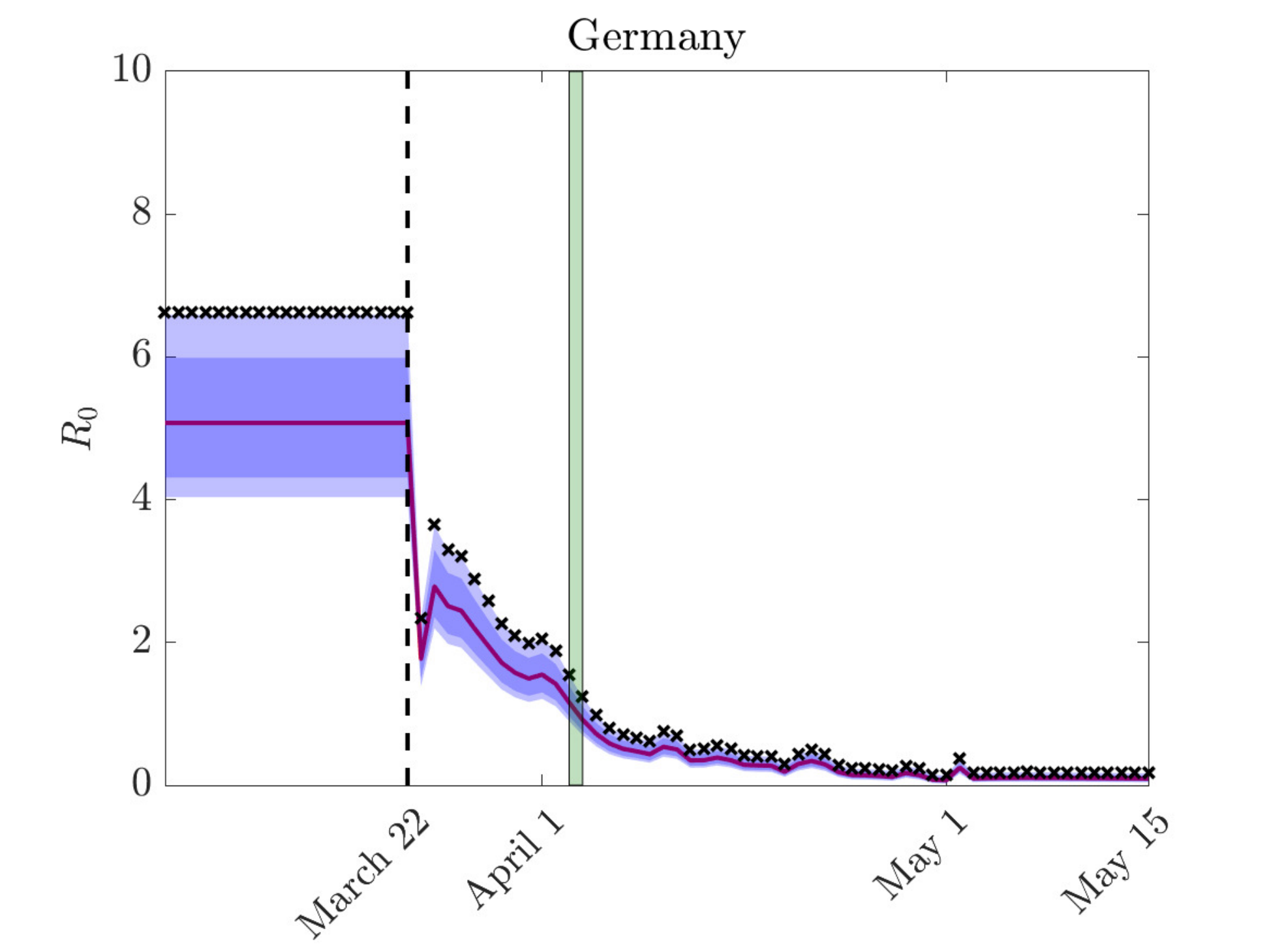}\hspace{-0.25cm}
\includegraphics[scale =.22]{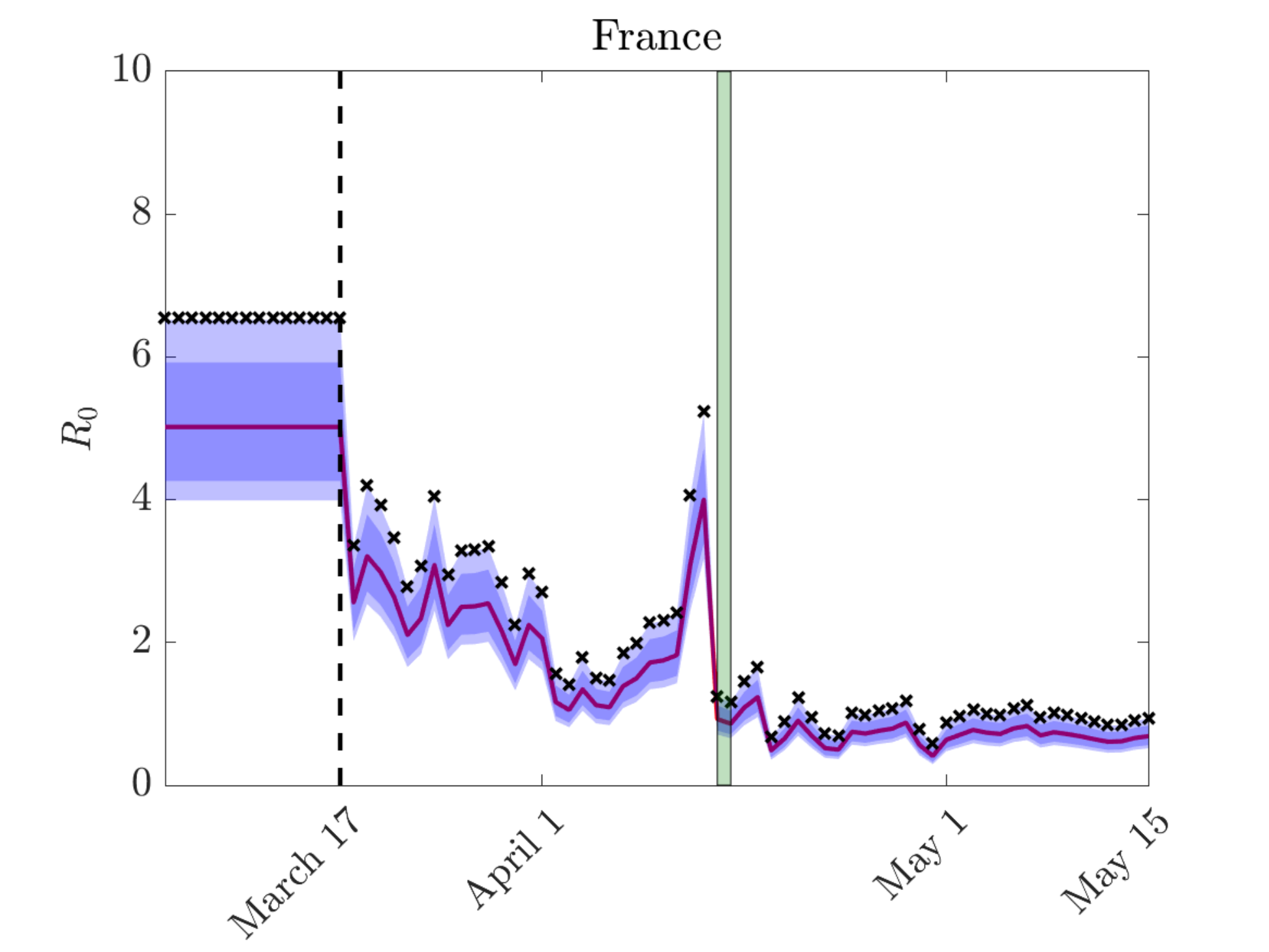}\hspace{-0.25cm}
\includegraphics[scale =.22]{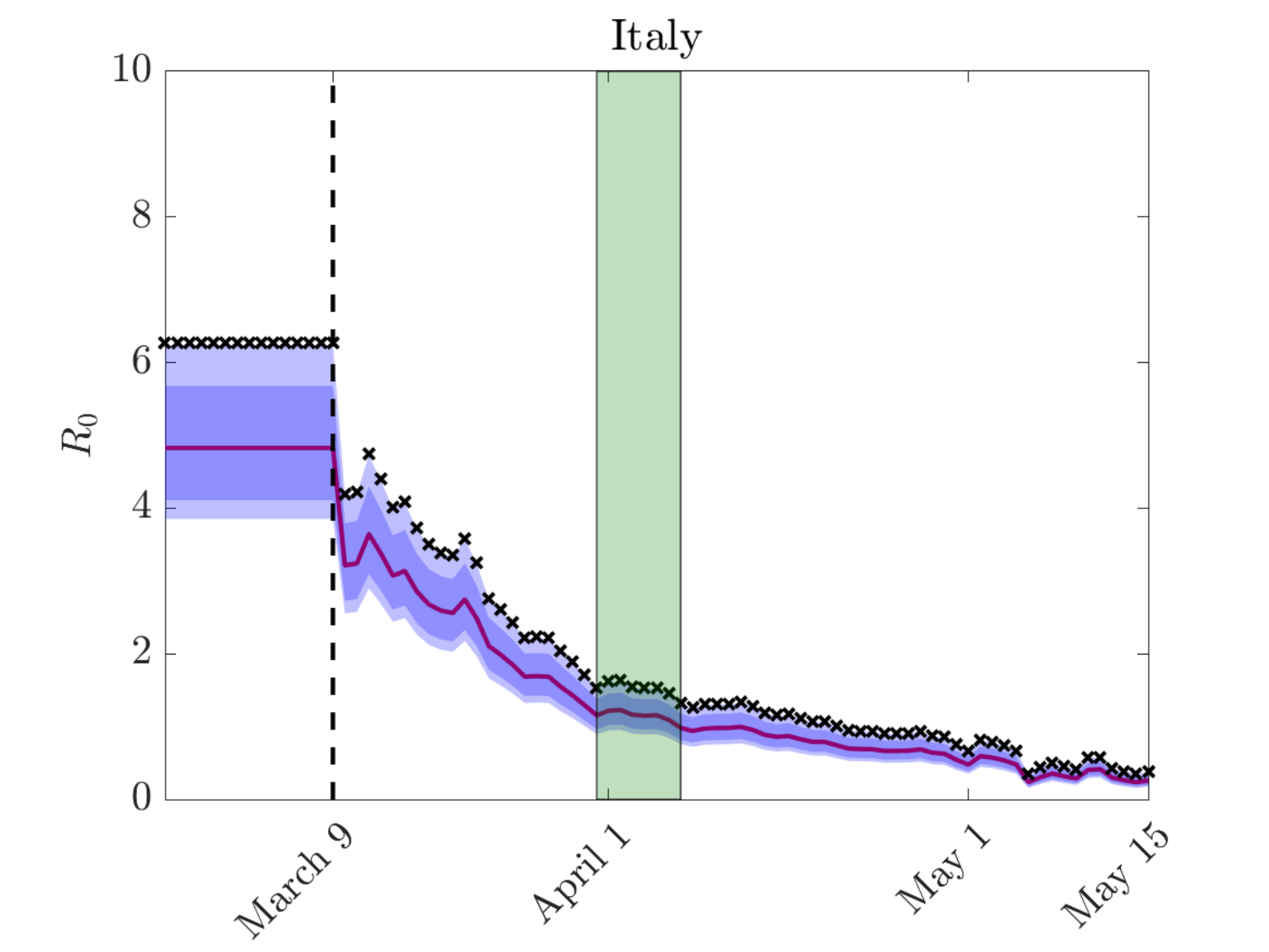} \\ \hspace{-0.25cm}
\includegraphics[scale =.22]{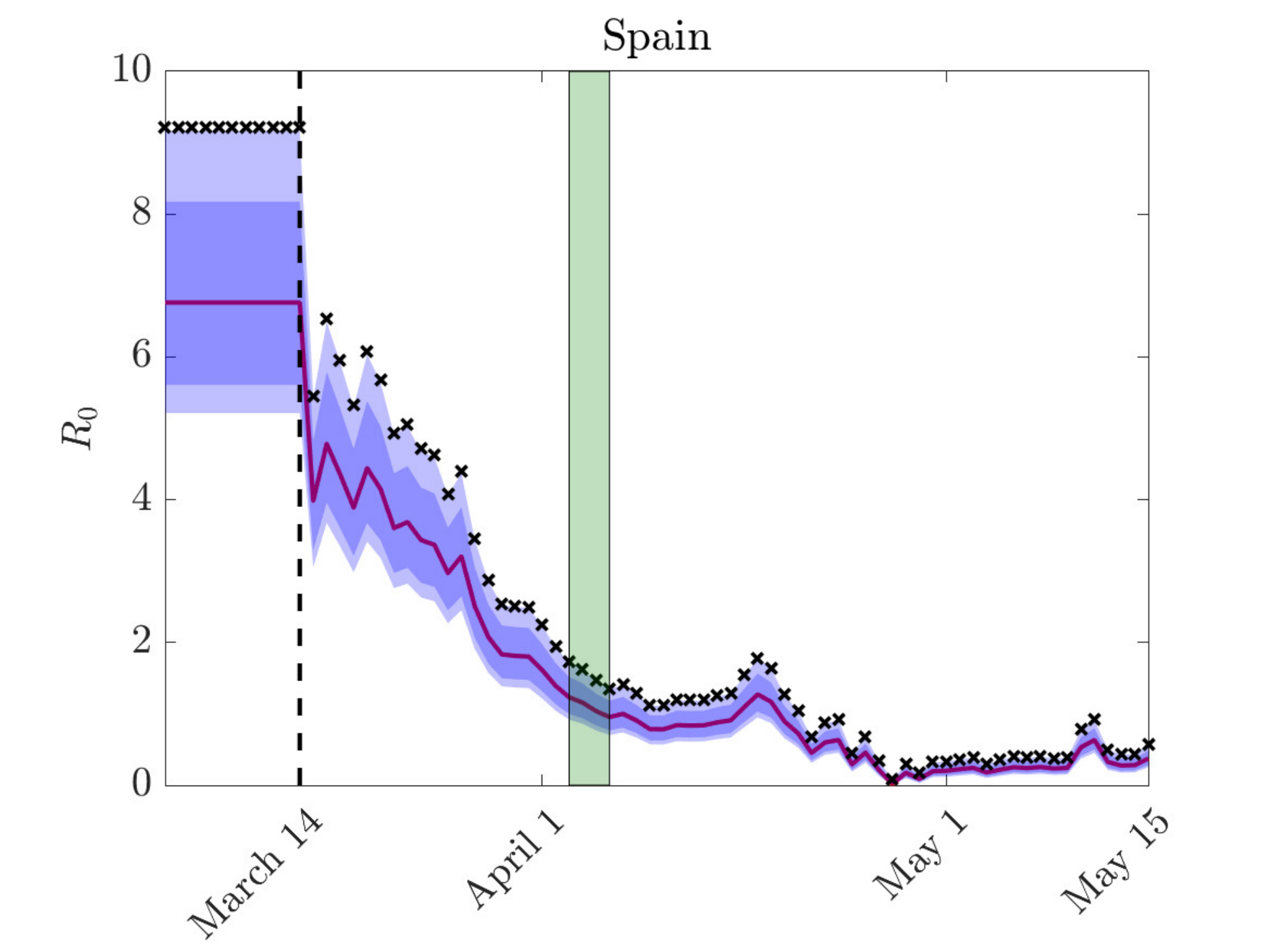}\hspace{-0.25cm}
\includegraphics[scale =.22]{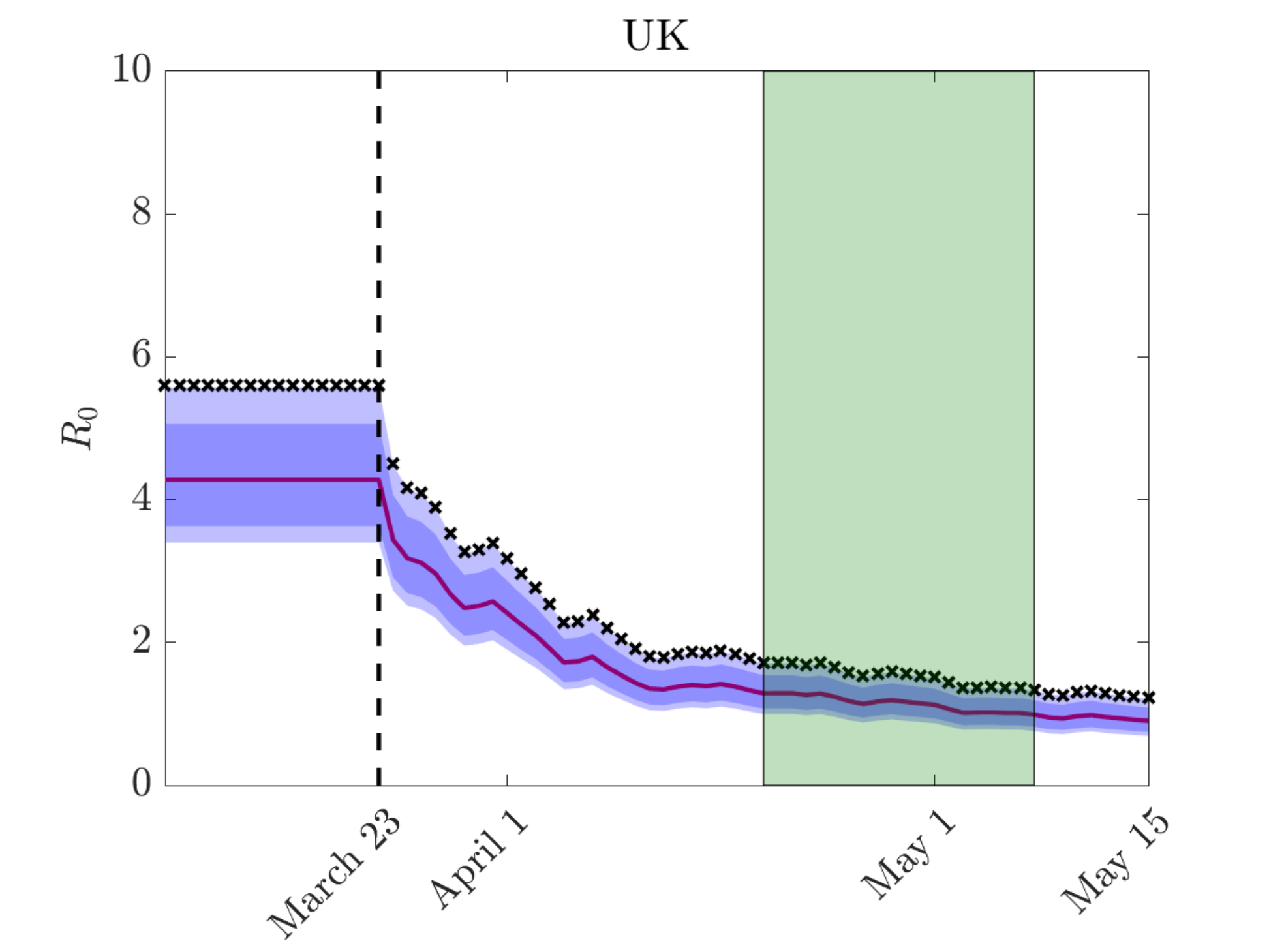}\hspace{-0.25cm}
\includegraphics[scale =.22]{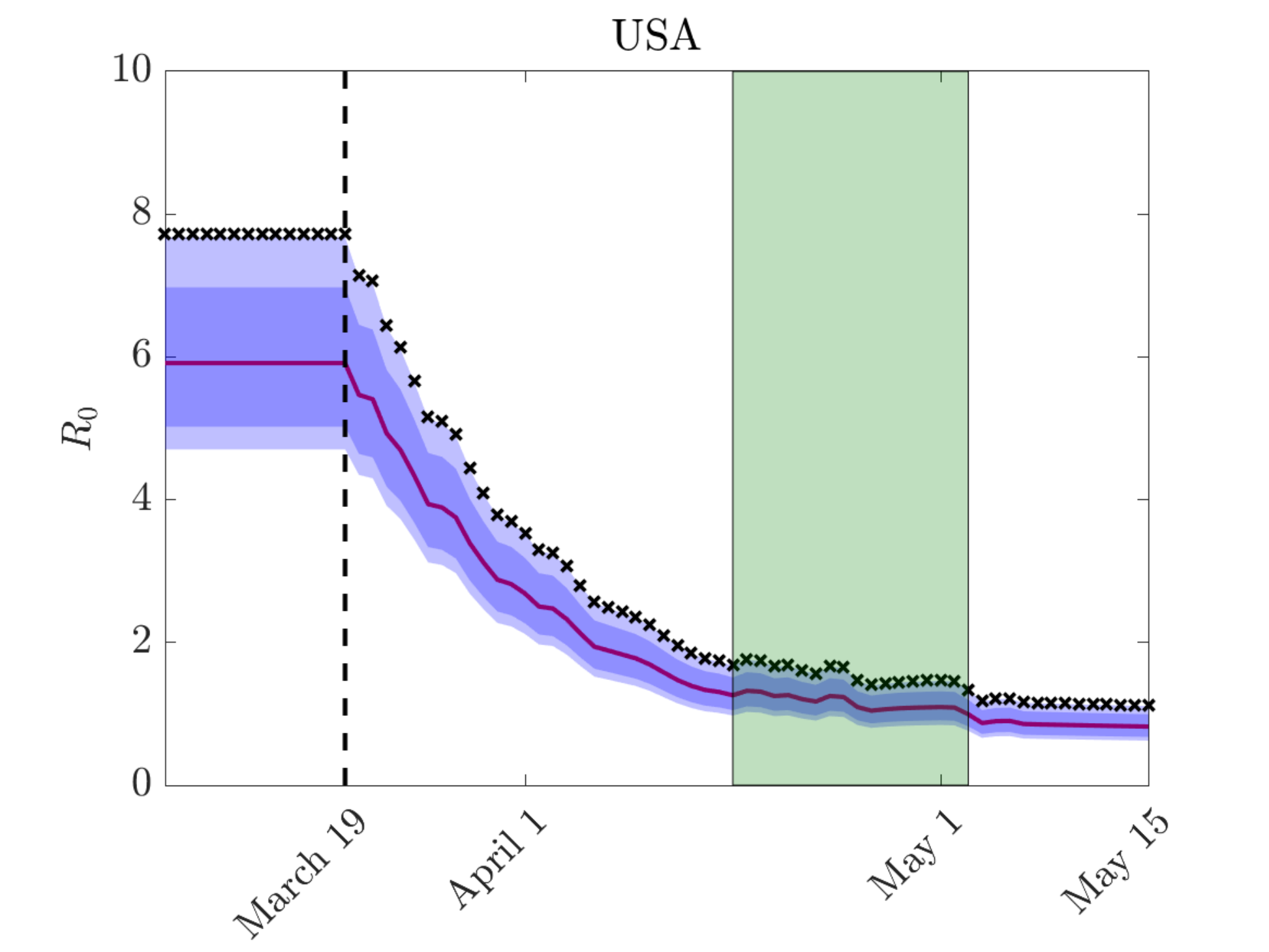}
\caption{Evolution of estimated reproduction number $R_0$ and its confidence bands for uncertain data in as in \eqref{eq:beta_z2}. The $95\%$ and $50\%$ confidence levels  are represented as shaded and darker shaded areas respectively. The green zones denote the interval between the first day  the $50\%$ confidence band and the expected value fall below $1$. }
\label{fig:R0}
\end{figure}

In Figure \ref{fig:R0} we report the evolution of reproduction number $R_0$ for the considered countries under the uncertainties in \eqref{eq:beta_z2} obtained with $\alpha_\beta=0.03$, $\alpha_\gamma=0.05$ and $z_2\sim B(2,2)$. It has been reported, in fact, that deterministic methods based on compartmental models overestimate the effective reproduction number \cite{Liu}.
The reproduction number is estimated from 
\[
R_0(z_2,t) = \dfrac{\beta(z_2) - u(t) \chi(t>\bar t)}{\gamma(z_2)}, 
\]
being the control $u(t)$ defined in \eqref{eq:r0e} and $\bar t$ is the country-dependent lockdown time. The estimated reproduction number relative to data  is reported with x-marked symbols and represents an upper bound for $R_0(z_2,t)$. The first day that the $50\%$ confidence interval and the expected value fall below $1$ is highlighted with a shaded green region. We can observe how the model estimates that for most countries in the first days of April the reproduction number $R_0$ has fallen below the threshold of $1$. On the other hand, in the UK and the US the same condition was reached between the end of April and the beginning of May. In realistic terms these dates should be considered as overestimates as they are essentially based on observations without taking into account the delay in the data reported. 

\subsection{Relaxing control on the various social activities}
We analyze the effects of the inclusion of age dependence and social interactions in the above scenario. The number of contacts per person generally shows considerable variability depending on age, occupation, country, in relation to the social habits of the population. However, some universal features can be extracted, which emerge as a function of specific social activities. 

\begin{figure}
	\centering
	\includegraphics[scale = 0.3]{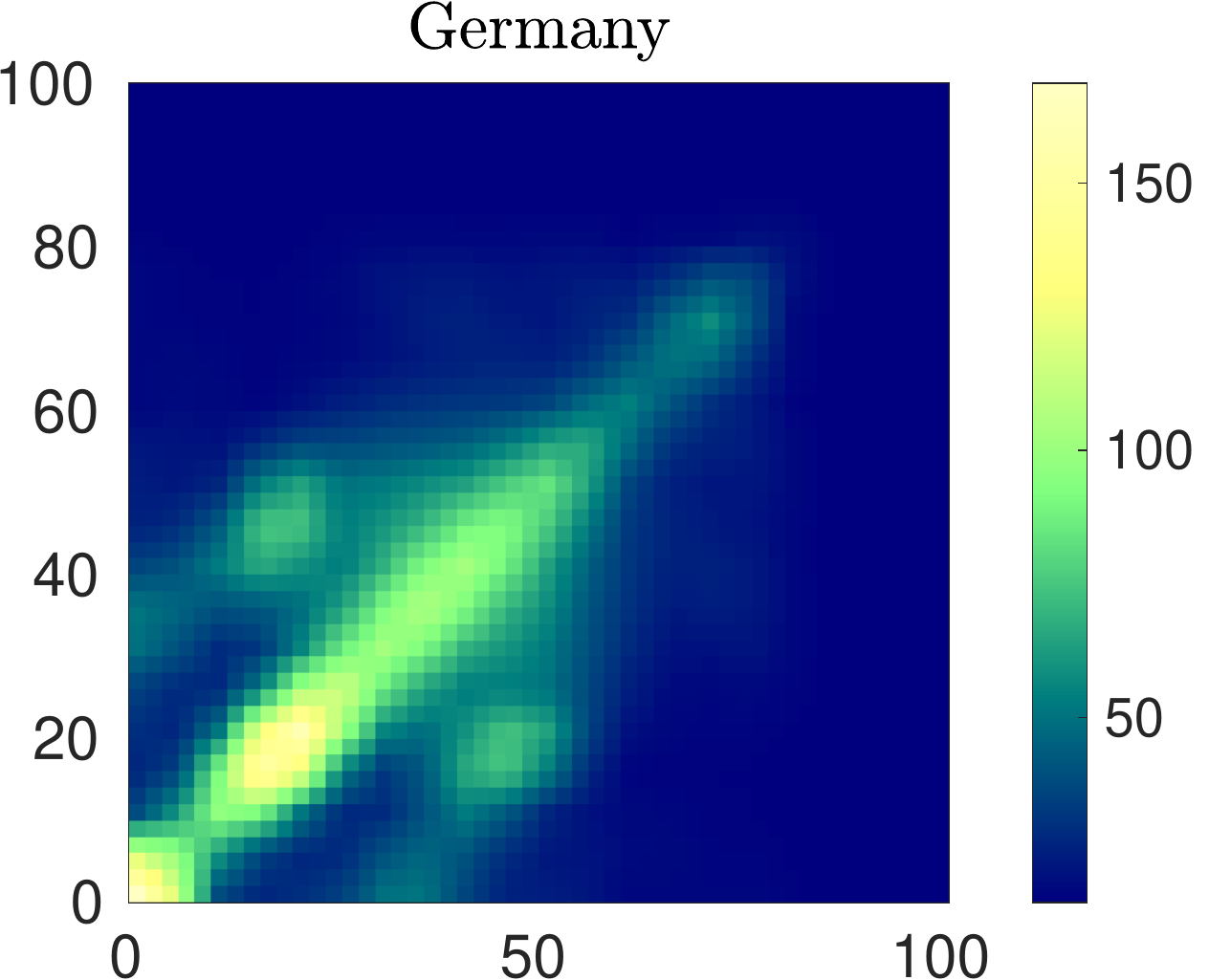}
	\includegraphics[scale = 0.3]{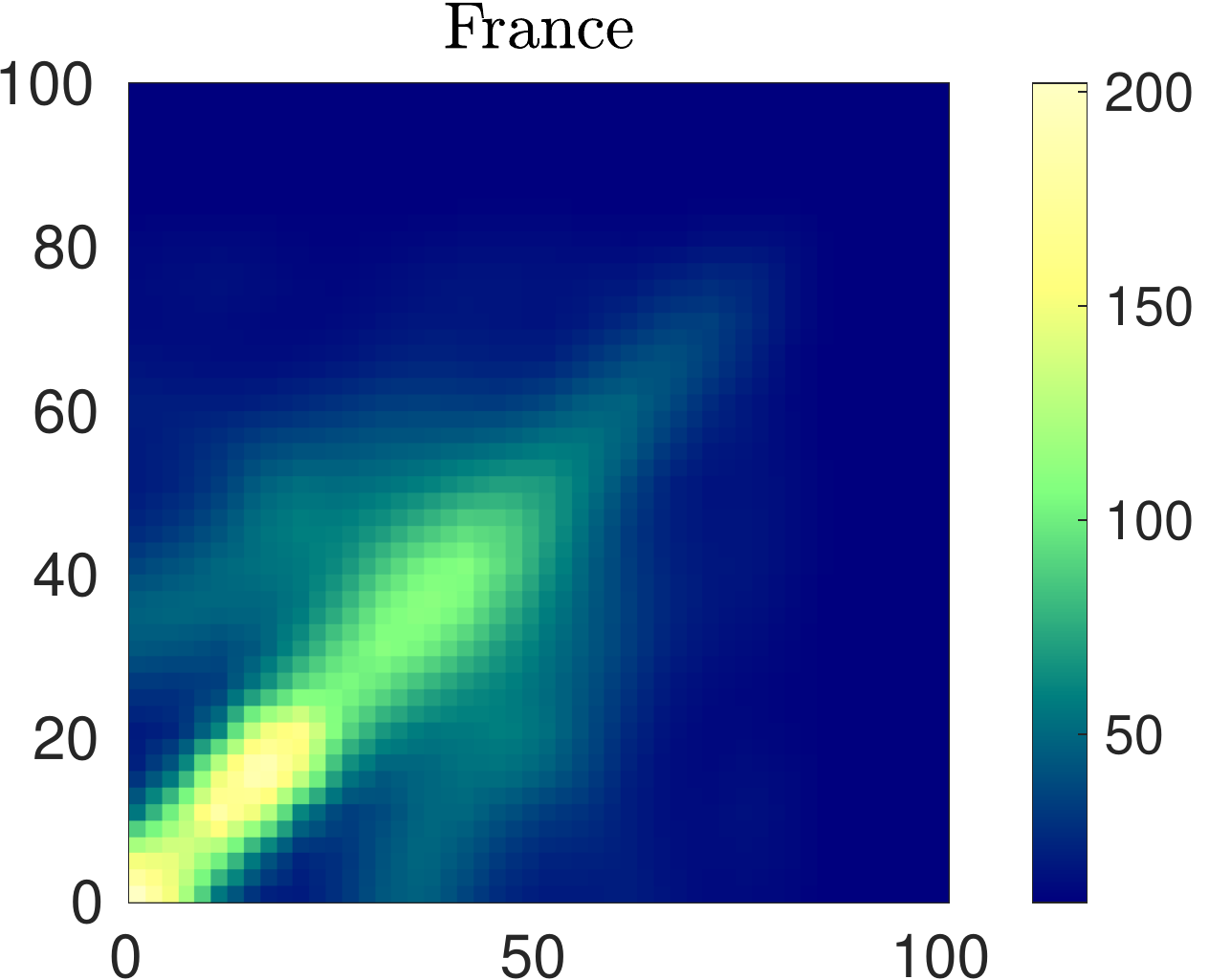}
	\includegraphics[scale = 0.3]{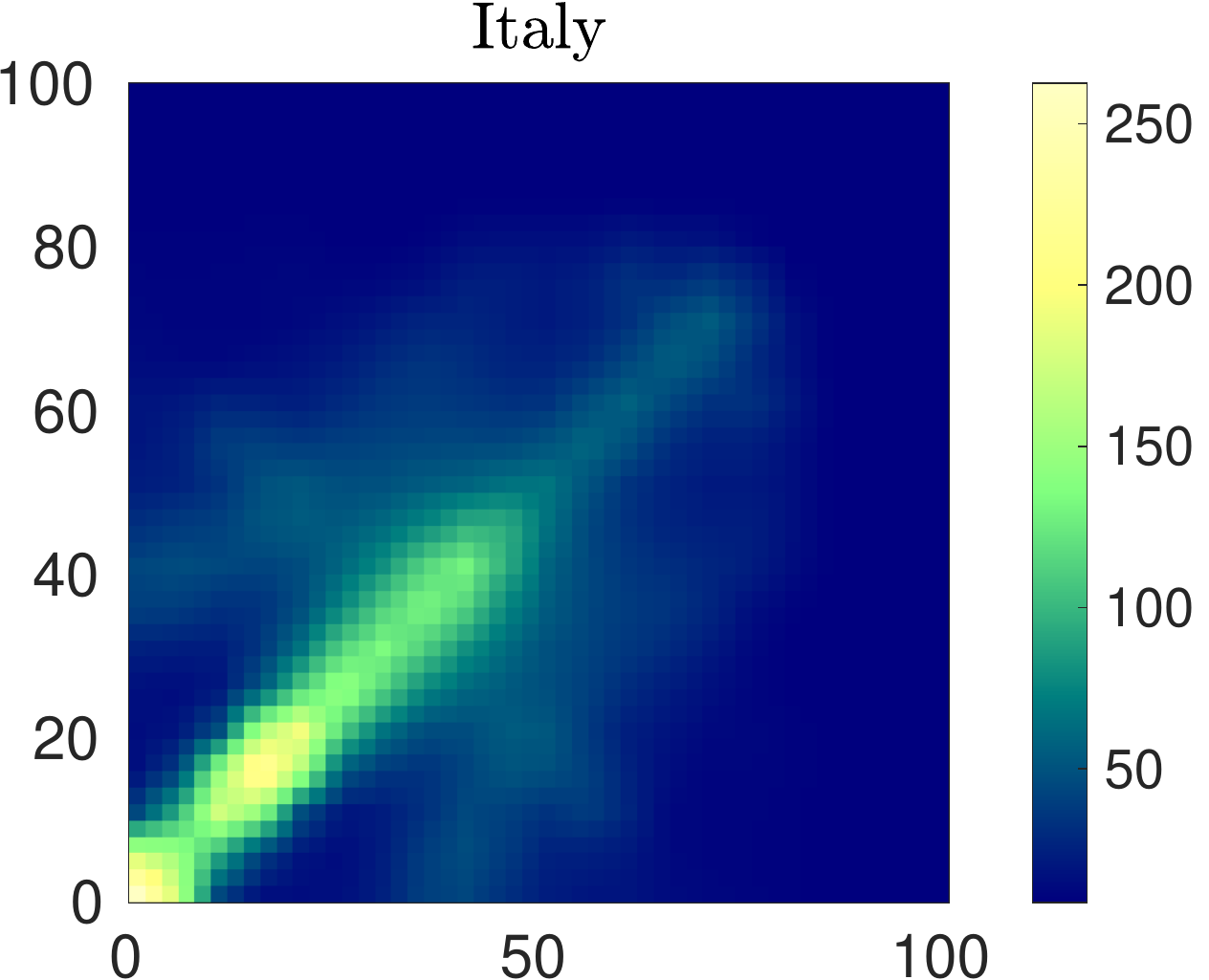}\\
	\vskip .2cm
	\includegraphics[scale = 0.3]{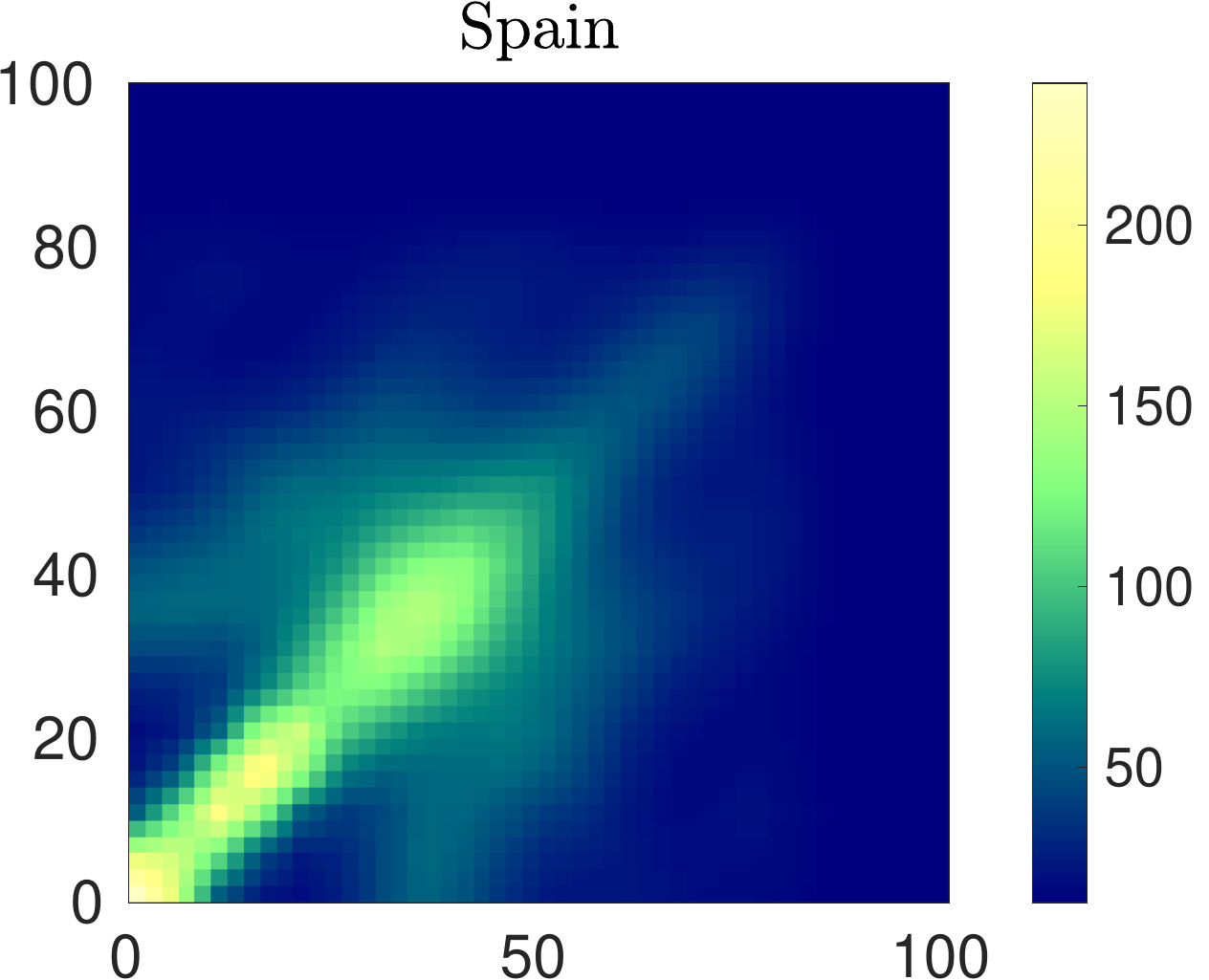}
    \includegraphics[scale = 0.3]{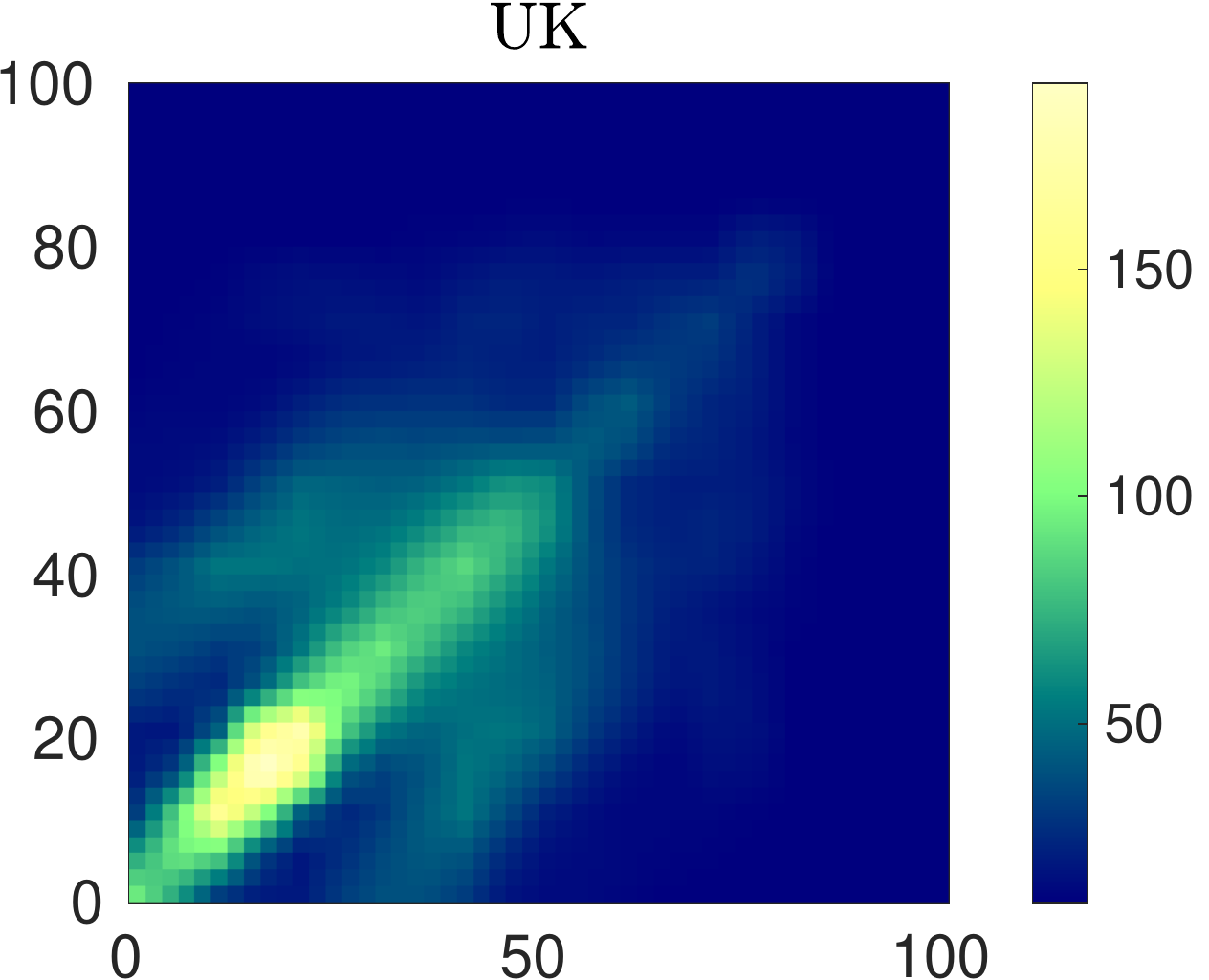}
	\includegraphics[scale = 0.3]{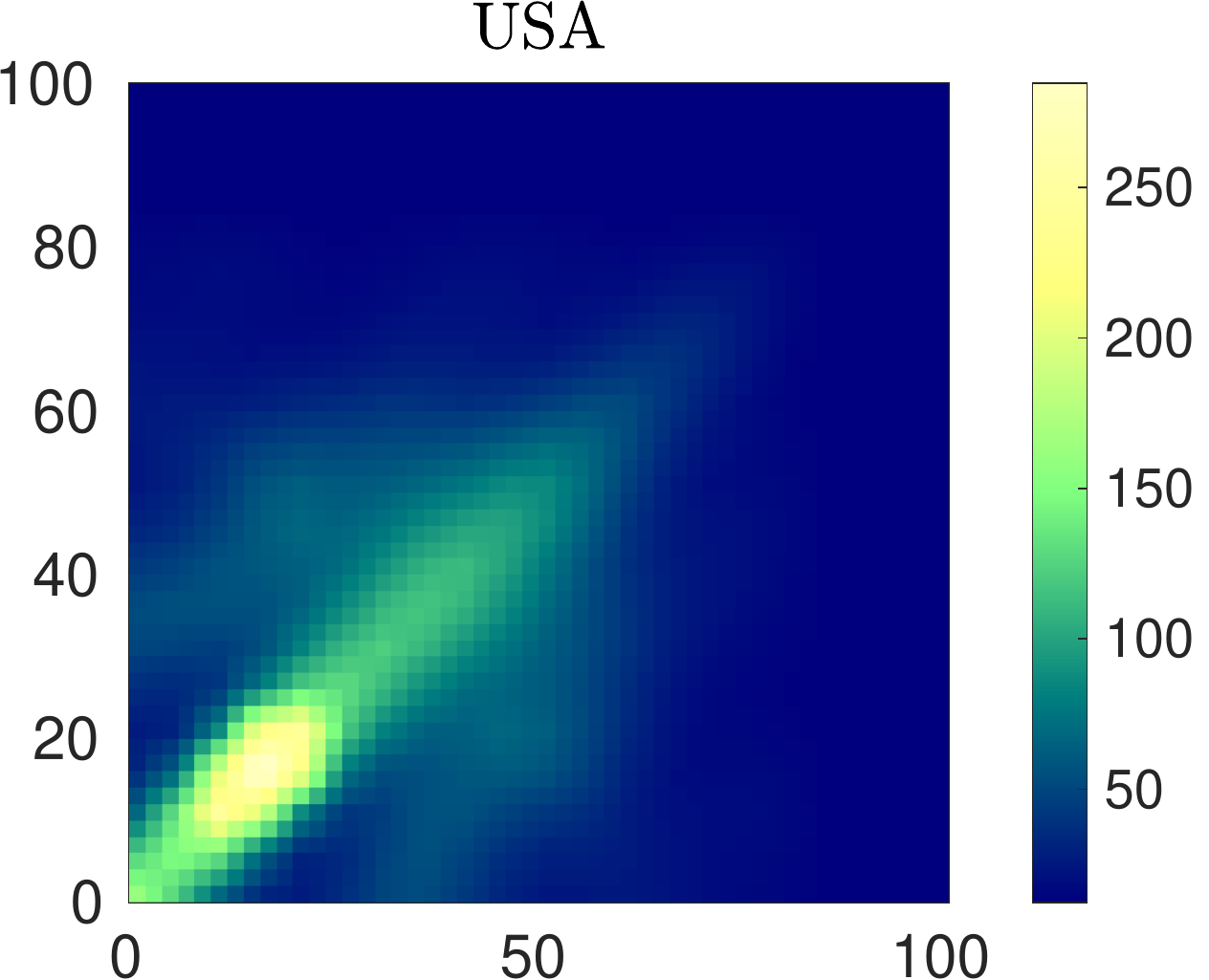}
	\vskip .2cm
		\caption{The total contact interaction function $\beta=\beta_F+\beta_E+\beta_P+\beta_O$ taking into account the contact rates of people with different ages. Family and home contacts are characterized by $\beta_F$, education and school contacts by $\beta_E$, professional and work contacts by $\beta_P$, and other contacts by $\beta_O$.}
	\label{fg:matrices}
\end{figure}

More precisely, we consider the social interaction functions corresponding to the contact matrices in \cite{PCJ} for the various countries. As a result we have four interaction functions characterized by $\mathcal A = \{F,E,P,O\}$, where we identify family and home contacts with $\beta_F$, education and school contacts with $\beta_E$,  professional and work contacts with $\beta_P$, and other contacts with $\beta_O$. These 
functions have been reconstructed over the age interval $\Lambda=[0,a_{\max}]$, $a_{\max}=100$ using linear interpolation. We report in Figure \ref{fg:matrices}, as an example, the total social interaction functions for the various countries. The functions share a similar structure but with different scalings accordingly to the country specific features identified in \cite{PCJ}.

In order to match the age-structured model with the homogeneous mixing model the social functions were normalized using the previously estimated parameters $\beta_e$ and $\gamma_e$ in accordance with
\be\label{eq:bg}
\beta_e = \frac1{a^2_{\max}L}\sum_{j\in\mathcal{A}}\int_{\LL\times \LL} \beta_j(a,a_*)\,da\,da_*,\qquad \gamma_e = \frac1{a_{\max}}\int_{\LL} \gamma(a)\,da.
\ee
We considered both a uniform and an age-related recovery rate \cite{GammaAge,Zanellaetal} as a decreasing function of the age in the form    
\be\label{eq:gamma_a}
\gamma(a) =  \gamma_e + C e^{-ra},  
\ee
with $r=5$ and $C\in\mathbb R$ such that \eqref{eq:bg} holds. Clearly, this choice involves a certain degree of arbitrariness since there are not yet sufficient studies on the subject, nevertheless, as we will see in the simulations, it is able to reproduce more realistic scenarios in terms of age distribution of the infected without significantly altering the behaviour relative to the total number of infected.  

In a similar spirit, to match the single control applied in the extrapolation of the penalization term $\kappa(t)$ to age dependent penalization factors $\kappa_j(a,t)$ we
redistribute their values as
\be
\kappa_j(a,t)^{-1}=\frac{w_j(t)\int_{\LL} \beta_j(a,a_*)\,da_*}{\sum_{j\in\mathcal{A}}w_j(t)\int_{\LL\times \LL} \beta_j(a,a_*)\,da\,da_*}\kappa(t)^{-1},\quad j\in \mathcal{A}
\ee
where $w_j(t)\geq 0$, are weight factors denoting the relative amount of control on a specific activity. In the lockdown period accordingly to other studies \cite{PCJ} we assume $w_E=1.5$, $w_H=0.2$, $w_P=0.5$, $w_O=0.6$, namely the largest effort of the control is due to the school closure which as a consequence implies more interactions at home. Work and other activities are equally impacted by the lockdown. In particular, these initial lockdown choices make it possible to have a good correspondence between the infectivity curves expected in the age dependent case and in the homogeneous mixing case. Therefore, these values have been set homogeneously for each country and correspond to the situation in the first lockdown period. We will discuss possible changes to these choices following a relaxation of the lockdown in the different scenarios presented below. 

\begin{figure}
\centering
\hspace{-0.4cm}
\includegraphics[scale =.3]{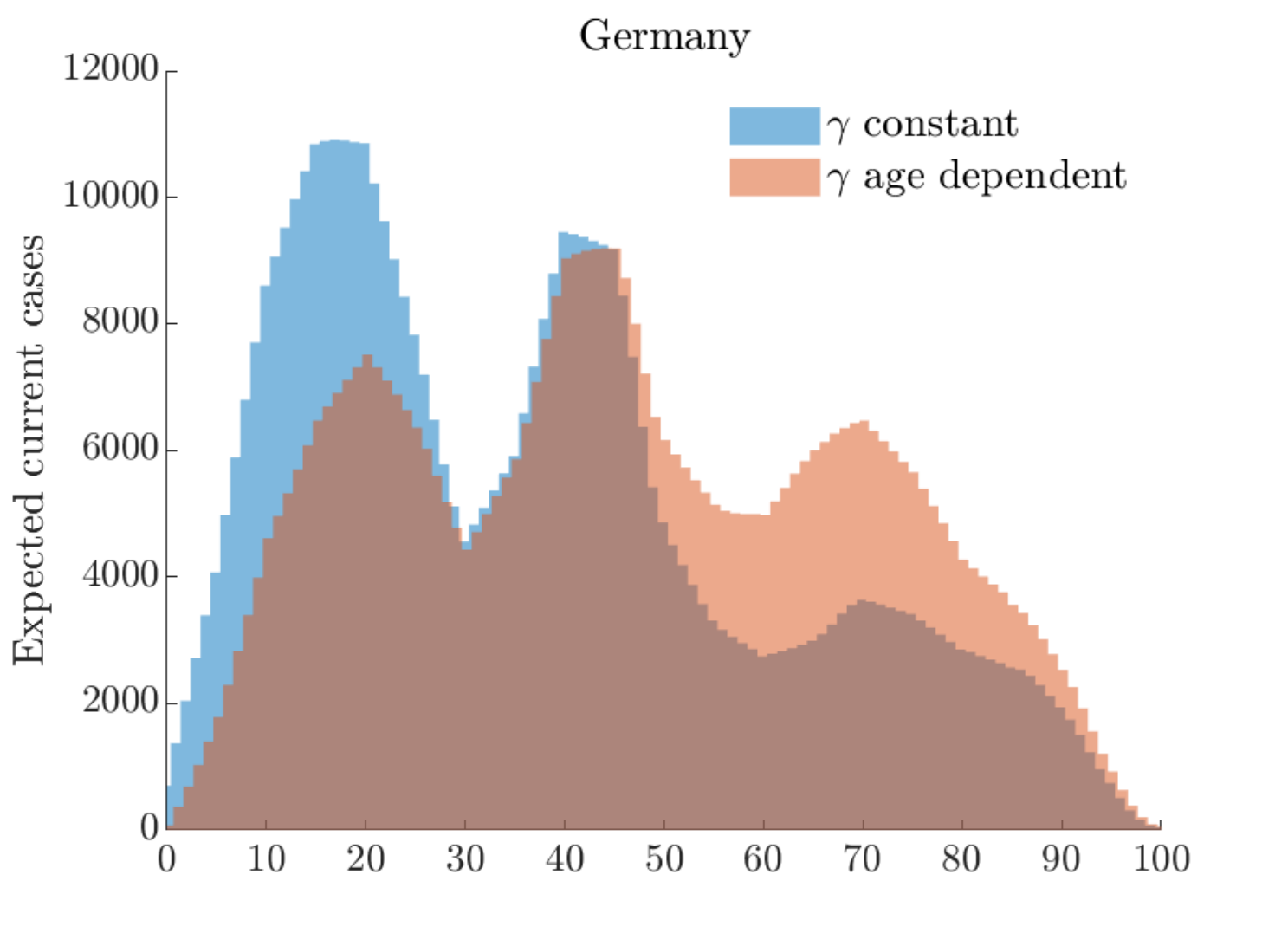}\hspace{-0.45cm}
\includegraphics[scale =.3]{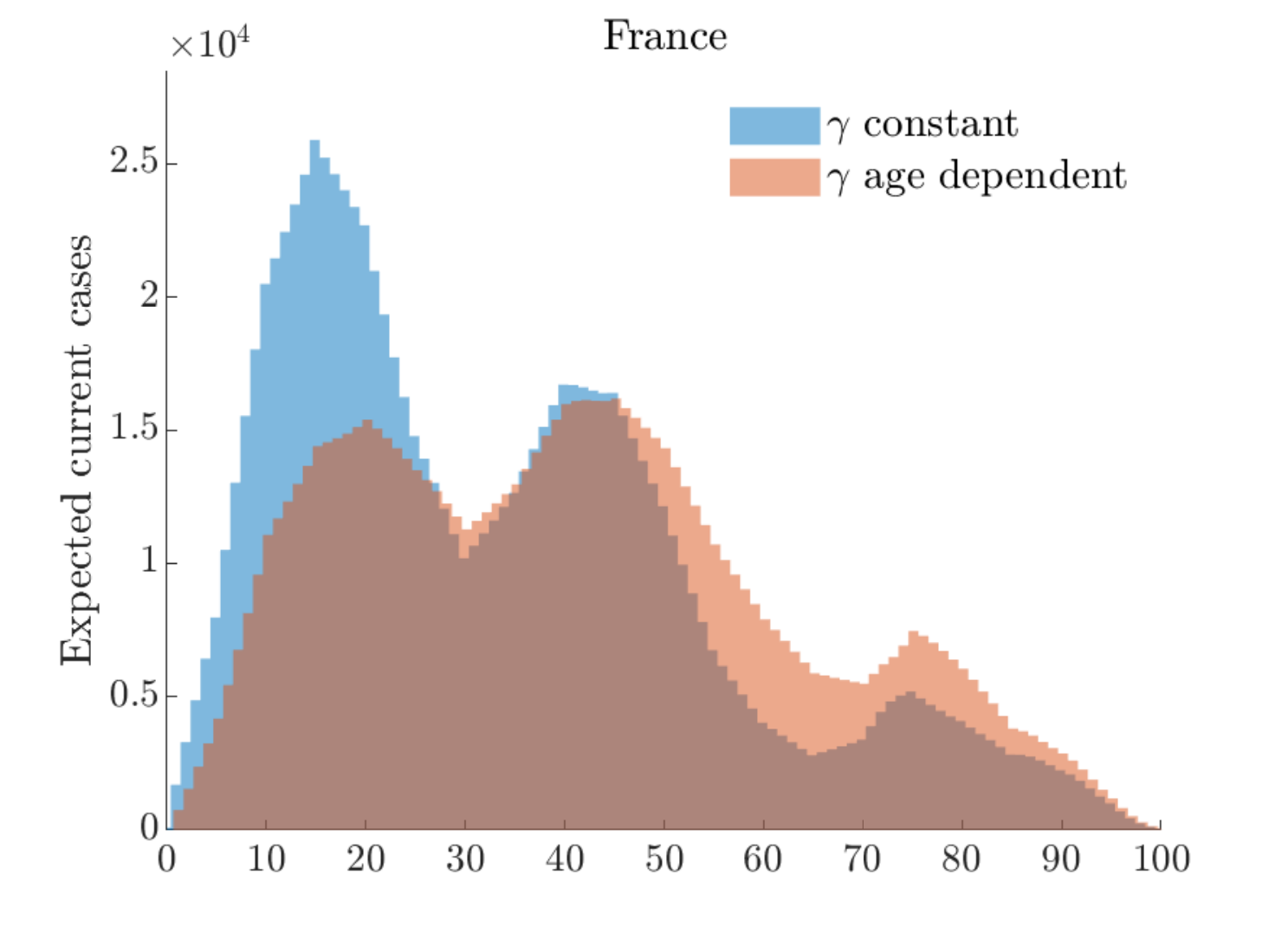}\hspace{-0.45cm}
\includegraphics[scale =.3]{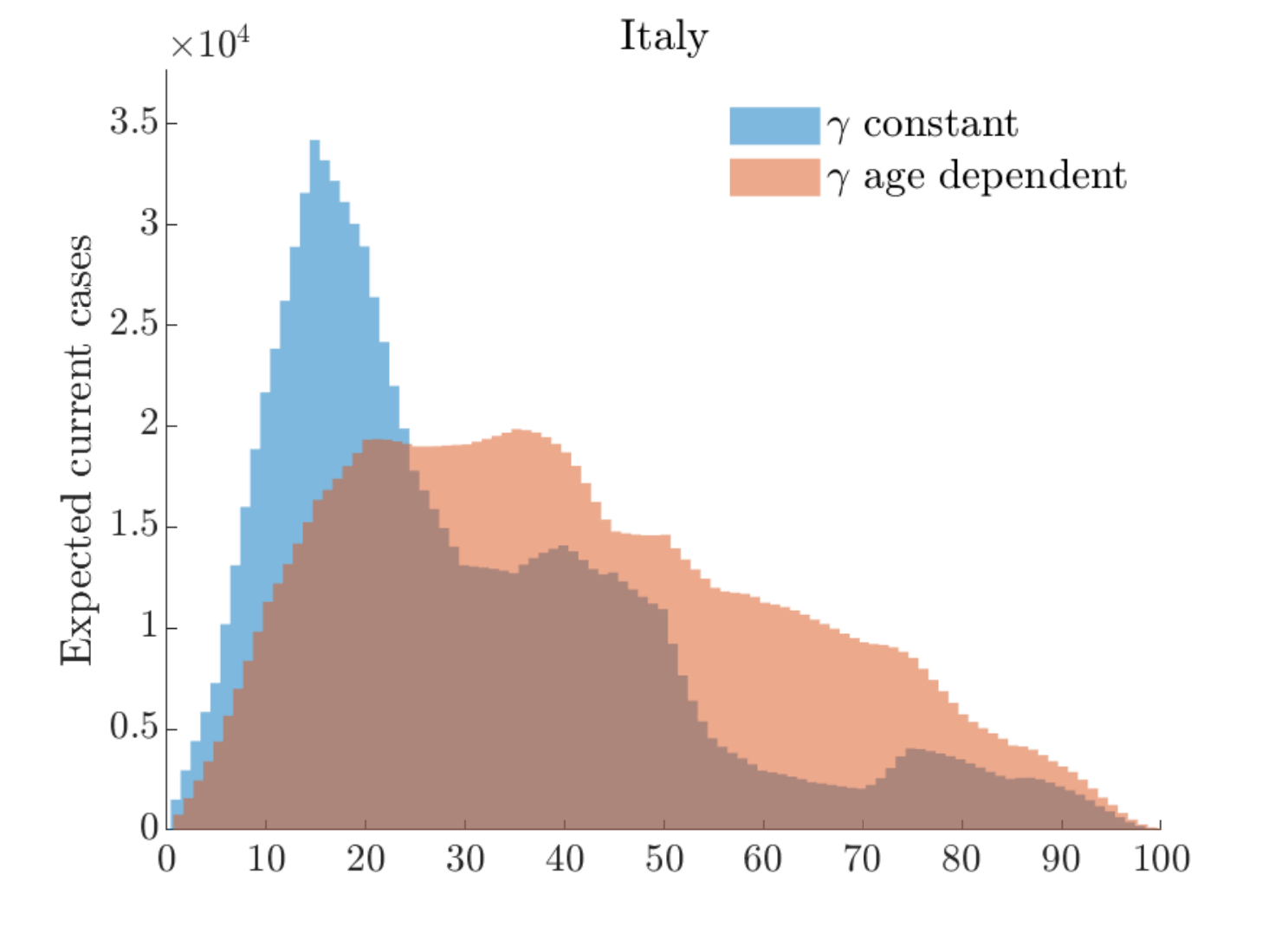} \\
\hspace{-0.5cm}
\includegraphics[scale =.3]{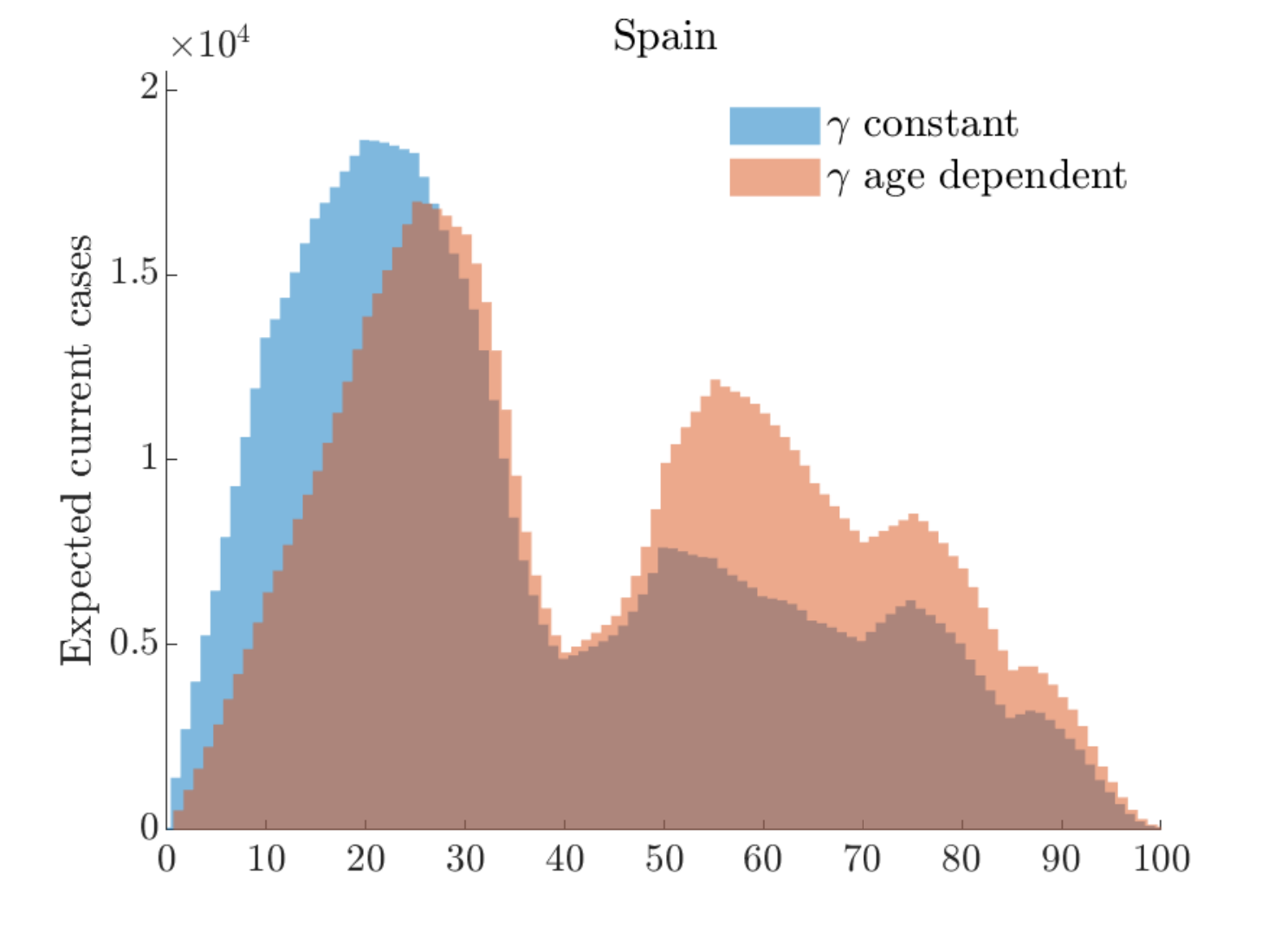}\hspace{-0.45cm}
\includegraphics[scale =.3]{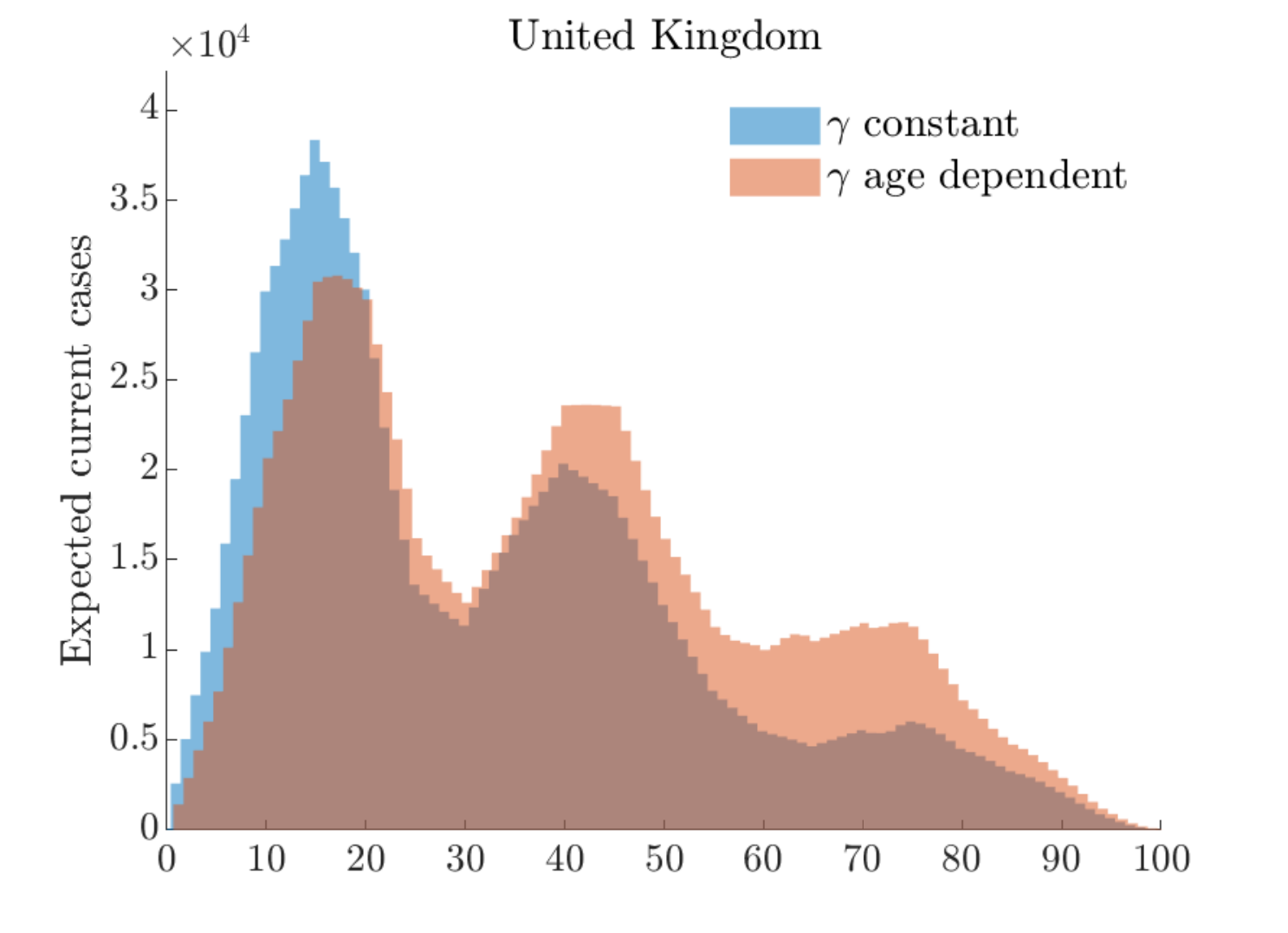}\hspace{-0.45cm}
\includegraphics[scale =.3]{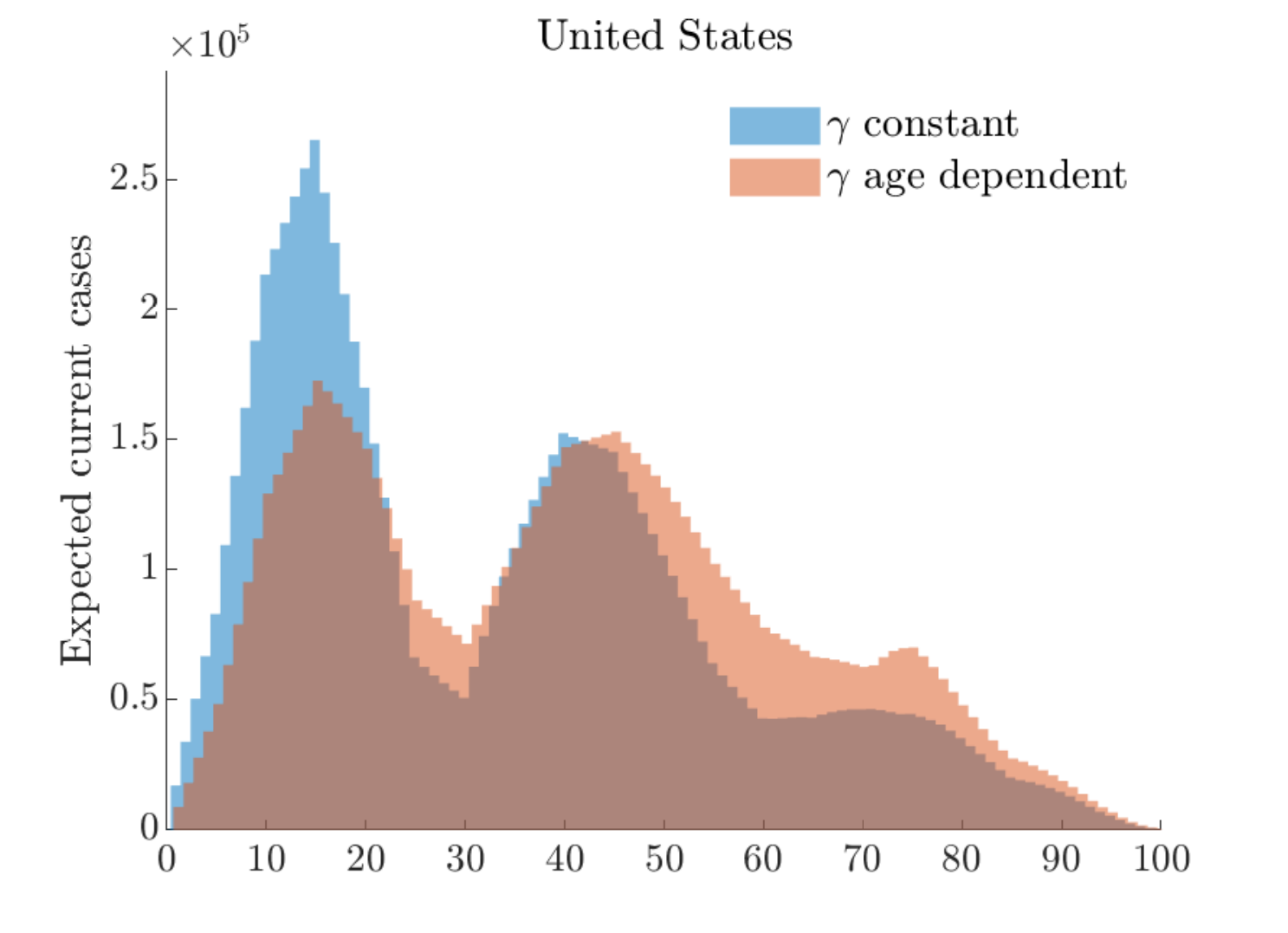}
\caption{Age distribution of infected using constant and age dependent recovery rates as in \eqref{eq:gamma_a} at the end of the lockdown period in different countries.}
\label{fig:age}
\end{figure}

We divided the computation time frame into two zones and used different models in each zone, in accordance with the policy adopted by the various countries. The first time interval defines the period without any form of containment, the second the lockdown period. In the first zone we adopted the uncontrolled model with homogeneous mixing for the estimation of epidemiological parameters. Hence, in the second zone we compute the evolution of the feedback controlled age dependent model \eqref{eq:ic1} with matching (on average) interaction and recovery rates \eqref{eq:bg} and with the estimated control penalization  
$\kappa(t)$.  
The initial values for the age distributions of susceptible have been taken from the specific demographic distribution of each country. 
More difficult is to get the same informations for the infected, since reported data are rather heterogeneous for the various country and the initial number of individuals is very small (we selected a time frame where the reported number of infectious is larger than $200$). Therefore, we tested the available data against a uniform distribution. As there were no particular differences in the results, we decided to adopt a uniform initial distribution of the infected for all countries. In Figure \ref{fig:age} we report the age distribution of infected computed for each country at the end of the lockdown period using an age dependent recovery and a constant recovery. The differences in the resulting age distributions are evident. In subsequent simulations, to avoid an unrealistic peak of infection among young people, we decided to adopt an age-dependent recovery \cite{GammaAge}.

\begin{figure}
\centering
\includegraphics[scale =.4]{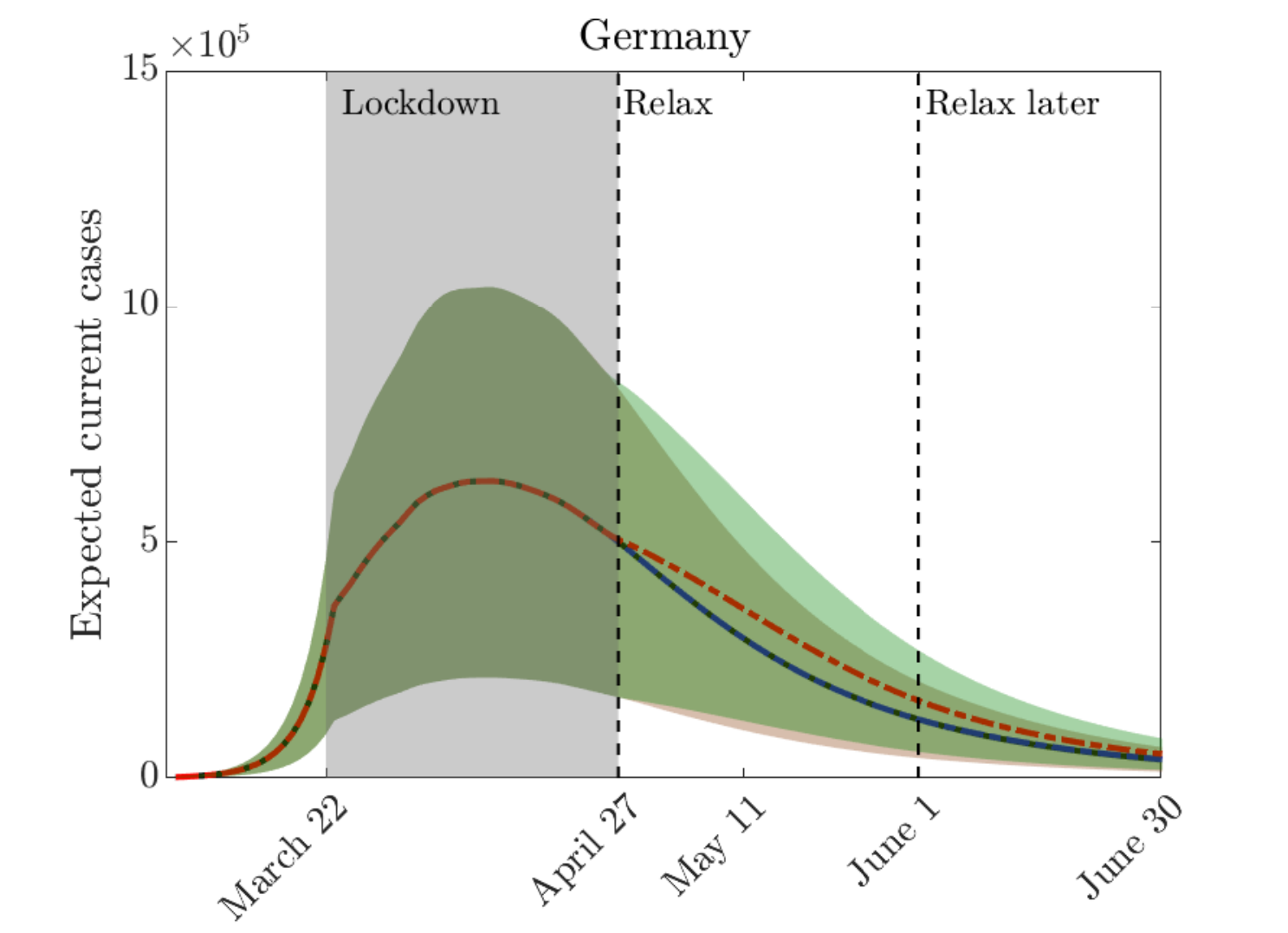}\hskip -.5cm
\includegraphics[scale =.4]{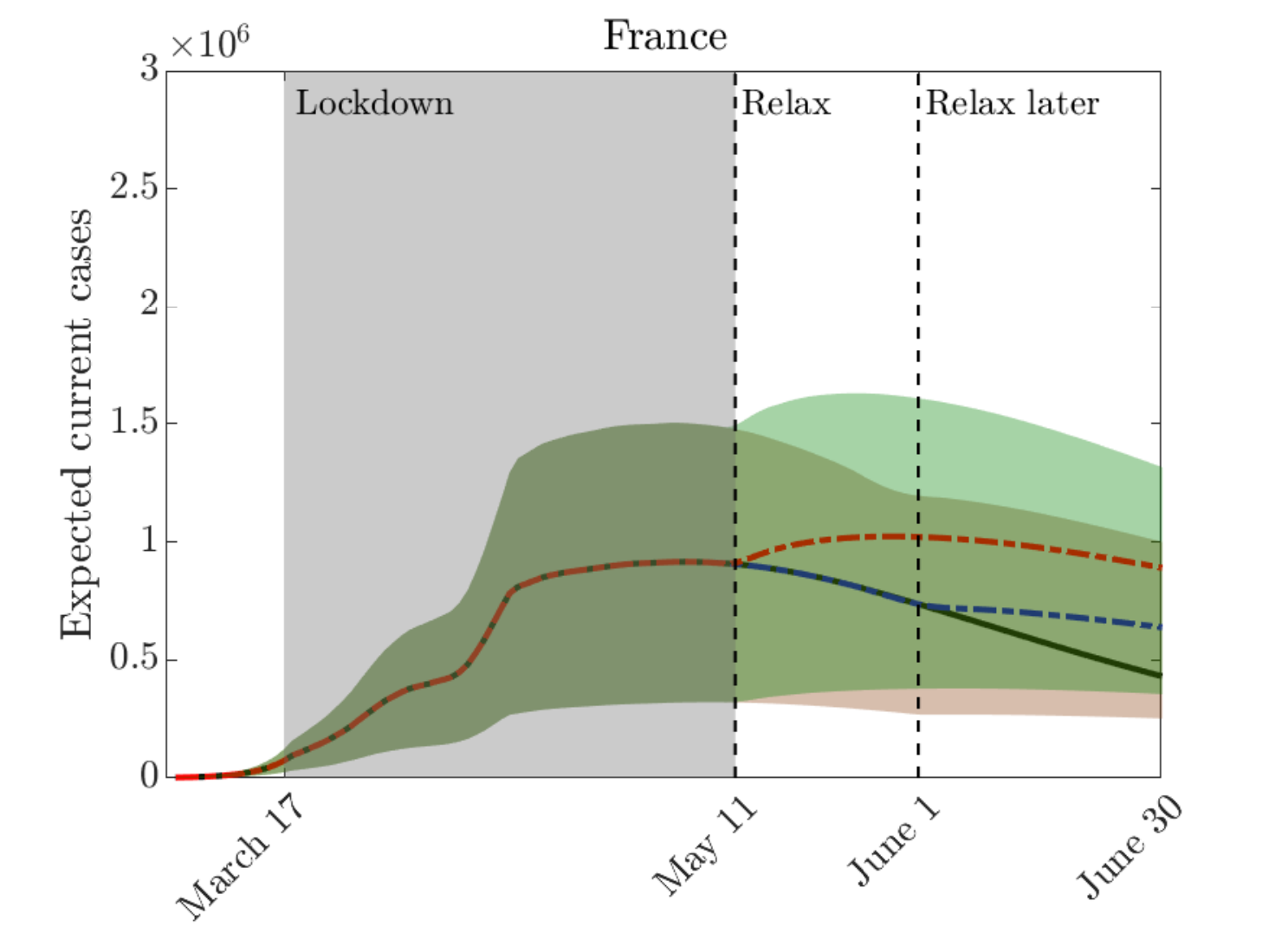}\\
\includegraphics[scale =.4]{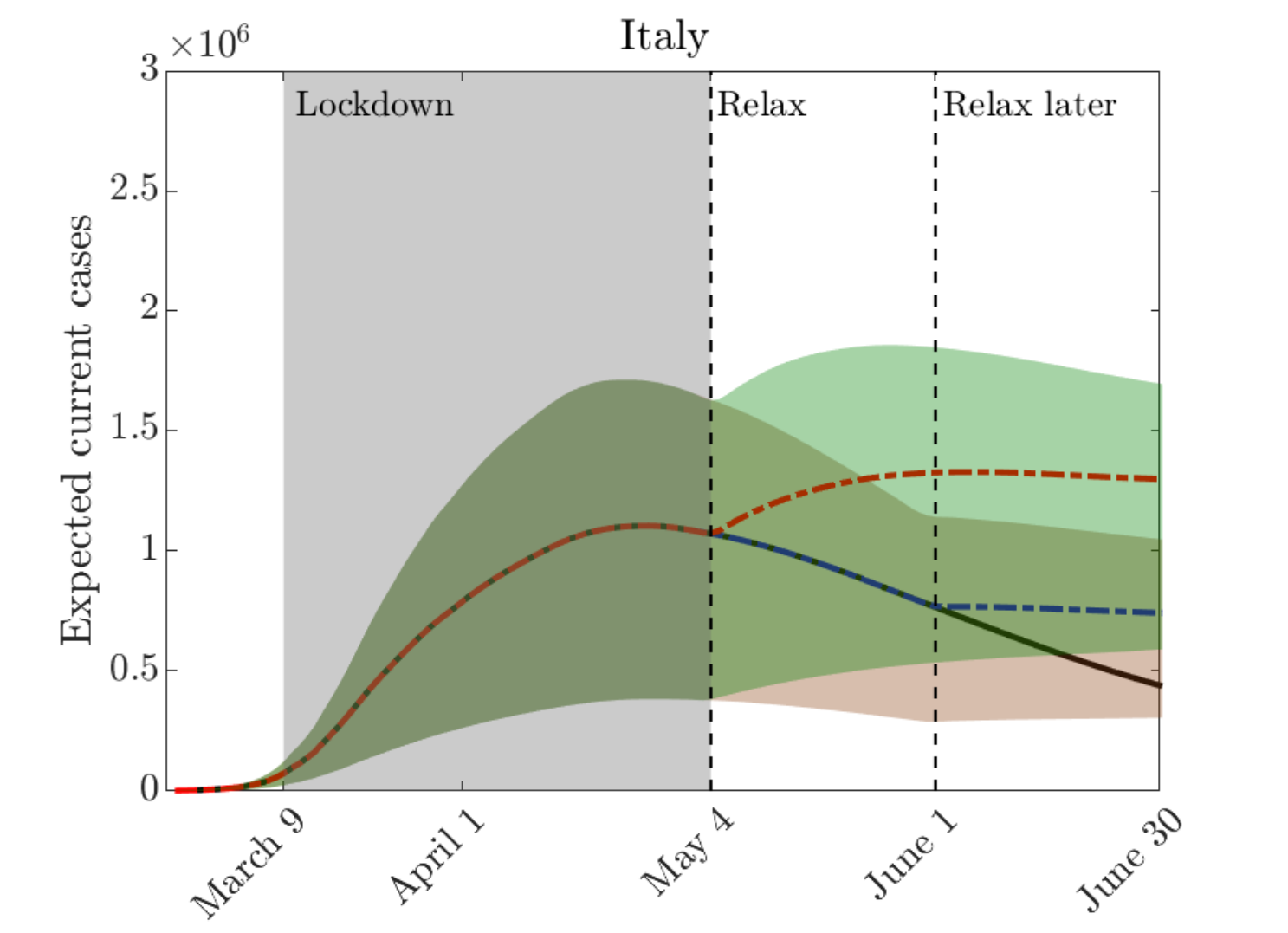} \hskip -.5cm
\includegraphics[scale =.4]{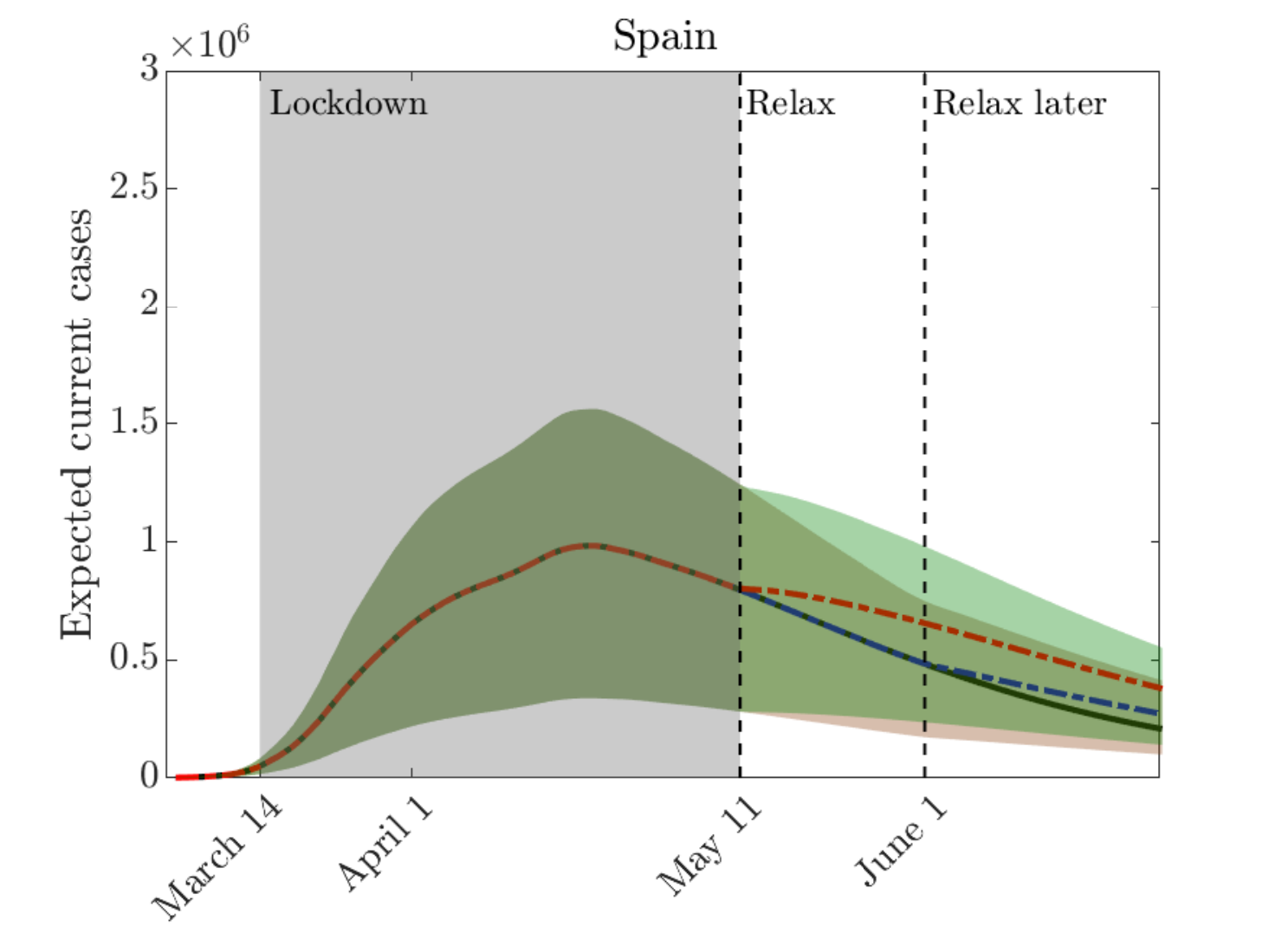}\\
\includegraphics[scale =.4]{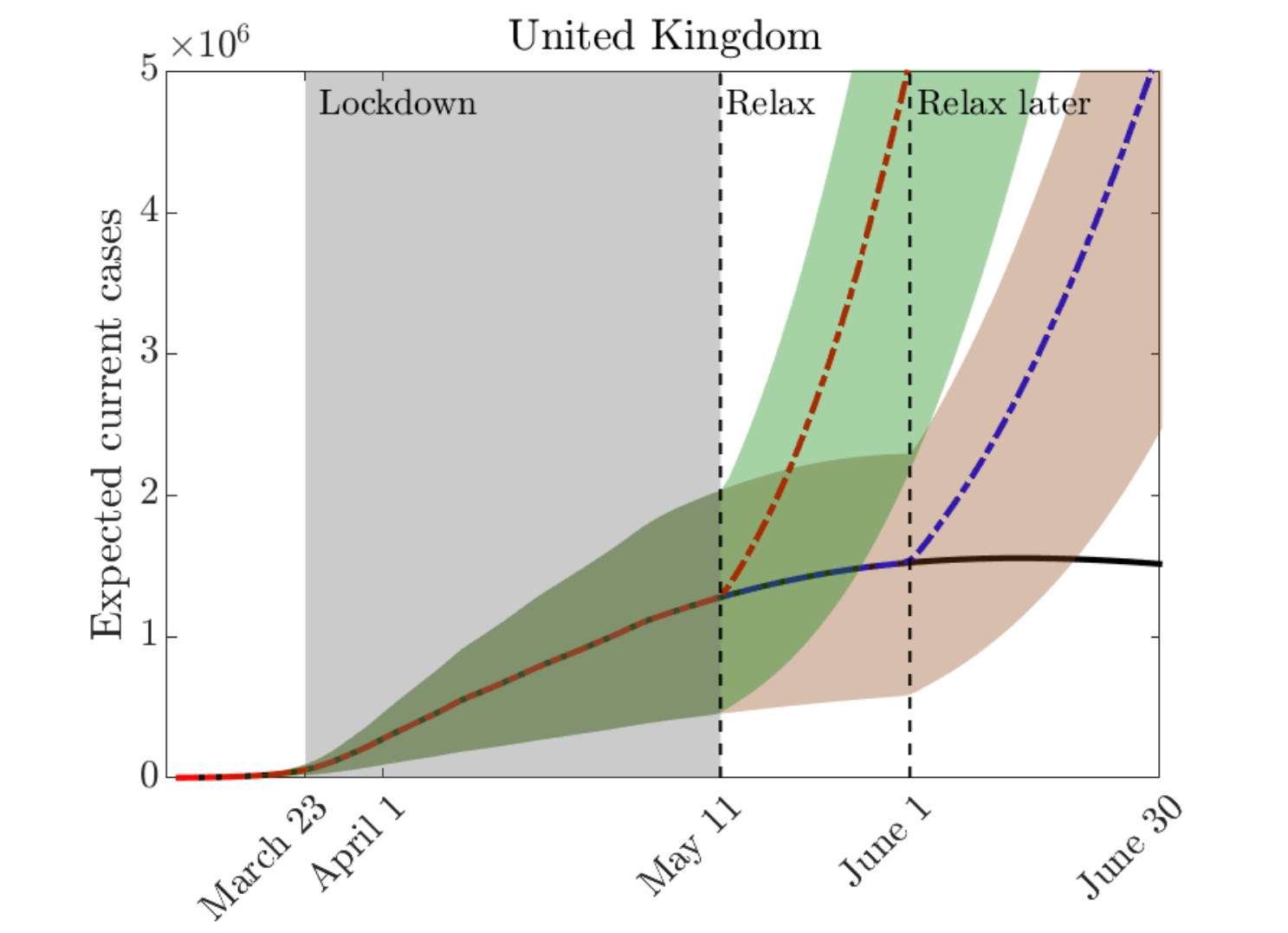}\hskip -.5cm
\includegraphics[scale =.4]{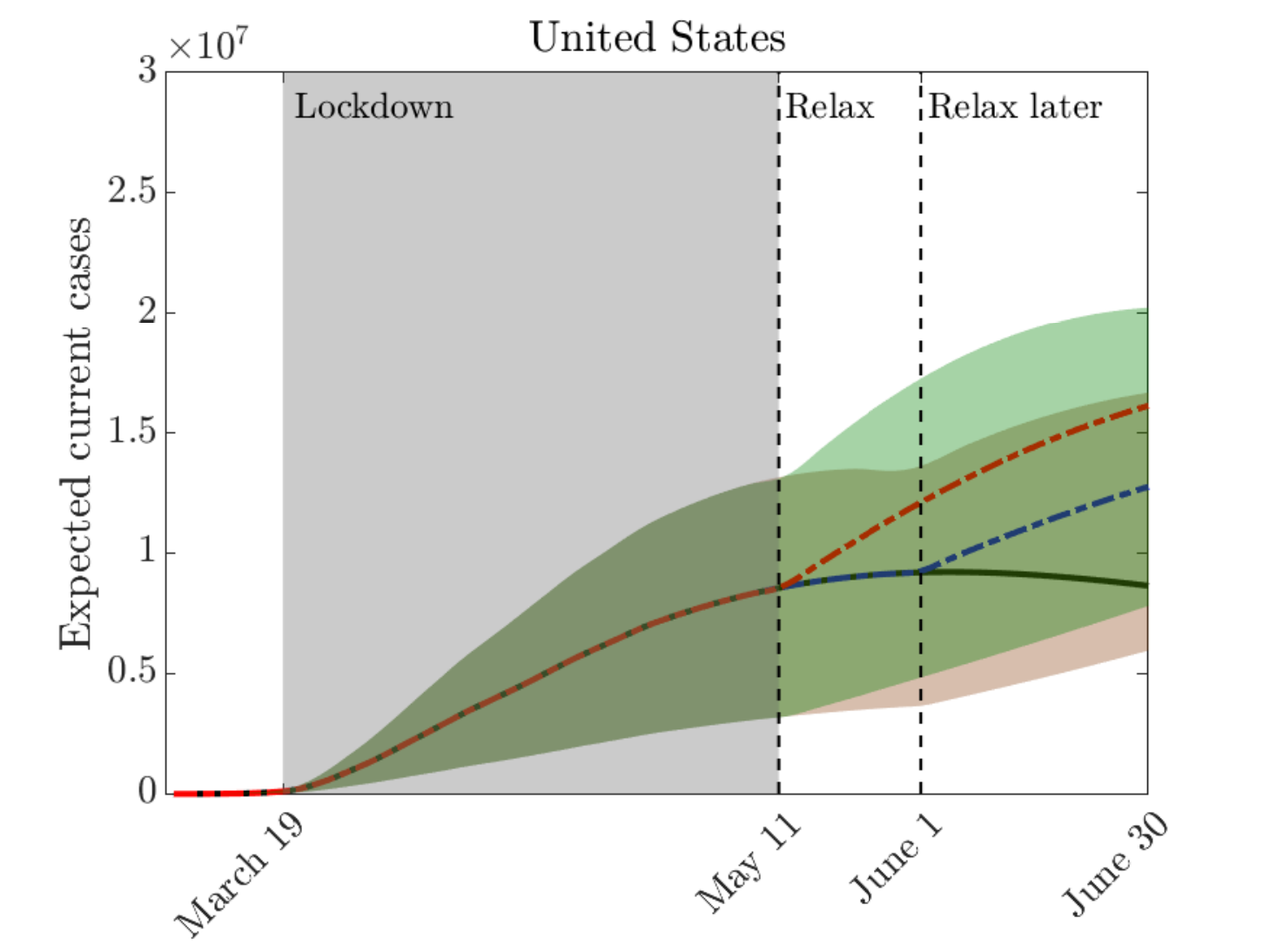}
\caption{{\bf Scenario 1:} Effect on releasing containment measures in various countries at two different times. In all countries after lockdown we assumed a reduction of individual controls on the different activities by $20\%$ on family activities, $35\%$ on work activities and $30\%$ on other activities by keeping the lockdown over the school.}
\label{fig:release1}
\end{figure}

\subsubsection{Scenario 1: Relaxing lockdown measures at different times}
In the first scenario we analyze the effects on each country of the same relaxation of the lockdown measures at two different times. The first date is country specific accordingly to current available informations, the second is June 1st for all countries.
For all countries we assumed a reduction of individual controls on the different activities by $20\%$ on family activities, $35\%$ on work activities and $30\%$ on other activities without changing the control over the school. The behaviors of the curves of infected people together with the relative $95\%$ confidence bands are reported in Figure \ref{fig:release1}. 

The results show well the substantial differences between the different countries, with a situation in the UK and US that highlight that the relaxation of lockdown measures could lead to a resurgence of the infection. On the contrary, Germany and, to some extent Spain, were in the most favorable situation to ease the lockdown without risking a new start of the infection.

\subsubsection{Scenario 2: Impact of school and work activities}\label{scenario2}

In order to highlight the differences in the infection dynamics according to the choices related to specific activities, such as school and work, we have considered the effects of a specific lockdown relaxation in these directions. Precisely for each country we have identified a range for such loosening which gives an indication of the maximum allowed opening of the activities before a strong departure of the infection. 

\begin{figure}
\centering
\includegraphics[scale =.3]{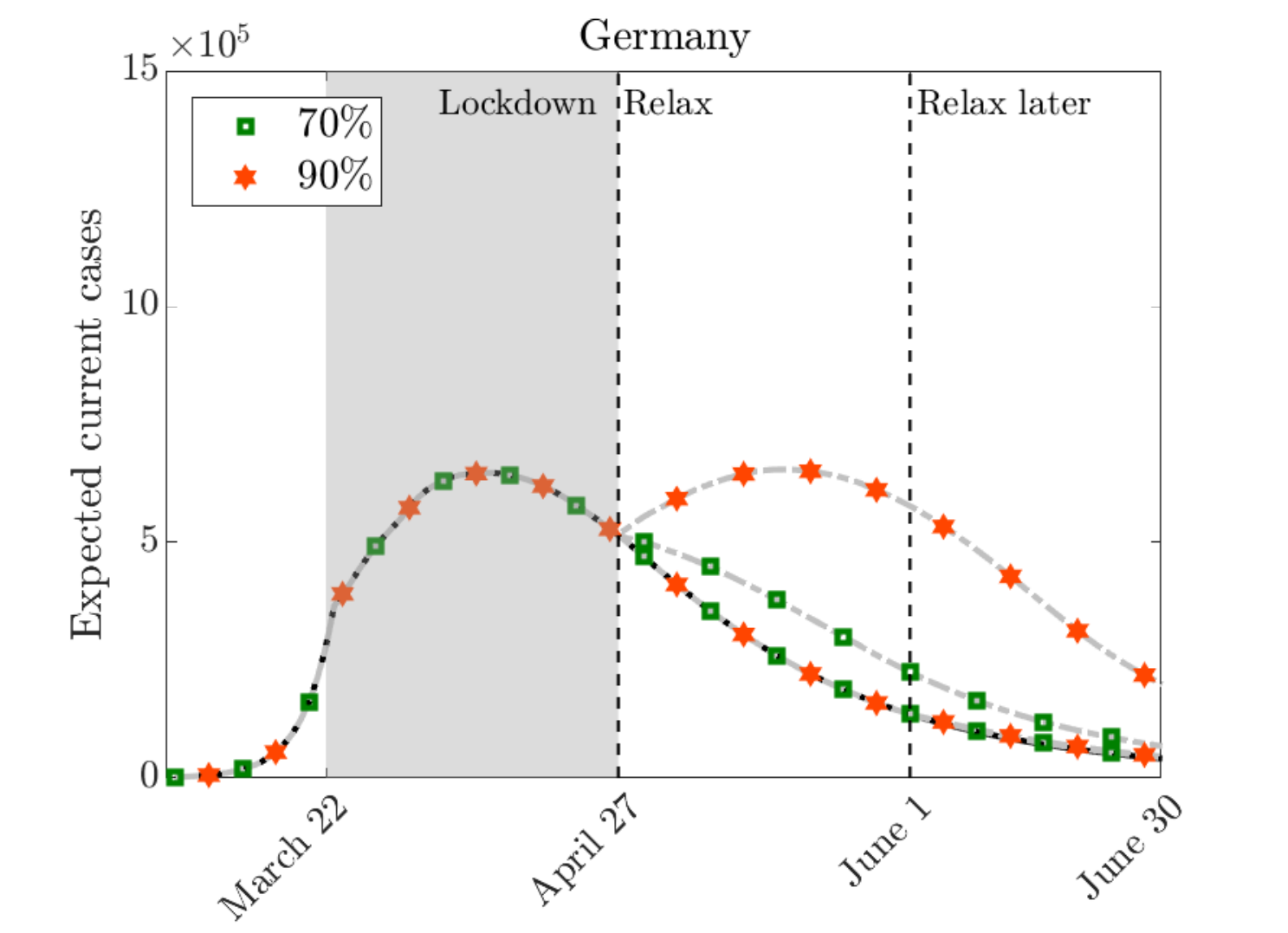}\hskip -.45cm
\includegraphics[scale =.3]{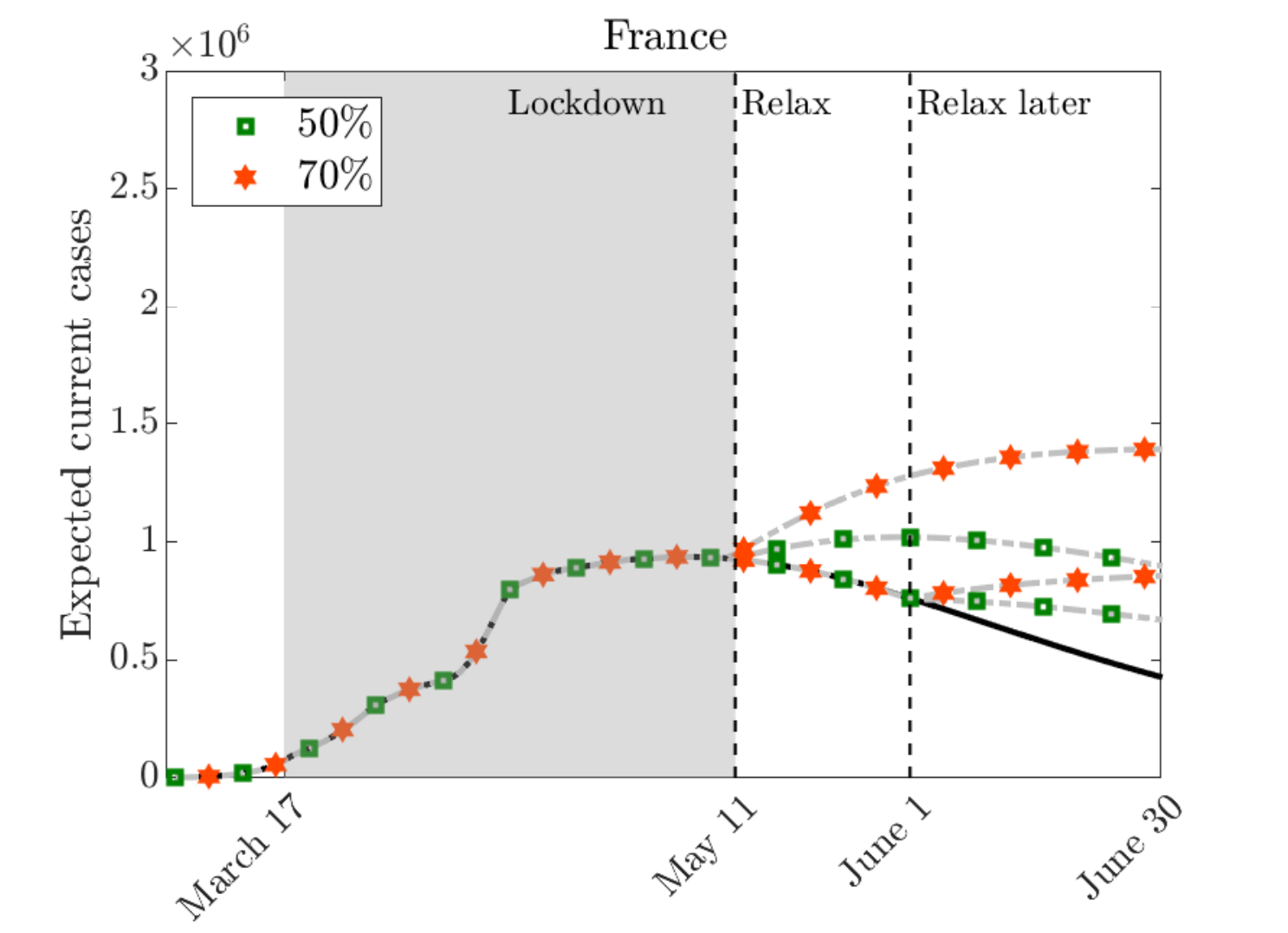}\hskip -.45cm
\includegraphics[scale =.3]{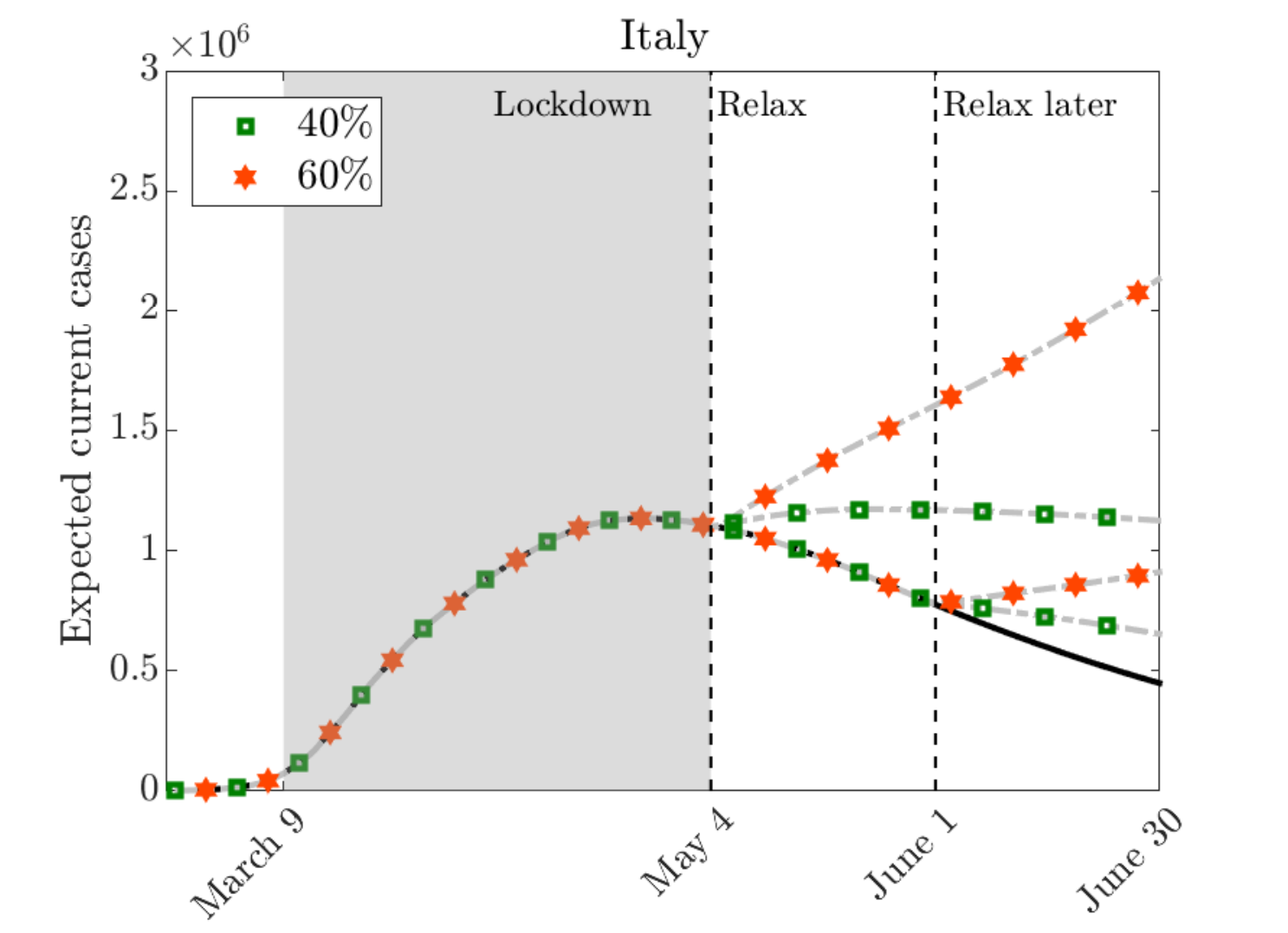}\hskip -.45cm \\
\includegraphics[scale =.3]{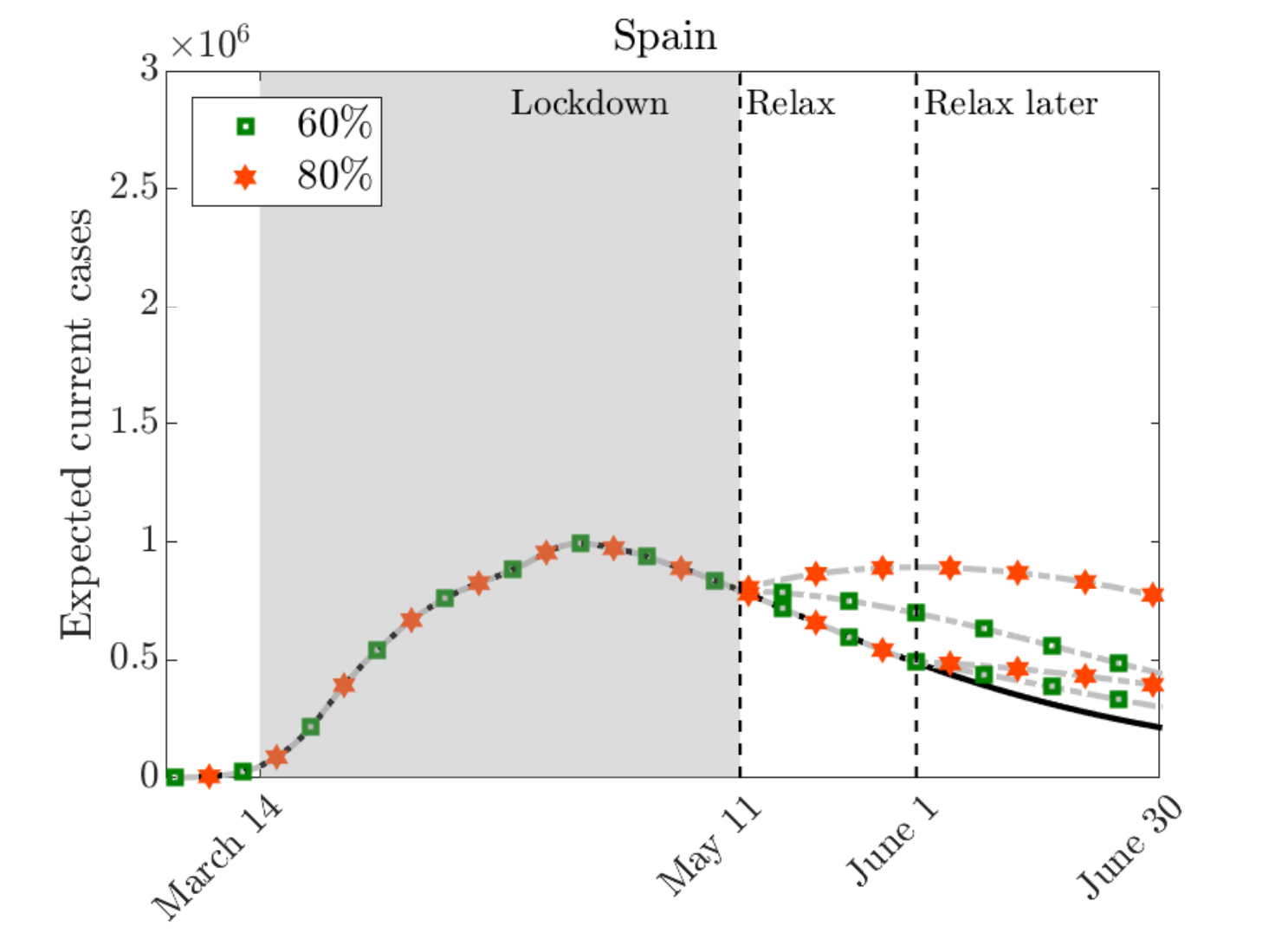}\hskip -.45cm
\includegraphics[scale =.3]{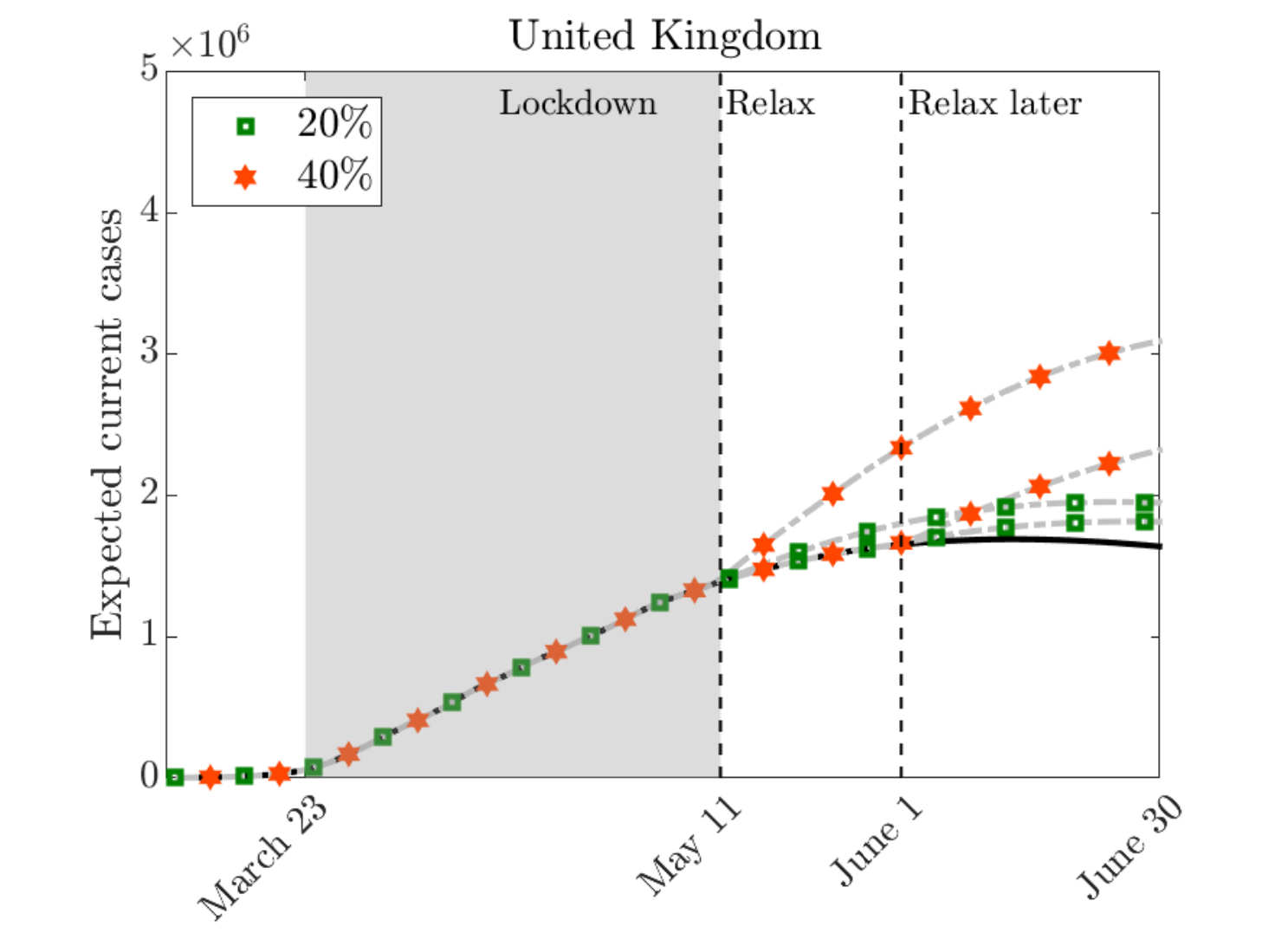}\hskip -.45cm
\includegraphics[scale =.3]{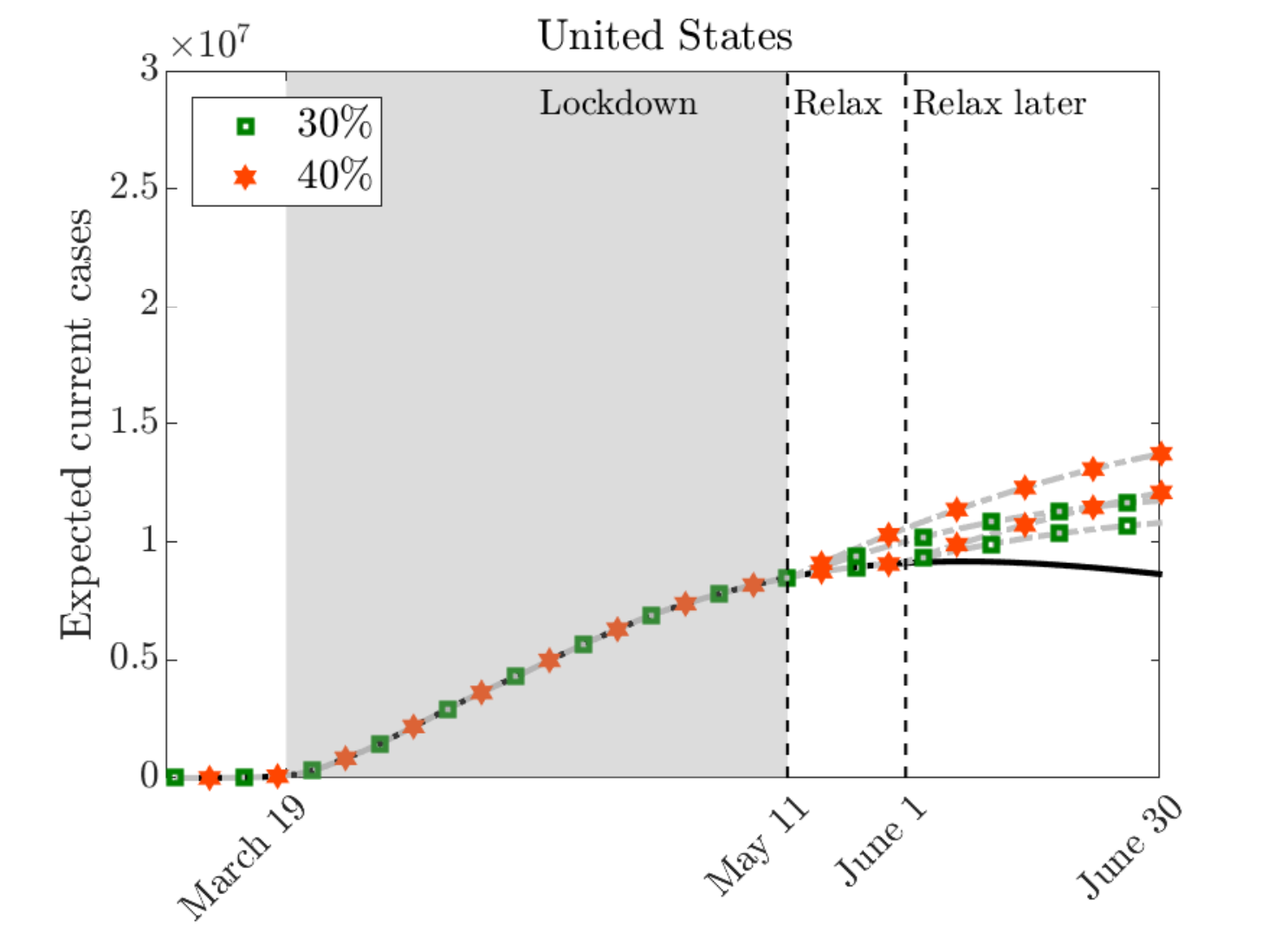}\hskip -.45cm
\caption{{\bf Scenario 2 - school:} Effect on releasing containment measures for school activities in various countries at two different times. Family, work and other activities are relaxed by $5\%$ for each $10\%$ release of the school activity. }
\label{fig:release2}
\end{figure}

\begin{figure}
\centering
\includegraphics[scale =.3]{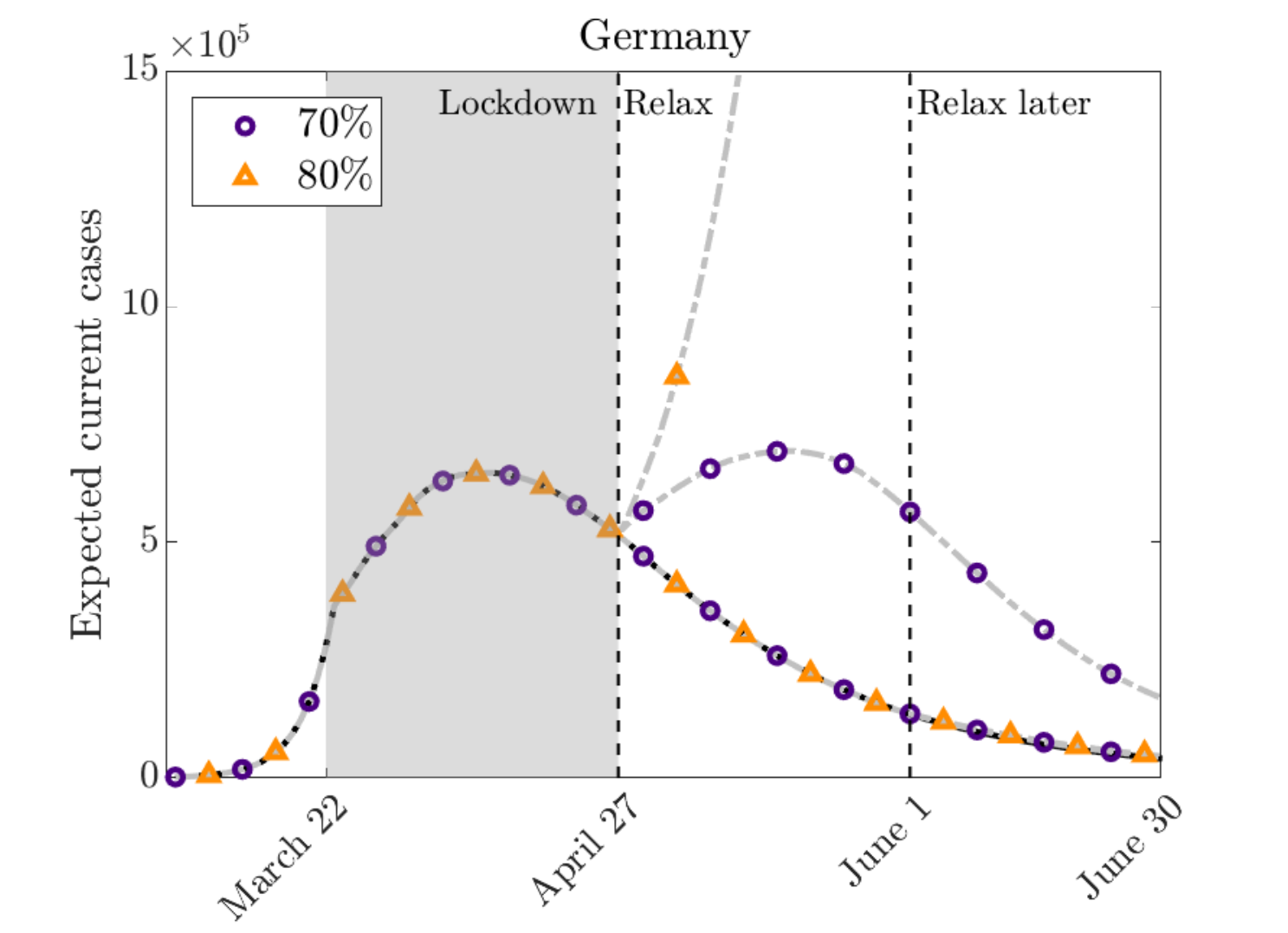}\hskip -.45cm
\includegraphics[scale =.3]{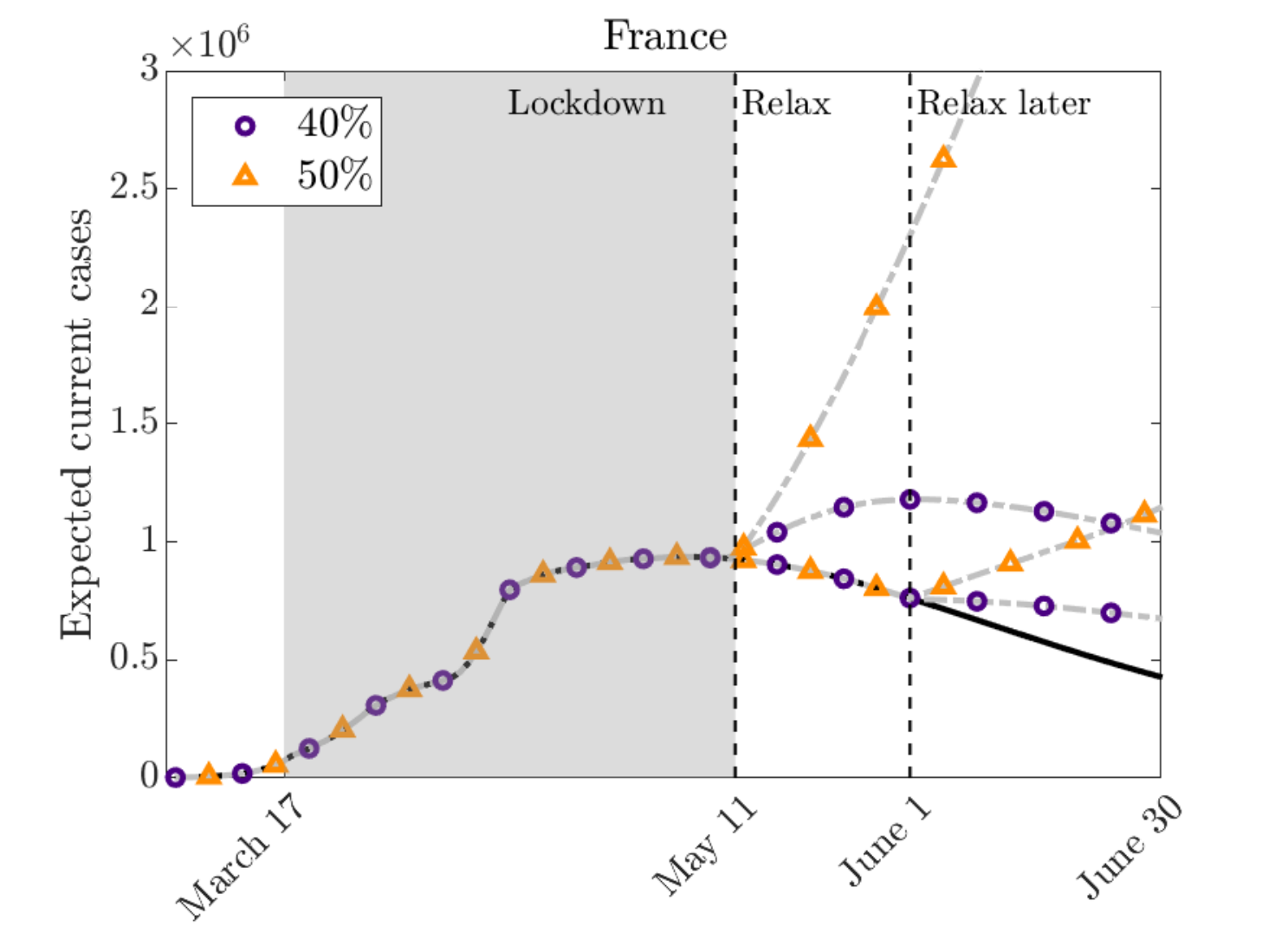}\hskip -.45cm
\includegraphics[scale =.3]{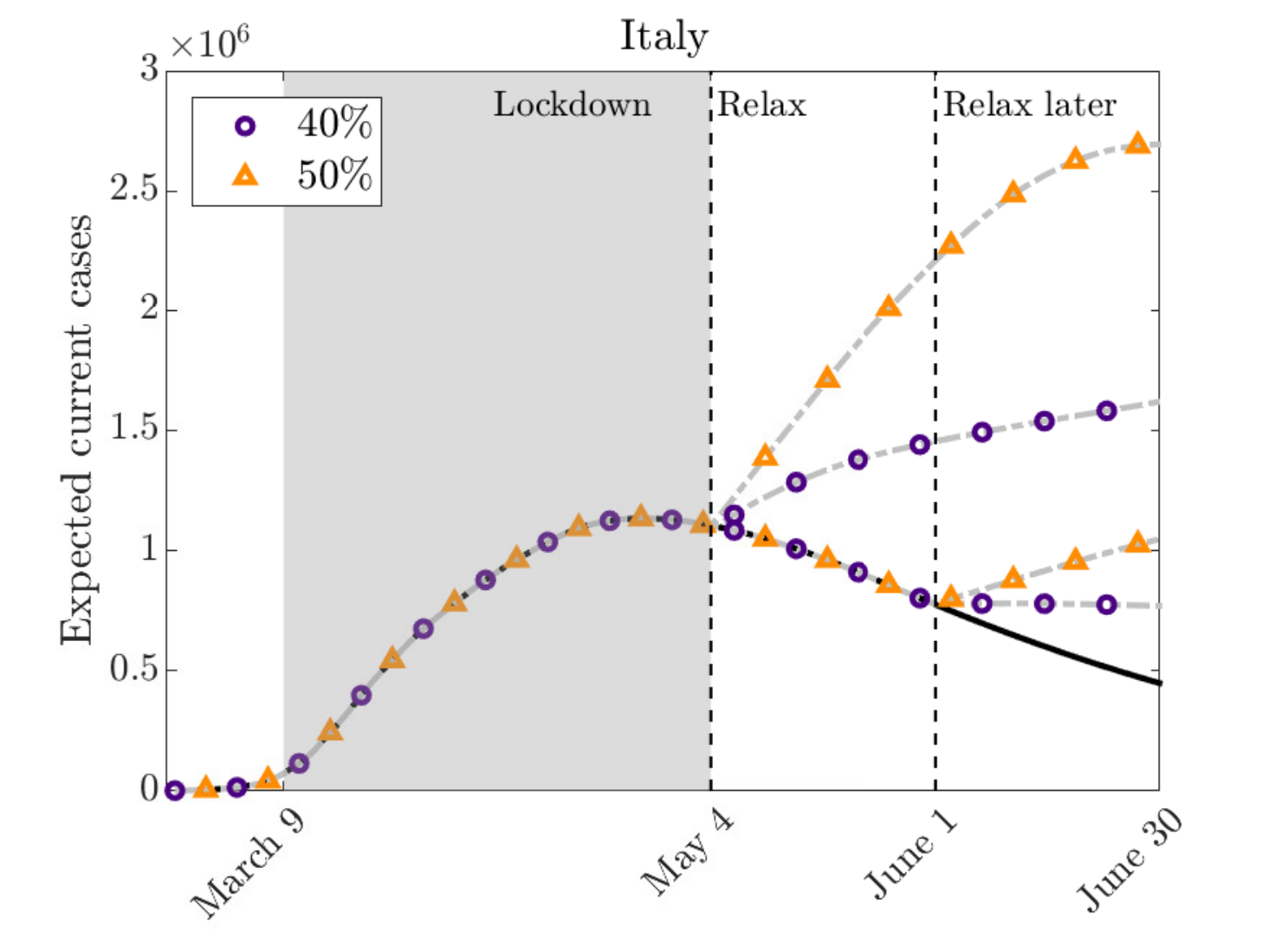}\hskip -.45cm \\
\includegraphics[scale =.3]{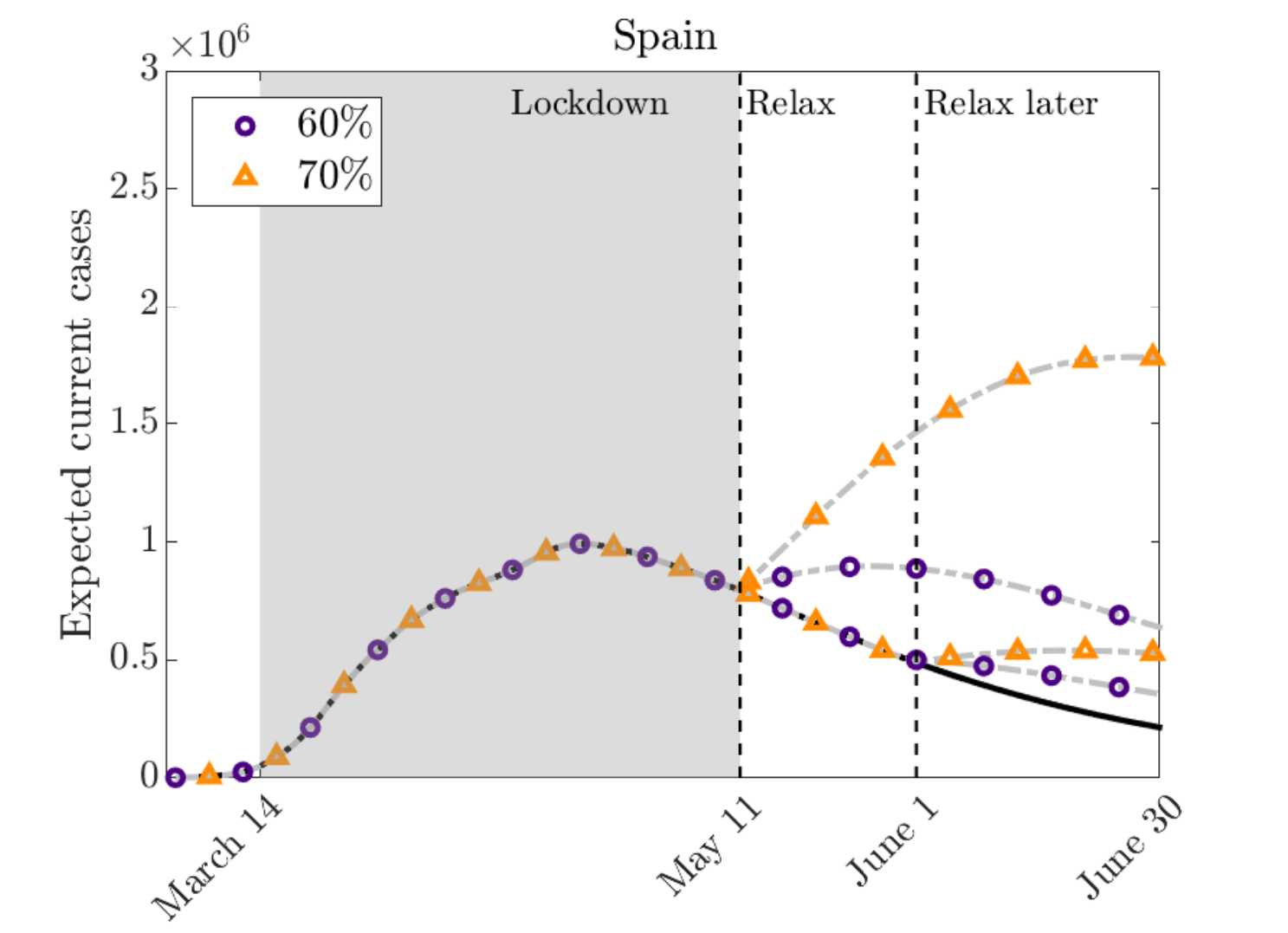}\hskip -.45cm
\includegraphics[scale =.3]{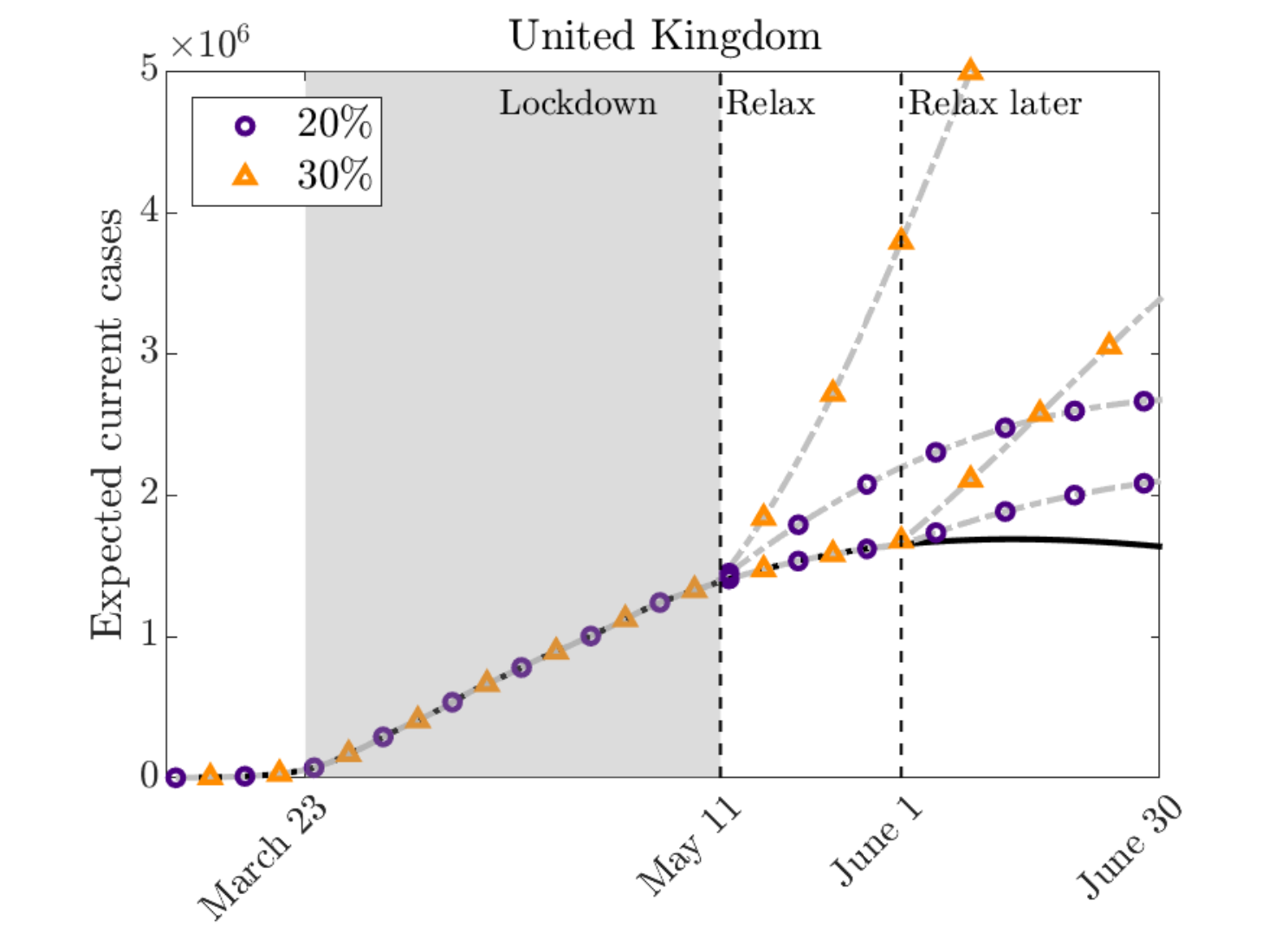}\hskip -.45cm
\includegraphics[scale =.3]{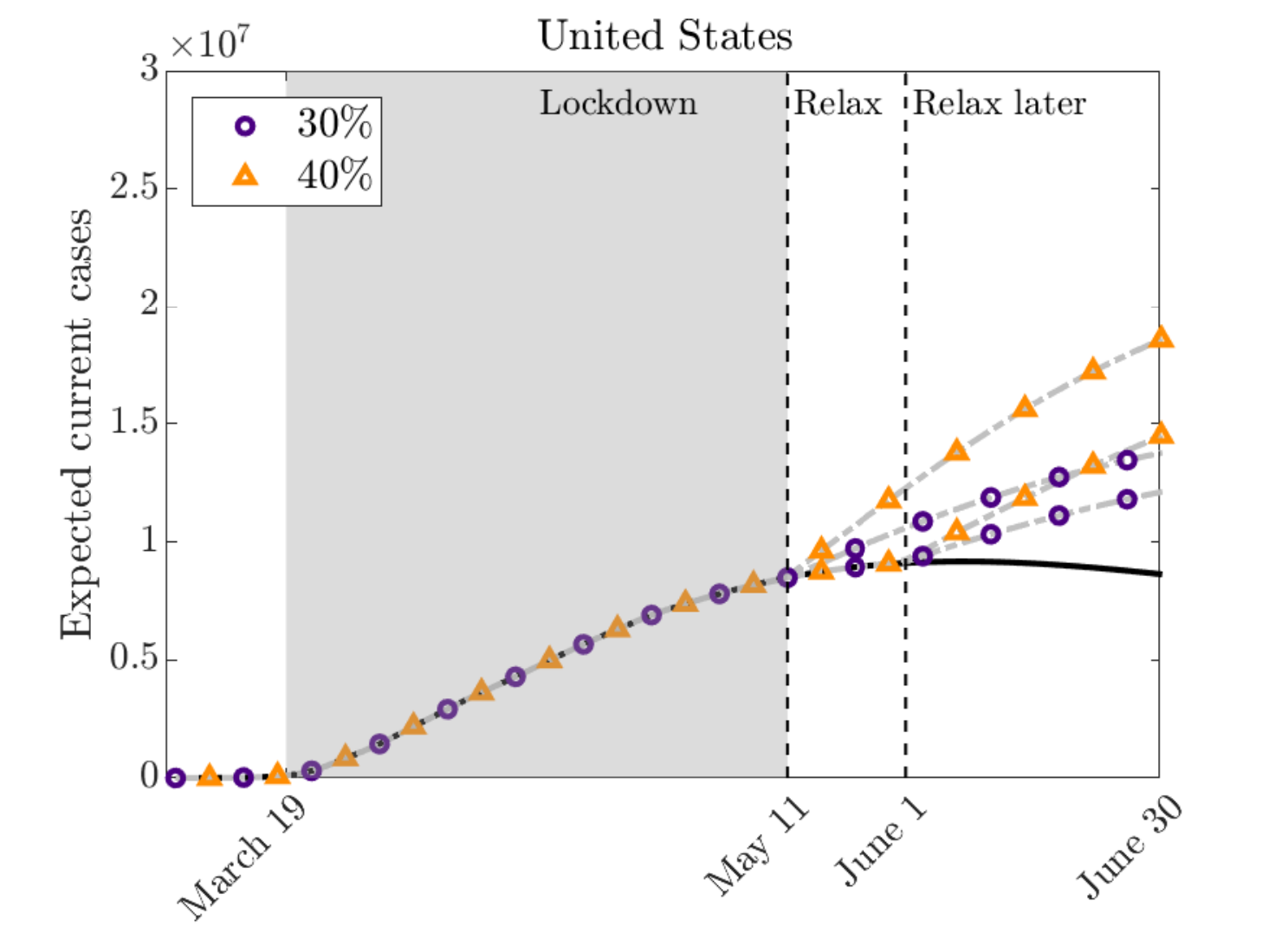}\hskip -.45cm
\caption{{\bf Scenario 2 - work:} Effect on releasing containment measures for productive activities in various countries at two different times. School is kept in lockdown. Family, and other activities are relaxed by $5\%$ for each $10\%$ release of the productive activity.}
\label{fig:release3}
\end{figure}

It was assumed to relax the lockdown of the school with a mild resumption of family, work and other activities interactions by $5\%$ for each $10\%$ release of the school. The results are reported in Figure \ref{fig:release2}.
Next, we perform a similar relaxation process oriented towards productive activities with a reduction of control on such activities at various percentages. Here we assumed no impact on school activities and a mild impact on family and other activities with a loosening at $5\%$ for each $10\%$ release of the work. The results are given in Figure \ref{fig:release3}. In both cases, the results show different infection dynamics in the selected countries as a consequence of the relaxation of lockdown policies. In particular, in the UK and USA any relaxation could determine a strong restart of the epidemic.

\subsubsection{Scenario 3: Restarting activities while keeping the curve under control}

One of the major problems in the application of very strong containment strategies, like lockdown measures, is the difficulty in maintaining them over a long period, both for the economic impact and for the impact on the population from a social point of view.

The results presented in Section \ref{scenario2} that the impact of relaxation policies may strongly differ one country from another. 
 
\begin{figure}
\centering
\includegraphics[scale =.4]{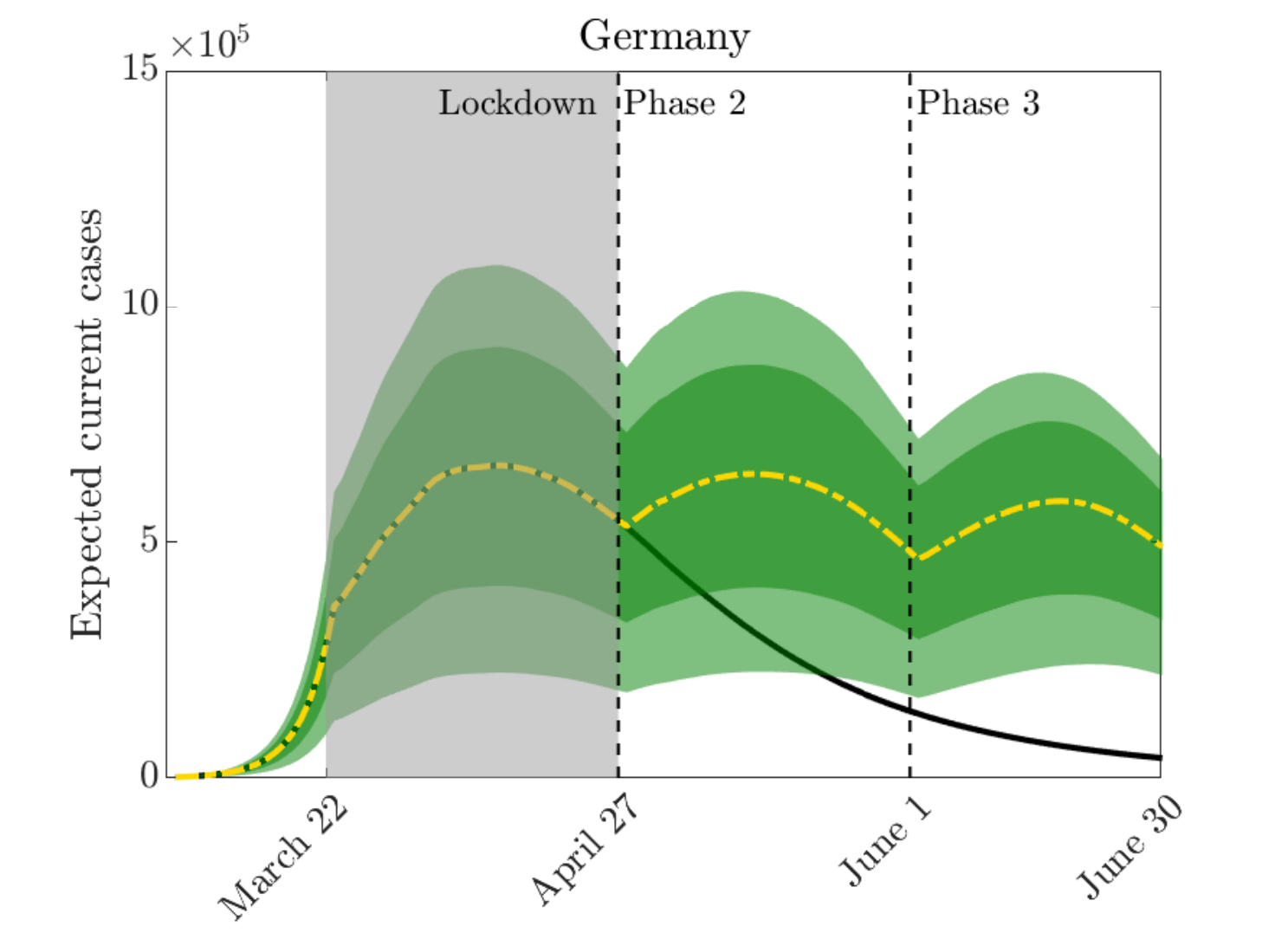}\hskip -.5cm
\includegraphics[scale =.4]{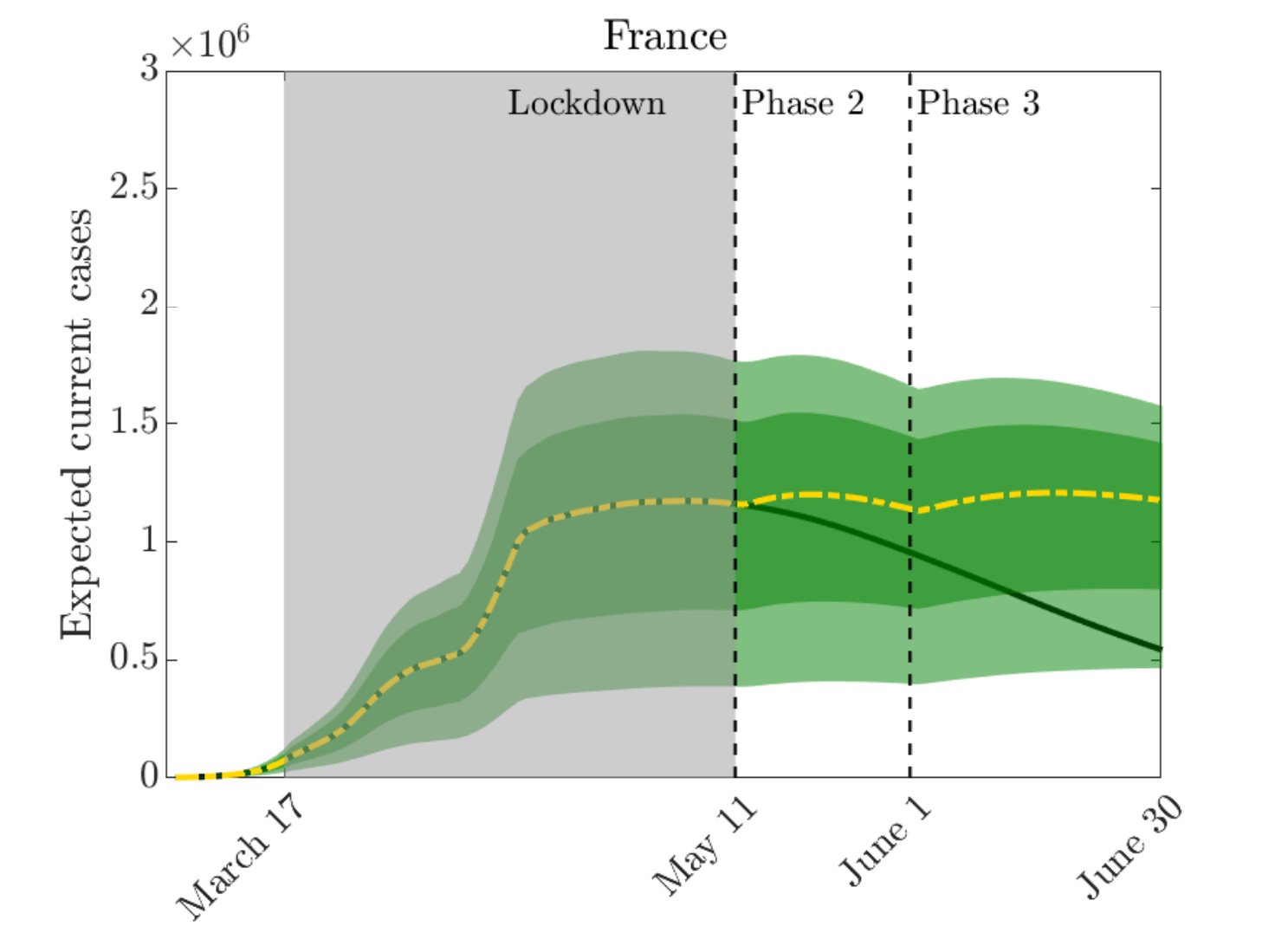}\\
\includegraphics[scale =.4]{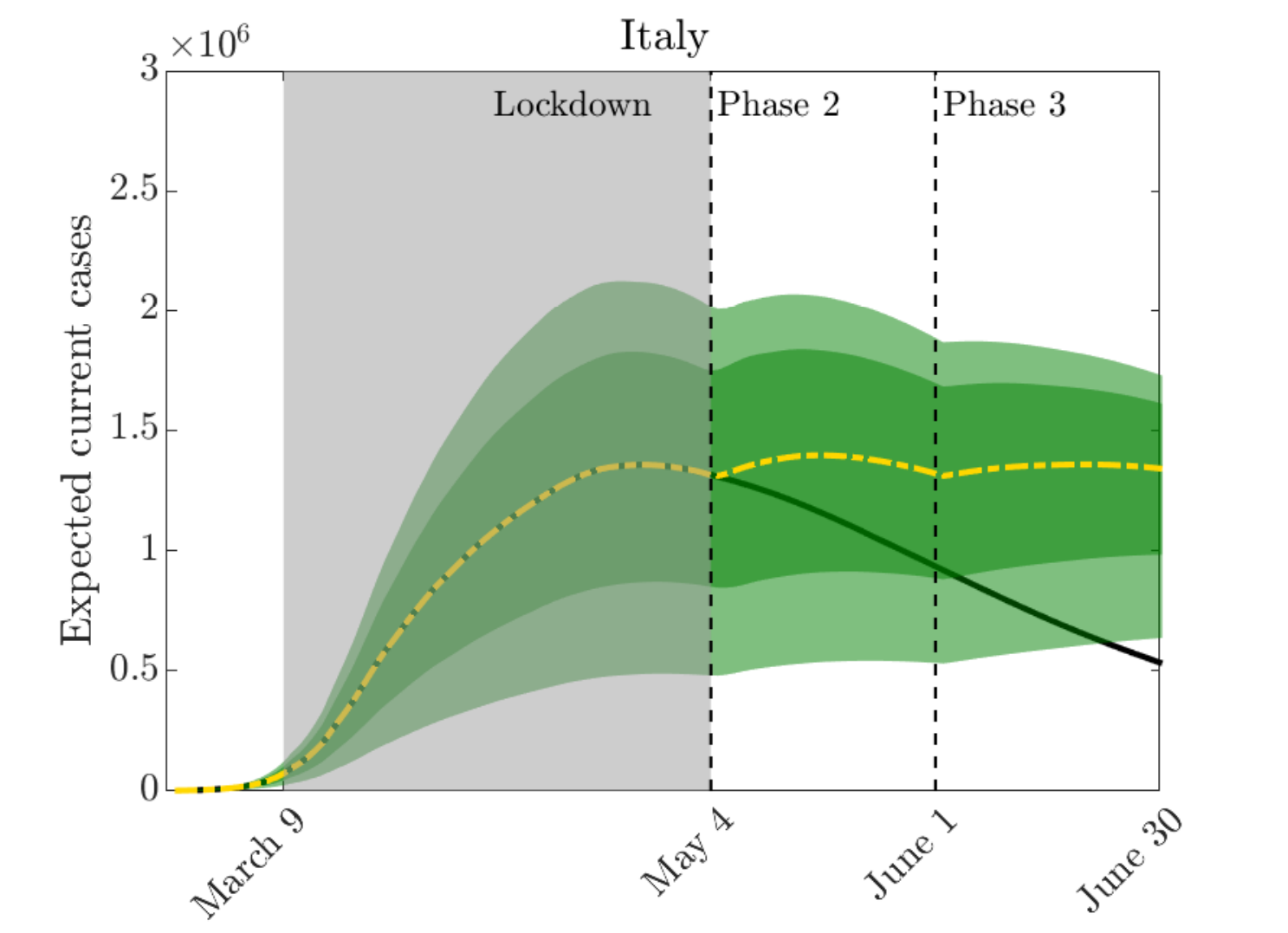} \hskip -.5cm
\includegraphics[scale =.4]{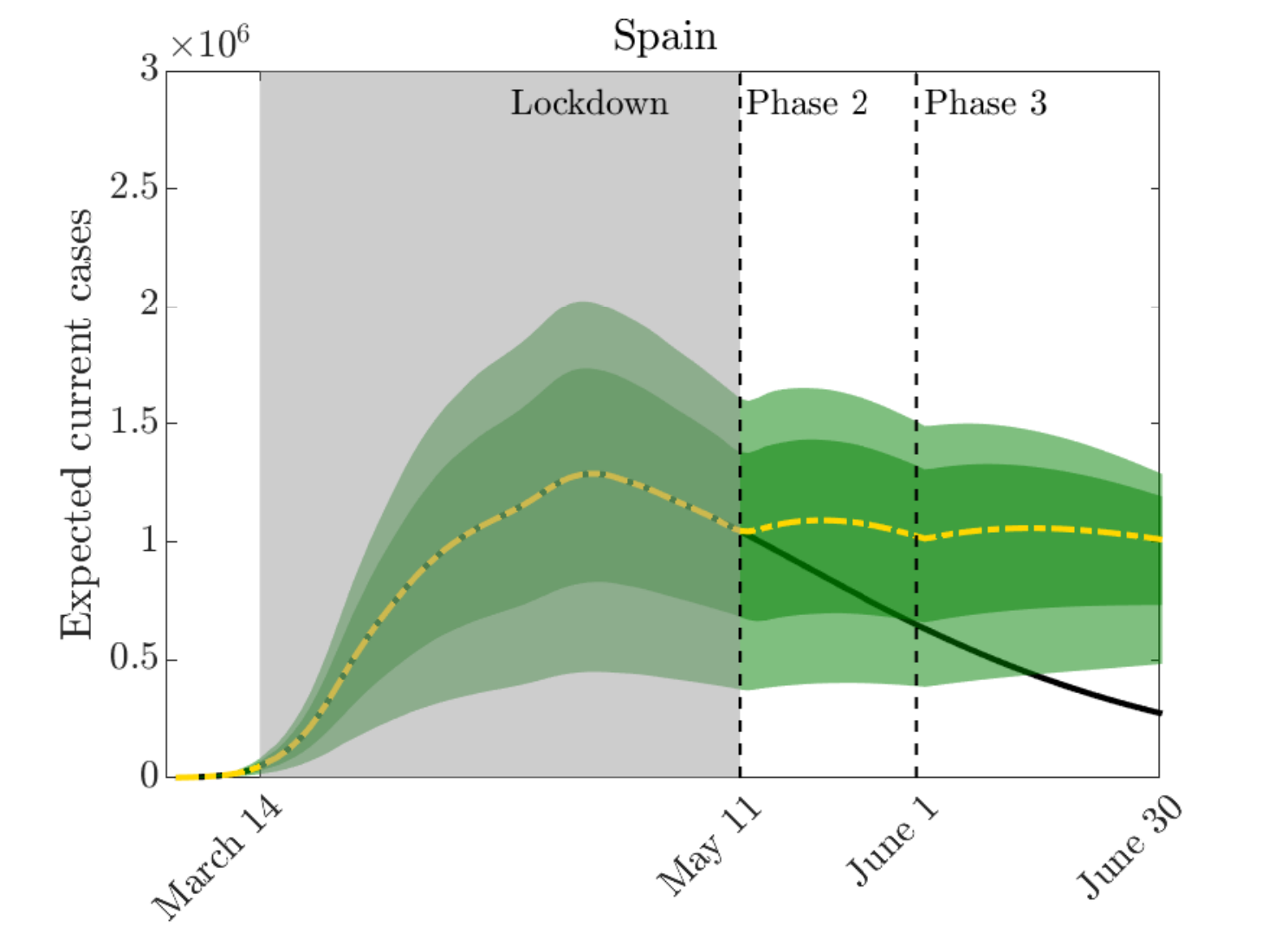}
\caption{{\bf Scenario 3:} Relaxing lockdown measures in a progressive way in two subsequent phases while keeping the epidemic peak under control. In the second phase only productive activities are restarted and partially home interactions and other activities. In a third phase school activities are also partially reopened (see Table \ref{tab:relaxpeak}).}
\label{fig:relaxpeak}
\end{figure}

 In this latter scenario, we consider a strategy based on a two-stage opening of the blocking measures with a progressive approach. This possibility is analysed for the four countries where the infection curve appears less sensitive to relaxation policies, i.e. Germany, Spain, France, and Italy. 
For each country we have selected a progressive lockdown relaxation focused mainly on the opening of productive activities in the second phase and with a partial reprise of school activities in the third phase. The reduction of the controls are now country specific and the values are reported in Table \ref{tab:relaxpeak}. In Figure \ref{fig:relaxpeak} we plot the resulting behavior for the expected number of current infectious. The simulations show that for all these countries, the relaxation of containment measures was possible while keeping the infection curve under control. However, timing and intensity of the relaxation choices play a fundamental rule in the process.

\begin{table}[t]
\begin{center}
\begin{tabular}{c | c | c |  c | c}
& $\underset{\textrm{Phase 2 - Phase 3}}{\textrm{Germany}}$ & $\underset{\textrm{Phase 2 - Phase 3}}{\textrm{France}}$& $\underset{\textrm{Phase 2 - Phase 3}}{\textrm{Italy}}$ & $\underset{\textrm{Phase 2 - Phase 3}}{\textrm{Spain}}$ \\ 
\hline\hline
& & & & \\[-.3cm]
Home & $30\%$-$60\%$ & $10\%$-$25\%$ & $10\%$-$15\%$ & $20\%$-$25\%$\\
School & $\,\,\,0\%$-$60\%$ & $\,\,\,0\%$-$30\%$ & $\,\,\,0\%$-$20\%$ & $\,\,\,0\%$-$40\%$\\
Work & $70\%$-$80\%$ & $35\%$-$45\%$ & $40\%$-$50\%$ & $60\%$-$70\%$\\
Other & $30\%$-$60\%$ & $10\%$-$25\%$ & $10\%$-$20\%$ & $20\%$-$45\%$
\end{tabular}
\end{center}
\caption{{\bf Scenario 3:} Progressive relaxation of lockdown measures for different countries as specific control reduction percentages. Results are reported in Figure \ref{fig:relaxpeak}.}
\label{tab:relaxpeak}
\end{table}

\section{Conclusions}

In order to contain epidemic dynamics, it is essential to have models capable of describing the impact of non pharmaceutical interventions, such as lockdown policies, based on specific social characteristics of the country and the containment actions implemented. In this work, aware of the complexity of the problem, we have tried to provide a suitable modeling context to describe possible scenarios in this direction. More precisely, with the aid of compartmental models incorporating specific feedback controls on social interactions capable to describe the selective action of a government in opening certain activities such as home, work, school and other activities, we can simulate their impact with respect to the epidemic trend. In particular, in an effort to take into account the high uncertainty in the data, the model has been formalized in the presence of uncertain input parameters that allow to explore hypothetical scenarios with appropriate confidence bands. 
Applications to the first wave of the COVID-19 pandemic to different countries, including Germany, France, Italy, Spain, the United Kingdom and the United States, has been considered. 
The results, in accordance with the observations, show situations with different levels of sensitivity to a hypothetical reopening of certain activities
Further studies are being conducted on geographical dependence through spatial variables. This would make it possible to characterize control measures on a local rather than global basis.

\section*{Acknowledgements}
This work has been written within the
activities of GNFM and GNCS groups of INdAM (National Institute of
High Mathematics).
G. Albi and L. Pareschi acknowledge the support of MIUR-PRIN Project 2017, No. 2017KKJP4X “Innovative numerical methods for evolutionary partial differential equations and applications” and G. Albi partial support of RIBA 2019, No. RBVR199YFL ``Geometric Evolution of Multi Agent Systems". M. Zanella was partially supported by the MIUR - ``Dipartimenti di Eccellenza'' Program (2018-2022) -- Department of Mathematics ``F. Casorati'', University of Pavia.

\end{document}